\documentclass[onecolumn]{IEEEtran}

 \usepackage{amsfonts,amsmath,amssymb,amsthm, graphicx}
\usepackage{setspace}

\usepackage{graphicx}
\usepackage{subfigure}

\usepackage{algorithmic}
\usepackage{algorithm}
\usepackage{cite}
\usepackage{url}
\usepackage{epstopdf}
\usepackage{enumitem}
\usepackage{algorithm}

\usepackage{color}
\usepackage{multicol}
\usepackage{multirow}
\usepackage{xspace}
\usepackage{comment}
\usepackage{color}
\usepackage{textcomp}

\newtheorem{theorem}{Theorem}
\newtheorem{lem}[theorem]{Lemma}
\newtheorem{prop}[theorem]{Proposition}
\newtheorem{corollary}[theorem]{Corollary}
\newtheorem{assumption}{Assumption}
\newtheorem{definition}{Definition}

\newcommand{\xref}[1]{\S\ref{#1}}

\newcommand{\OurAlg}{{\sf ColoEDR}\xspace}
\newcommand{\ouralg}{{\sf ColoEDR}\xspace}
\newcommand{\ouralgvdr}{{\sf ColoEDR}\xspace}

\newcommand{\pone}{$\mathsf{SCM}$\xspace}
\newcommand{\edr}{EDR\xspace}
\newcommand{\edrp}{EDR\xspace}
\newcommand{\lse}{LSE\xspace}

\newcommand{\pb}{p(\mathbf{b})}

\newcommand{\mz}{\frac{\partial^+ c_n(z)}{\partial z}}

\newcommand{\lmc}{\frac{\partial^- c_n(s_n)}{\partial s_n}}
\newcommand{\rmc}{\frac{\partial^+ c_n(s_n)}{\partial s_n}}
\newcommand{\mrmc}{\frac{\partial^+ \hat{c}_n(s_n)}{\partial s_n}}
\newcommand{\brange}{\frac{1}{2}\left(\frac{\alpha\delta}{N} + \sqrt{\frac{\alpha\delta}{N}(\frac{\alpha\delta}{N} + 4\Sigma_{m\ne n}b_m)}\right)}

\newcommand{\eps}{\varepsilon}

\usepackage{ifthen}
\newboolean{showcomments}
\setboolean{showcomments}{true}
\newcommand{\adam}[1]{\ifthenelse{\boolean{showcomments}}
{ \textcolor{red}{(Adam says:  #1)}}{}}
\newcommand{\shaolei}[1]{\ifthenelse{\boolean{showcomments}}
{ \textcolor{red}{(Shaolei says:  #1)}}{}}
\newcommand{\niangjun}[1]{\ifthenelse{\boolean{showcomments}}
{ \textcolor{red}{(Niangjun says:  #1)}}{}}
\newcommand{\xiaoqi}[1]{\ifthenelse{\boolean{showcomments}}
{ \textcolor{red}{(Xiaoqi says:  #1)}}{}}
\newcommand{\addcites}[0]{\ifthenelse{\boolean{showcomments}}
{ \textcolor{green}{(add citation(s))}}{}}
\newcommand{\addref}[0]{\ifthenelse{\boolean{showcomments}}
{ \textcolor{green}{(add ref)}}{}}
\newcommand{\todo}[1]{\ifthenelse{\boolean{showcomments}}
{ \textcolor{red}{(To do:  #1)}}{}}

\DeclareMathOperator*{\argmin}{arg\,min}
\DeclareMathOperator*{\argmax}{arg\,max}

  \renewcommand{\baselinestretch}{.95}

  \makeatletter

  \def\thebibliography#1{%
     \section[References]{{References} 
         \@mkboth{{\refname}}{{\refname}}%
     }%
     \list{[\arabic{enumi}]}{%
         \settowidth\labelwidth{[#1]}%
         \leftmargin\labelwidth
         \advance\leftmargin\labelsep
         \parsep=0pt\itemsep=1pt 
         \usecounter{enumi}
     }%
     \let\newblock\@empty
     \raggedright 
     \sloppy
     \sfcode`\.=1000\relax
  }

\begin{document}

\date{}


\title{ \huge Greening Multi-Tenant Data Center Demand Response}
\author{Niangjun Chen, Xiaoqi Ren, Shaolei Ren, Adam Wierman}


\maketitle

\abstract

Data centers have emerged as promising resources for demand response, particularly for emergency demand response (\edr), which saves the power grid from incurring blackouts during emergency situations. However, currently, data centers typically participate in \edr by turning on backup (diesel) generators, which
is both expensive and environmentally unfriendly.
In this paper, we focus on ``greening'' demand response in multi-tenant data centers, i.e., colocation data centers, by designing a pricing mechanism through which the data center operator can efficiently extract load reductions from tenants during emergency periods to fulfill energy reduction requirement for \edr.  In particular, we propose a pricing mechanism for both mandatory and voluntary \edr programs, \ouralg, that is based on parameterized supply function bidding and provides provably near-optimal efficiency guarantees, both when tenants are price-taking and when they are price-anticipating. In addition to analytic results, we extend the literature on supply function mechanism design, and evaluate \ouralg using trace-based simulation studies.  These validate the efficiency analysis and conclude that the pricing mechanism is both beneficial to the environment and to the data center operator (by decreasing the need for backup diesel generation), while also aiding tenants (by providing payments for load reductions).

\section{Introduction}\label{sec:introduction}

Data centers have emerged as a promising demand response opportunity.  However, data center demand response today is not environmentally friendly since data centers typically participate by turning on backup (diesel) generators.  In this paper, we focus on designing a pricing mechanism for a crucial class of data centers for demand response -- multi-tenant colocation data centers -- that allows the data center operator to encourage load shedding among tenants in response to demand response signals; thus greening data center demand response by reducing the need for use of backup (diesel) generation.

\textbf{Data center demand response.}
Power-hungry data centers have been quickly expanding in both number and scale to support the exploding IT demand, consuming 91 billion kilowatt-hour (kWh) electricity in 2013 in the U.S. alone  \cite{NRDC_Colocation_2014}. 
While traditionally viewed purely as a negative, the massive energy usage of data centers has recently begun to be recognized as an opportunity.  In particular, because the energy usage of data centers tends to be flexible, they are promising candidates for \emph{demand response}, which is a crucial tool for improving grid reliability and incorporating renewable energy into the power grid. From the grid operator's perspective, a data center's flexible power demand serves as a valuable energy buffer, helping balance grid power's supply and demand at runtime \cite{AdamWierman_DataCenterDemandResponse_Survey_IGCC_2014}.


To this point, data center is a promising, but still largely under-utilized opportunity for demand response.  However, this is quickly changing as data centers play an increasing role in emergency demand response (\edr) programs. \edr is the most widely-adopted demand response program in the U.S.,
representing 87\% of demand reduction capabilities across all reliability regions  \cite{EDR_Market_Overview}.
Specifically,
during emergency events (e.g., extreme weather or natural disasters),
\edr coordinates many large energy consumers, including data
centers, to shed their power loads, serving as the last protection
against cascading blackouts that could potentially result in economic losses of billions of dollars \cite{pjm_emergency_demand_response_Performance,Demand_response_Evidence_Blackouts_Canada_US}.
The U.S. EPA  has identified data centers as critical resources for \edr \cite{Demand_response_US_EPA_EnerNOC}, which was
attested to by the following example: on July 22,
2011, hundreds of data centers participated in \edr by cutting their electricity
usage before a large-scale blackout would have occurred \cite{Demand_response_Evidence_Blackouts_Canada_US}.

While data centers are increasingly contributing to \edr,  they typically  participate by turning on
their on-site backup diesel generators, which is neither cost effective
nor environmentally friendly. For example, in California (a major
data center market), a standby diesel generator often produces 50-60 times more nitrogen oxides (a smog-forming pollutant) compared to a typical power plant for each kWh of electricity, and diesel particulate represents
the state's most significant toxic air pollution problem \cite{Diesel_Pollution_Reference}.

In addition, relying on diesel generation for \edr
presents emerging challenges which, if left unaddressed, may forfeit data center's \edr capability.
{First}, as \edr is becoming more frequent, the
current financial compensation offered by power grid to data centers (for committed
energy reduction during \edr) may not be enough to cover
the growing cost of diesel generation.
{Second}, data center operators are aggressively cutting the huge capital
investment in their power infrastructure (e.g., 10-15\$/watt \cite{Hoelzle_datacenter_book_2013,LimKansalLiu_ATC2011}),
by down-sizing the capacity
of diesel generator and uninterrupted power supply (UPS) system \cite{Wang:2014:UBP:2541940.2541966}.
Such under-provisioning of diesel generator may compromise data center's \edr capability.
Therefore, to retain and encourage data center's participation in \edr without contaminating the environment, it is critical and urgent that data centers seek alternative
ways to shed load.

Consequently, modulating
server energy for green \edr (as well as other demand response programs such as regulation service \cite{Xiaorui_2013data_frequency_regulation})
has received an increasing amount
of attention in recent years, e.g., \cite{AksanliRosing14_ProvidingRegulationServicesManagingDataCenterPeakPower,
Chen_PowerControl_Regulation_CDC_2013,aikema2012data_ancillary_IGCC_2012,
Liu:2014:PDC:2591971.2592004,hamed_datacenter_ancillary_smartgridcom_2012,Xiaorui_2013data_frequency_regulation,
DataCenterDemandResponsePreliminary_Feedback_2012,AdamWierman_DataCenterDemandResponse_Survey_IGCC_2014}.
These studies leverage various  widely-available IT computing knobs
(e.g., server turning on/off and workload migration)
in data centers and provide algorithms to optimize them for participation in demand response markets. Importantly, these are not simply theoretical studies.  For example, a field study by Lawrence Berkeley National Laboratory (LNBL) has illustrated that data centers can reduce energy consumption by 10-25\% in response to demand response signals, without noticeably impacting data center's normal operation \cite{DataCenterDemandResponse_Report_Berkeley}.

\textbf{Demand response in collocation data centers.}
While existing studies on data center demand response show promising progress, they are primarily focused on owner-operated data centers (e.g., Google) whose operators have full control over both servers and facilities.  
Unfortunately, such companies may actually be the least likely to participate in demand response programs, because many of their workloads are extremely delay sensitive and their data centers have been optimized for delay.

In this paper, we focus on another type of data centers ---
multi-tenant colocation data centers (e.g., Equinix).  These have been investigated much less frequently, but are actually better targets for demand response then owner-operated data centers. In a colocation data center (simply called ``colocation'' or ``colo''),
multiple tenants deploy and keep full control of their own physical servers in a shared space, while the colo operator only provides facility support (e.g., high-availability power and cooling). Colos are less studied than owner-operated data centers, but they are actually more common in practice.  Colos offer
data center solutions to many industry sectors, and serve as physical home to many private clouds,
medium-scale public clouds (e.g., VMware) \cite{colocation_cloud_in_SuperNAP_Switch_LasVegas_2014},
and content delivery providers (e.g., Akamai).  Further, a recent study shows that colos consume nearly 40\% data center energy in the U.S.,
while Google-type data centers collectively account for less than 8\%, with the remaining
going to enterprise in-house data centers \cite{NRDC_Colocation_2014}.

In addition to consuming a significant amount of energy (more than Google-type data centers), colos are often located in places more useful for demand response. While many
mega-scale owner-operated data centers are built in rural areas, 
colos are mostly located in metropolitan areas (e.g.,
Los Angeles, New York) \cite{colocation_usa_datacentermap},
which are the very places where \edr
is most needed. Further, workloads in colos are highly heterogenous, and
many tenants run non-mission-critical workloads (e.g., lab computing \cite{Colocation_Symantec_Megawwat_Lease_2015}) that have very high scheduling flexibilities, different delay sensitivities, peak load periods, etc., which is ideal for demand response participation.

For all these reasons, colos are key participants in \edr programs.
Compared to owner-operated data centers that can leverage various computing knobs, however,
greening colos' participation in \edr by reducing reliance on diesel generator is significantly more challenging,
because of colo operators' lack of control over their tenants' servers.
On the other hand, many tenants in colos run servers hosting highly-flexible
and non-critical workloads with a great potential for shedding loads when called upon \cite{Colocation_Symantec_Megawwat_Lease_2015}.
Thus, tenants' load shedding potentials, if appropriately exploited,
can altogether form a green alternative to diesel generation for colo \edr.
 Nonetheless, tenants manage their own servers independently and may not have
incentive to cooperate with the operator for \edr, thus raising
the research question: how can a colo operator \emph{efficiently} incentivize its tenants' load shedding for \edr?\footnote{Tenants receive UPS-protected power from colo operator and share
cooling systems. In other words, tenants' total energy consumption is not directly
provided by grid and includes non-separable cooling
energy, which makes tenants ineligible for direct participation in \edr \cite{pjm_emergency_demand_response_Performance}.}

\textbf{Contributions of this paper.} In this paper, we focus on ``greening'' colocation
demand response by extracting load reduction from tenants instead of relying on backup diesel generation.
We study both
\emph{mandatory} \edr, a type of \edr program in which participants
sign contracts and are obliged to reduce loads when requested \cite{pjm_emergency_demand_response_Performance}, and \emph{voluntary} \edr, where participants voluntarily reduce loads for financial compensation upon grid request.
In both cases, we propose a new pricing mechanism with which colo operators can extract load shedding from tenants.  In particular, our proposed approach, called \ouralg, can
effectively provide incentives for tenants to reduce energy consumption during \edr events, complementing (and even substituting for) the high-cost and environmentally-unfriendly diesel generation.

\ouralg works as follows. After an \edr signal arrives at the colo operator, tenants bid using a parameterized supply function, and then the colo operator announces a market clearing price which, when plugged into the bids, specifies how much energy tenants will reduce and how much they will be paid.  Participation by the tenants is straightforward, since they are required to bid only one parameter, which can be viewed as a proxy of how much flexibility in energy usage they have at that moment.  This participation can be automated and so can be easily incorporated into current practice, and mimics the way generation resources participate in electricity markets more broadly.  For example, colo operators at Verizon Terremark already communicate with tenants in preparation for an \edr event.

The main technical contribution of the paper is the analysis of the efficiency of the supply function mechanism proposed in \ouralg.  In particular, while there is a large literature studying supply function bidding \cite{johari2011,day2002, baldick2004, green1992, green1996}, our setting here is novel and different because the colo operator can either satisfy the \edr request using flexibility from the tenants (as in prior supply funding literature) or through its backup diesel generator.  Thus, the diesel generator is an outside option that allows for \emph{elasticity} in the amount of response extracted from the tenants.  Further, the colo operator can combine and balance between its two options (i.e., tenant load shedding and backup generator) in order to minimize costs.  This creates a multi-stage game and adds a considerable complexity as compared to the standard setting without an outside option, e.g., \cite{johari2011}.

Despite the added complexity, our analysis precisely characterizes the equilibrium outcome, both when tenants are price-taking and when they are price-anticipating.  In both cases, our results highlight that \ouralg suffers little performance loss compared to the socially optimal outcome, both from the operator's and the tenants' perspectives.  However, our analysis does highlight one possible drawback of \ouralg.  In the worst case, it is possible that \ouralg may result in
using significantly more on-site diesel generation than would the socially optimal.  However, this bad event occurs only in cases where one tenant has an overwhelmingly dominant amount of servers and has a unit cost (for energy reduction) just below that of on-site diesel generation.  Such an exploitation of market power is unlikely to be possible in practical multi-tenant colocation data centers.

In addition to our theoretical analysis, we investigate a case study of colocation demand response in \xref{sec:simulation} using trace-based experiments. The results further validate the design of \OurAlg, and show that it achieves the mandatory energy reduction for \edr while benefiting tenants through financial incentives and decreasing the operator's cost. Moreover, our simulation study shows that the efficiency loss in practical settings is even lower than what is suggested by the analytic bounds.  This is especially true for the amount of on-site generation, which the analytic results suggest can (in the worst-case) be significantly larger than socially optimal but in realistic settings is very close to the social optimal.



\section{Problem Formulation}\label{sec:model}

Our focus is to design a mechanism for a colo operator to extract tenant load reductions in response to to an \edr signal.   Thus, we need to begin by describing a model for a colo operator. 

Recall that the colo operator is responsible for non-IT facility support (e.g., high-availability power, cooling). We capture the non-IT energy consumption using Power Usage Effectiveness (PUE) $\gamma$, which is
the ratio of the total colocation energy consumption to the IT energy consumption. Typically, $\gamma$ ranges from 1.1 to 2.0, depending on factors such as outside temperature.

When the operator receives an \edr signal from the \lse, it has two options for satisfying the load reduction.  First, without involving the tenants, the colo operator can use its on-site backup diesel generator.\footnote{Other alternatives, e.g., battery \cite{Wang:2014:UBP:2541940.2541966}, usually
only last for $<5$ minutes. So, diesel
generation is the typical method \cite{Demand_response_US_EPA_EnerNOC}.}
We denote the amount of energy reduction by diesel generation by $y$ and
the cost per kWh of diesel generation (e.g., for fuels) by $\alpha$.

Alternatively, the colo operator could try to extract IT load reductions from the tenants.  We consider a setting where there are $N$ tenants, $i \in \mathcal{N}=\{1,2, \cdots,N\}$.
When shedding energy consumption, a tenant $i$ will incur some
costs and we denote the cost from shedding $s_i$ by a function $c_i(s_i)$.  These costs could be due to wear-and-tear, performance degradation, workload shifting, etc.  For the purposes of our model, we do not specify which technique reduces the IT load, only its cost.  For details on how one might model such costs, see \cite{ong2010impacts, fan2007power, andrew2010optimality, wierman2009power}.
A standard, natural assumption on the costs is the following.

\begin{assumption} For each $n$, the cost function $c_n(s_n)$ is continuous, with $c_n(s_n) = 0$ if $s_n \le 0$. Over the domain $s_n \ge 0$, the cost function $c_n$ is convex and strictly increasing.
\label{asn: cost_convexity}
\end{assumption}

Intuitively, convexity follows from the conventional assumption that the unit cost increases as tenants reduce more energy (e.g., utilization becomes higher  when servers are off, leading to a faster increase in response time of tenants' workloads).




\section{Pricing Tenant Load Shedding in Mandatory \edr}

\edr is the last line of protection against cascading power failures, and represents 87\% of demand reduction capabilities across all the U.S. reliability regions  \cite{EDR_Market_Overview}. In general, there are two types of \edr programs: mandatory and voluntary (also called economic) \cite{pjm_emergency_demand_response_Performance}. We focus on mandatory \edr first, and return to voluntary \edr in Section \ref{sec: voluntary_EDR}.

For mandatory \edr, participants typically sign contracts with a load serving entity (\lse) in advance (e.g., 3 years ahead in PJM \cite{pjm_emergency_demand_response_Performance}) and receive financial rebates for their committed energy reduction even if no \edr signals are triggered during the participation year, whereas non-compliance (i.e., failure to cut load as required during \edr) incurs heavy penalty \cite{pjm_emergency_demand_response_Performance}.  If an \lse anticipates that an emergency will occur, participants are notified, usually at least 10 minutes in advance, and obliged to fulfill their contracted amounts of energy reduction for the length of the event, which may span a few minutes to a few hours.

In mandatory \edr, the colo operator has two options for obtaining load reductions  in response to an \edr signal
that specifies the reduction amount -- tenants or on-site generation.
 Thus, it must balance between paying tenants for reduction and using on-site generation in order to minimize cost. Note that tenants' load
reduction can also reduce the usage of diesel generator, mitigating environmental impacts. Nonetheless,
the challenge is that the operator does not know the tenant cost functions, and so cannot determine the cost-minimizing price.

Consequently, the operator has two options: (i) predict the tenant supply function and compute prices based on the predictions, or (ii) allow tenants to supply some information about their cost functions through bids.  Clearly, there is a tradeoff here between the accuracy of predictions and the manipulation possible in the bids.
Both of these approaches have been looked at in the literature \cite{mohsenian2010, Liu:2014:PDC:2591971.2592004,johari2011,day2002, anderson2008}, though not in the context of colo demand response.  In general, the broad conclusion is that approach (i) is appropriate when predictions are accurate and one bidder has market power (e.g., is significantly larger than other bidders).  While market power is a considerable issue for the participation of owner-operated data centers in demand response programs due to their large size compared
to other participants, it is not an issue within a specific colo that houses
multiple tenants (typically of comparable sizes), and so we adopt approach (ii) in this paper.

Specifically, we design a mechanism, named \ouralg, where tenants bid using parameterized supply functions and then, given the bids, the operator decides how much load to shed via tenants and how much to shed via on-site generation.  In the following, we describe the mechanism and then contrast our approach with other potential alternatives.

Note that, throughout this paper, we focus on one \edr event, and thus we omit the time index. In the case of multiple consecutive \edr events,
\ouralg will be executed once at the beginning of each event, as is standard in the literature \cite{Liu:2014:PDC:2591971.2592004,Shaolei_Colocation_ICAC_2014}.

\subsection{An overview of ColoDR}
The operation of \ouralg is summarized below, and then discussed in detail in the text that follows.
\begin{enumerate}
\item The colo operator receives an \edr reduction target $\delta$ and broadcasts the supply function $S(b_n, p)$ to tenants according to \eqref{eqn: supply_function};
\item Participating tenants respond by placing their bids $b_n$;
\item The colo operator decides the amount of on-site generation $y$ and market clearing price $p$ to minimize its cost, using equations \eqref{eqn: price} and \eqref{eqn: local_gen1};
\item \edr is exercised. $\forall n\in\mathcal{N}$, tenant $n$ sheds $S(b_n, p)$, and receives $pS(b_n, p)$ reward.
\end{enumerate}

Given the overview above, we now discuss each step in more detail.

\emph{Step 1.} Upon receiving an \edr notification of an energy reduction target $\delta$, the colo operator broadcasts a parameterized supply function $S(b,p)$ to tenants (by, e.g., signalling
to the tenants' server control interfaces, which are widely existing
today). The form of $S(b, p)$ is the following parameterized family\footnote{The supply function allows tenants to have negative supply, i.e., tenants consume more energy intentionally, which is neither profit maximizing nor practical. We show in \xref{sec: eff_analysis} that energy reduction of each tenant is always nonnegative in both equilibrium and social optimal outcomes.}:
\begin{equation}
S(b_n, p) = \delta - \frac{b_n}{p}.
\label{eqn: supply_function}
\end{equation}
where $p$ is offered reward for each kWh of energy reduction and $b_n$ is the bidding values that can be chosen by tenant $n$. This form is inspired by \cite{johari2011}, where it is shown that by restricting the supply function to this parameterized family, the mechanism can guide the firms in the market reach to an equilibrium with desirable properties.\footnote{\cite{johari2011} studies the case where firms bid to supply an inelastic demand, which is equivalent to fixing the diesel generation $y=0$ in our case. Allowing the operator to choose $y$ in a cost-minimizing manner leads to significantly different results, as will be shown in \xref{sec: price-taking} and \xref{sec: price-anticipating}.} Note that, to be consistent with the supply function literature, we exchangeably use ``price'' and ``reward rate'' wherever applicable.

\emph{Step 2.} Next, according to the supply function, each participating tenant submits its
bid $b_n$ to the colo operator.  This bid specifies that, at each price $p$, it is willing to reduce $S(b_n, p)$ unit of energy. The bid is chosen by tenants individually to maximize their own utility and can be interpreted as the amount of IT service revenue that tenant $n$ is willing to forgo. Note that $b_n$ can be chosen to ensure that tenant $n$ will not be required to reduce more energy than its capacity. To see this, note that since the operator is cost-minimizing, $p(\mathbf{b}, y) \le \alpha$ always holds, i.e., the market clearing price is lower than the unit cost of diesel generation. Hence, if $K_n$ is the capacity of reduction for tenant $n$, as long as $b_n \ge \alpha (\delta - K_n)$, then
\begin{equation*}
S(b_n, p) = \delta - \frac{b_n}{p} \le \delta - \frac{b_n}{\alpha} \le K_n.
\end{equation*}
An important note about the tenant bids is that the supply function is likely of a different form than the true cost function $c_n$, and so it is unlikely for the tenants to reveal their cost functions truthfully. This is necessary in order to provide a simple form for tenant bids. Bidding their true cost functions is too complex
and intrusive. However, a consequence of this is that one must carefully analyze the emergent equilibrium to understand the efficiency of the pricing mechanism.  We study both the cases of price-taking and price-anticipating equilibrium in \xref{sec: eff_analysis}.

\emph{Step 3.} After tenants have submitted their bids, the colo operator
decides the amount of energy $y$ to produce via on-site generation and the clearing price $p$. Given $y$, the market clearing price has to satisfy~ $\Sigma_n S(p(\mathbf{b}), b_n) + y = \delta$, thus 
\begin{equation}
p(\mathbf{b}, y) = \frac{\sum_n b_n}{(N-1)\delta + y}.
\label{eqn: price}
\end{equation}
To determine the amount of local generation $y$, the operator minimizes the cost of the two
load-reduction options, i.e.,
\begin{align}
y = \argmin_{0 \le y \le \delta} (\delta - y)\cdot p(\mathbf{b}, y) + \alpha y.
\label{eqn: local_gen1}
\end{align}

\emph{Step 4.} Finally, \edr is exercised and tenants receive financial compensation from
the colo operator via the realized price in \eqref{eqn: price}, shed load $S(p,b_n)$, and on-site generation produces \eqref{eqn: local_gen1}.

\subsection{Discussion}

To the best of our knowledge, this paper represents the first attempt to design a supply function bidding mechanism for colocation demand response. Although alternative
mechanisms may be applicable,
there are compelling advantages to the supply function approach.
First, bidding for the tenants is simple -- they only need to communicate one number, and it is already common practice for operators to communicate with tenants before \edr events, so the overhead is small. Second, the colo operator collects
just enough information (i.e., how much energy reduction
each tenant will contribute to \edr), while
tenants' private information (i.e., how
much performance penalty/cost each for energy reduction)
is masked by the form of the supply function and hence not solicited.
Third, \ouralg guarantees that the colo operator
will not incur a higher cost than the case where only
diesel generator is used. Further, \ouralg pays
a uniform price to all participating tenants and hence ensures fairness.

The most natural alternative design to supply function bidding is a VCG-based mechanism, as is suggested in \cite{Shaolei_Colo_TruthDR_Tech}.  While VCG-based mechanisms have the benefits of incentive compatibility, however, these mechanisms violate all the four properties discussed above. Under such approaches, tenants must submit very complex bids describing their precise cost functions, the true private cost of tenants is disclosed, payment made to tenants may be unbounded, and prices to different tenants are differentiated and thus raises unfairness issues.

Due to these shortcomings, VCG-based mechanisms are typically not adopted in complex resource allocation settings such as power markets, where supply-function based designs are common \cite{johari2011}.  In fact, nearly all generation markets use a variation of supply function bidding.


\newcommand{\yterm}{\frac{\alpha}{2N\delta}(y+(N-1)\delta)^2}

\section{Efficiency Analysis of \ouralg for Mandatory EDR}\label{sec: eff_analysis}

Given the \ouralg mechanism described above, our task now is to characterize its efficiency.  There are two potential causes of inefficiency in the mechanism: the cost minimizing behavior of the operator and the strategic behavior (bidding) of the tenants.  In particular, since the forms of the tenant's cost functions are likely more complex than the supply function bids, tenants cannot bid their true cost function even if they wanted to.  This means that evaluating the equilibrium outcome is crucial to understanding the efficiency of the mechanism.

Further, the equilibrium outcome that emerges depends highly on the behavior of the tenants -- whether they are \emph{price-taking}, i.e., they passively accept the offered market price $p$ as given when deciding their own bids; or \emph{price-anticipating}, i.e., they anticipate how the price $p$ will be impacted by
their own bids. We investigate both models, in \xref{sec: price-taking} and \xref{sec: price-anticipating}, respectively.  

In both cases, the goal of our analysis is to assess the efficiency of \ouralg.  To this end, we adopt a notion of a (socially) optimal outcome, and focus on the following social
cost minimization  problem.
\begin{subequations}
\label{eqn: edr1}
\begin{eqnarray}
\label{eqn:obj_edr}
\mathsf{SCM}:&& \min \;\; \alpha y + \sum_{i\in\mathcal{N}}c_i(s_i)\\
\label{eqn:constraint_edr}
\text{s.t.} && y +  \gamma\cdot\sum_{i\in\mathcal{N}}s_i = \delta\\
\label{eqn:nonnegative} \label{eqn:nonnegative_y}
&& s_i \ge 0,\;\forall i\in\mathcal{N} \text{, } \quad y \geq 0.
\end{eqnarray}
\end{subequations}
where $s_i$ and $c_i$ are tenant $i$'s energy reduction and corresponding
cost, respectively.

The objective in \pone can be interpreted as the tenants' cost plus the colo operator's cost.  Note that the internal payment transfer between the colo operator and tenants cancels, and does not impact the social cost.
Also, note that payment from the \lse to the colo operator is not included
in the social cost objective, since it is independent of how the operator obtains the amount of $\delta$ load reduction. Additionally, we do not include the option of ignoring the event and taking the penalty, since the penalties for lack of participation are typically extreme. Finally, the Lagrangian multiplier of \eqref{eqn:constraint_edr} can be interpreted as the social optimal price $p^*$, i.e., given this price as reward for energy reduction, each tenant will individually reduce their energy by $s_n$ that corresponds to the social cost minimization solution in \eqref{eqn: edr1}.


Before moving to the analysis, in order to simplify notation, we suppress the PUE $\gamma$ by, without loss of generality, setting $\gamma=1$.   This is equivalent to a change of notation $y' = y / \gamma$, $\delta' = \delta / \gamma$, and $\alpha' = \alpha \gamma$, i.e., translating the diesel generation, unit cost of diesel generation, and \edr energy reduction target into their respective equivalent amounts in terms of server energy.

\subsection{Price-Taking Tenants}
\label{sec: price-taking}
 When tenants are price-taking, they maximize their net utility, which is the difference between the payment they receive and the cost of energy reduction, given the assumption that they consider their action does not impact the price.
\begin{subequations}
\label{eqn: cost_taking}
\begin{eqnarray}
P_n(b_n, p) &= pS_n(b_n, p) - c_n(S_n(b_n, p)) \\
&= p\delta - b_n - c_n\left(\delta - \frac{b_n}{p}\right).
\end{eqnarray}
\end{subequations}
Here, the price-taking assumption implies that the variable $p$ is considered to be as is. The market equilibrium for price-taking tenants is thus defined as follows.
\begin{definition}
 A triple $(\mathbf{b}, p, y)$ is a (price-taking) market equilibrium if each tenant maximizes its payoff defined in \eqref{eqn: cost_taking}, market is cleared by setting price $p$ according to \eqref{eqn: price}, and the amount of on-site generation is decided by \eqref{eqn: local_gen1}, i.e.,
 \begin{align} \label{def: ne1_taking}
P_n(b_n; p) &\ge P_n(\bar{b}_n; p)\quad  \forall \bar{b}_n \ge 0, \quad n=1, \ldots, N .\quad \\
\label{def: price_ne1}
 p &= \frac{ \sum_{i\in\mathcal{N}}b_i}{(N-1)\delta+y}.\\
 \label{def: local_gen_ne1}
 y &= \argmin_{0 \le y \le \delta} (\delta - y)\cdot p(\mathbf{b}, y) + \alpha y.
 \end{align}
 \end{definition}

\subsubsection{Market Equilibrium Characterization}

The key to our analysis is the observation that the equilibrium can be characterized by an optimization problem.  Once we have this optimization, we can use it to characterize the efficiency of the equilibrium outcome.  This approach parallels that used in  \cite{johari2011}; however, the optimization obtained has a different structure due to local diesel generation. Additionally, though we use an optimization to characterize the equilibrium, the game is not a potential game.

Our first result highlights that, given any choice for on-site generation, a unique market equilibrium exists for the tenants, and can be characterized via a simple optimization.

\begin{prop}
 Under Assumption \ref{asn: cost_convexity}, when tenants are price-taking, for any on-site generation level $0 \le y < \delta$, there exists a market equilibrium, i.e., a vector $\mathbf{b}^t = (b_1^t, \ldots, b_N^t)\ge 0$ and a scalar $p > 0$ that satisfies \eqref{eqn: price}, and the resulting allocation $s_n = S(b_n, p)$ is the optimal solution of the following
 \begin{subequations}
 \begin{align}
 \label{eqn: p1-1}
 \min_{\mathbf{s}}  & \quad\sum_{i\in \mathcal{N}} c_i(s_i) \\
 \label{eqn: p1-2}
 s.t. & \quad   \sum_{i\in \mathcal{N}} s_i = (\delta - y), \\
 \label{eqn: p1-3}
 &\quad s_i \ge 0,\  \forall i \in \mathcal{N}.
 \end{align}
 \label{eqn: p1}
 \end{subequations}
 \label{prop: ec1}
 \end{prop}

This result is a key tool for understanding the overall market outcome.
Intuitively, the operator running \OurAlg is more likely (than the social optimal) to use on-site generation, since this reduces the price paid to tenants.  The following proposition quantifies this statement.

\begin{prop}
Under Assumption \ref{asn: cost_convexity}, it is optimal for price-taking tenants to use on-site generation if and only if
\begin{equation}
 \alpha < \frac{(\Sigma_n b_n)}{(N-1)\delta},\footnote{We adopt the convention that $\frac{0}{0} = 0$ and $\frac{x}{0} = +\infty$ when $x>0$. Therefore, when $N=1,$ unless the bid is 0, the condition is always satisfied.}
 \label{eqn: cheap_on-site}
\end{equation}
However, when the operator is profit maximizing, it will turn on on-site generation if and only if
\begin{equation}
 \alpha < \frac{N}{N -1} \frac{(\Sigma_n b_n)}{(N-1)\delta}.
  \label{eqn: cheap_on-site2}
 \end{equation}
 \label{prop: on-site_gen}
\end{prop}

This proposition is an important building block because the most interesting case to consider is when it is optimal to use some on-site generation and some tenant load shedding, i.e., $\delta>y^* >0 $. Otherwise the EDR demand should be entirely fulfilled by tenants, and the analysis reduces to the case of an inelastic demand, as studied in \cite{johari2011}. Thus, subsequently, we make the following assumption, which ensures that on-site generation is valuable.

\begin{assumption}
The unit cost of on-site generation is cheap enough that the optimal on-site generation is non-zero, i.e., $\alpha$ satisfies \eqref{eqn: cheap_on-site}.
\label{asn: cheap_on-site}
\end{assumption}
Note that, when Assumption~\ref{asn: cheap_on-site} holds, by first-order optimality condition of \eqref{eqn: local_gen1} we have
\begin{equation}
y = \sqrt{\frac{(\Sigma_{i\in \mathcal{N}}b_i) N\delta}{\alpha }} - (N-1)\delta,
\label{eqn: on-site_gen}
\end{equation}
and so the market clearing price for the tenants given on-site generation is
\begin{equation}
p =\frac{ \sum_{i\in\mathcal{N}}b_i}{(N-1)\delta+y} = \sqrt{\frac{(\Sigma_{i\in \mathcal{N}}b_i)\alpha}{N\delta}}.
\label{eqn: price2}
\end{equation}

Using these allows us to prove a complete characterization of the market equilibrium under price-taking tenants.
This theorem is the key to our analysis of market efficiency.

\begin{theorem}
 When Assumptions \ref{asn: cost_convexity} and \ref{asn: cheap_on-site} hold there is a unique {market equilibrium}, i.e., a vector $\mathbf{b}^t = (b_1^t, \ldots, b_N^t) \ge 0$, $y^t > 0$ and a scalar $p^t>0$ that satisfies \eqref{def: ne1_taking}-\eqref{def: local_gen_ne1}, and the resulting allocation $(\mathbf{s}^t, y^t)$ where $s_n^t = S(b_n^t, p^t)$ is the optimal solution of the following problem
 \begin{subequations}
 \label{eqn: price-taking0}
 \begin{align}
 \label{eqn: price-taking1}
 \min_{\mathbf{s}, y}\quad& \sum_n c_n(s_n) + \yterm \\
 \label{eqn: price-taking2}
 s.t. \quad& \sum_n s_n = \delta - y, \\
 \label{eqn: price-taking3}
 & s_n \ge 0, \;\forall n, \quad y \ge 0.
 \end{align}
 \end{subequations}
 \label{thm: price-taking-characterization}
\end{theorem}

\subsubsection{Bounding Efficiency Loss}

We now use Theorem~\ref{thm: price-taking-characterization} to bound the efficiency loss due to strategic behavior in the market.  Denote the socially optimal on-site generation by $y^*$, the optimal price that leads to the optimal allocation $s_i, \forall i \in \mathcal{N}$ by $p^*$, and let $y^t$ and $p^t$ be the allocation under the price-taking assumption.

Our first result highlights that, due to the cost-minimizing behavior of the operator, the equilibrium outcome uses more on-site generation and pays a lower price to the tenants than the social optimal.

\begin{prop}
Suppose that Assumptions~\ref{asn: cost_convexity} and~\ref{asn: cheap_on-site} hold.  When tenants are price-taking, the operator running \OurAlg uses more on-site generation and pays a lower price for power reduction to its tenants than the social optimal. Specifically,
$y^t \ge y^*$ and~  $\frac{N-1}{N}p^* \le p^t \le p^*.$
\label{prop: ec2}
\end{prop}

Now, we move to more detailed comparisons.  There are three components of market efficiency that we consider: social welfare, operator cost, and tenant cost.

First, let us consider the social cost.


\begin{theorem}
Suppose that Assumptions \ref{asn: cost_convexity} and \ref{asn: cheap_on-site} hold. Let $(\mathbf{s}^t, y^t)$ be the allocation when tenants are price-taking, and $(\mathbf{s}^*, y^*)$ be the optimal allocation. Then the welfare loss is bounded by:
$\sum_n c_n(s_n^t) + \alpha y^t \le \sum_n c_n(s_n^*) + \alpha y^* + \alpha\delta/2N.$
\label{thm: welfare_loss_taking}
\end{theorem}

Importantly, this theorem highlights that the market equilibrium is quite efficient, especially if the number of tenants is large (the efficiency loss decays to zero as $O(1/N)$). However, the market could maintain good overall social welfare at the expense of either the operator or the tenants.  The following results show this is not true.

Let $\mathrm{cost}_o(p, y)$ be the operator's cost, i.e.,
 \begin{align}
 \mathrm{cost}_o(p, y) &= p(\delta - y) + \alpha y.
 \label{eqn: operator_cost}
 \end{align}
Then, we have the following results.


 \begin{theorem}
 Suppose that Assumptions \ref{asn: cost_convexity} and \ref{asn: cheap_on-site} are satisfied. The cost of colo operator with price-taking tenants is smaller than the cost in the socially optimal case. Further, we have
$\mathrm{cost}_o(p^*, y^*) -\alpha\delta/N\le\mathrm{cost}_o(p^t, y^t) \le \mathrm{cost}_o(p^*, y^*).$
 \label{thm: colo_cost1_taking}
 \end{theorem}
%
%
%
%
%

\subsection{Price-Anticipating Tenants}
\label{sec: price-anticipating}

In contrast to the price-taking model, price-anticipating tenants realize that they can change the market price by their bids, i.e., that $p$ is set according to \eqref{eqn: price2}, and adjust their bids accordingly.  Clearly, this additional strategic behavior can lead to larger efficiency loss.  But, in this section, we show that the extra loss is surprisingly small, especially when a large number of tenants participate in \OurAlg.

Given bids from the other tenants, each price-anticipating tenant $n$ optimizes the following cost over bidding value $b_n$
 \begin{align*}
 Q_n(b_n, \mathbf{b}_{-n}) &= p(\mathbf{b}) S_n(b_n, p) - c_n(S_n(b_n, p))
 \end{align*}
where we use $\mathbf{b}_{-n}$ to denote the vector of bids of tenants other than $n$; i.e., $\mathbf{b}_{-n} = (b_1, \ldots, b_{n-1}, b_{n+1}, \ldots, b_N)$. Thus, substituting \eqref{eqn: supply_function} and \eqref{eqn: price2}, we have
\begin{equation}
 Q_n(b_n; \mathbf{b}_{-n}) =  \sqrt{\frac{(\Sigma_n b_n)\alpha\delta}{N}}  - b_n - c_n\left(\delta - \frac{b_n}{\sqrt{\Sigma_m b_m}}\sqrt{\frac{N\delta}{\alpha}} \right).
 \label{eqn: cost_anticipating}
\end{equation}

Note that the payoff function $Q_n$ is similar to the payoff function $P_n$ in the price-taking case, except that the tenants anticipate that the colo operator will set the price $p$ according to $p = p(\mathbf{b}, y)$ from \eqref{eqn: price2}.

\begin{definition}
 A triple $(\mathbf{b}, p, y)$ is a (price-anticipating) market equilibrium if each tenant maximizes its payoff defined in \eqref{eqn: cost_anticipating}, the market is cleared by setting the price $p$ according to \eqref{eqn: price} and the amount of on-site generation is decided by \eqref{eqn: local_gen1}, i.e.,
 \begin{align}
 \label{def: ne1}
 Q_n(b_n; \mathbf{b}_n) &\ge Q_n(\bar{b}_n; \mathbf{b}_n)\quad  \forall \bar{b}_n \ge 0, \quad n=1, \ldots, N \\
 \label{def: ne2}
 p &= \frac{ \sum_n b_n}{(N-1)\delta + y}.\\
 \label{def: ne3_price_anticipating}
 y &= \argmin_{0 \le y \le \delta} (\delta - y)\cdot p(\mathbf{b}, y) + \alpha y.
 \end{align}
 \end{definition}

Note that our analysis in this section requires one additional technical assumption about the tenant cost functions.
 \begin{assumption}
 The marginal cost of all the tenants at 0 is greater than $\frac{\alpha}{2N}$, i.e.,
 $\frac{\partial^+ c_n(0)}{\partial s_n} \ge \frac{\alpha}{2N}, \ \forall n.$
\label{asn: mc_lowerbound}
 \end{assumption}
This assumption is quite mild, especially if the number of tenants $N$ is large.  Intuitively, it says that the unit cost of on-site generation is competitive with the cost of tenants reducing their server energy.

\subsubsection{Market Equilibrium Characterization}

Our analysis of market equilibria proceeds along parallel lines to the price-taking case.  We again show that there exists a unique equilibrium and, furthermore, that the tenants and operator behave in equilibrium as if they were solving an optimization problem of the same form as the aggregate cost minimization \eqref{eqn: edr1}, but with ``modified'' cost functions.

 \begin{theorem}
 Suppose that Assumption \ref{asn: cost_convexity}-\ref{asn: mc_lowerbound} are satisfied, then there exists a unique equilibrium of the game defined by \\$(Q_1, \ldots, Q_n)$ satisfying \eqref{def: ne1}-\eqref{def: ne3_price_anticipating}.  For such an equilibrium, the vector $\mathbf{s}^a$ defined by $s_n^a = S(p(\mathbf{b}^a), b_n^a)$ is the unique optimal solution to the following optimization:
 \begin{subequations}
 \label{eqn: mc0}
 \begin{align}
 \label{eqn: mc1}
\min \quad & \sum_n \hat{c}_n(s_n) + \frac{\alpha}{2N\delta}(y+(N-1)\delta)^2\\
 \label{eqn: mc2}
 \text{s.t.}  \quad &  \sum_n s_n = \delta - y \\
 \label{eqn: mc3}
 & y \ge 0,\  s_n \ge 0, \quad n = 1, \ldots, N,
 \end{align}
 \end{subequations}
 where, for $s_n \ge 0$,
 \begin{align}
 \hat{c}_n(s_n) = &\frac{1}{2}\left(c_n(s_n) + s_n \frac{\alpha}{2N}  \right)
  + \frac{1}{2}\int^{s_n}_0  \sqrt{ \left( \mz - \frac{\alpha}{2N}  \right)^2 + 2 \mz \frac{ z\alpha}{N\delta}} dz,
 \label{eqn: modified_cost}
 \end{align}
 and for $s_n <0, \quad \hat{c}_n(s_n) = 0.$
 \label{thm: modified_cost}
 \end{theorem}

Although the form of $\hat{c}_n(s_n)$ looks complicated, there is a simple linear approximation that gives useful intuition.

\begin{lem}
Suppose that Assumption \ref{asn: cost_convexity}-\ref{asn: mc_lowerbound} are satisfied. For all modified cost $\hat{c}_n, n \in 1, \ldots, N$, for any $0 \le s_n \le \delta$,
 \[ c_n(s_n) \le \hat{c}_n(s_n) \le c_n(s_n) + s_n\frac{\alpha}{2N}, \]
 Furthermore, when the left or right derivatives of $\hat{c}(\cdot)$ is defined, it can be bounded by
 \[ \lmc \le \frac{\partial^-\hat{c}(s_n)}{\partial s_n} \le \frac{\partial^+\hat{c}(s_n)}{\partial s_n} \le \rmc + \frac{\alpha}{2N}.\]
 \label{lem: mc_bound}
\end{lem}

The form of Lemma~\ref{lem: mc_bound} shows that the difference between the modified cost function in \eqref{eqn: modified_cost} and the true cost diminishes as $N$ increases, and this is the key observation that underlies our subsequent results upper bounding the efficiency loss of \OurAlg.


\subsubsection{Bounding Efficiency Loss}

We now use Theorem \ref{thm: modified_cost} to bound the efficiency loss due to strategic behavior.  Note that, by comparing to both the socially optimal and the price-taking outcomes, we can understand the impact of both strategic behavior by the operator and the tenants.

Our first result focuses on comparing the price-anticipating and price-taking equilibrium outcomes.  It highlights that price-anticipating behavior leads to tenants receiving higher price while providing less load shedding.

 \begin{theorem}
 Suppose Assumption \ref{asn: cost_convexity}-\ref{asn: mc_lowerbound} hold. Let $(p^t, y^t)$ be the equilibrium price and on-site generation when tenants are price-taking, and $(p^a, y^a)$ be those when tenants are price-anticipating, then we have,
 $y^t \le y^a \le y^t + \delta/2$ and~ $p^t \le p^a \le p^t + \alpha/2N.$
 \label{thm: diff-price-anticipate}
 \end{theorem}

Next, combining Theorem \ref{thm: diff-price-anticipate} and Proposition \ref{prop: ec2} yields the following comparison between the price-anticipating and socially optimal outcomes.


 \begin{corollary}
 Suppose Assumption \ref{asn: cost_convexity}-\ref{asn: mc_lowerbound} hold. When tenants are price-anticipating, an operator running \OurAlg uses more on-site generation and pays lower market price than in the socially optimal case, i.e.,
 $y^a \ge y^*$ and~$\frac{N-1}{N}p^* \le p^a \le p^*. $
 \label{cor: diff-price-anticipate2}
 \end{corollary}

Now, we move to more detailed comparisons.  There are three components of market efficiency that we consider: social welfare, operator cost, and tenant cost.

First, let us consider the social cost.

\begin{theorem} \label{thm: welfare_loss2}
 Suppose that Assumption \ref{asn: cost_convexity}-\ref{asn: mc_lowerbound}  hold. Let $(\mathbf{s}^a, y^a)$ be the allocation when tenants are price-anticipating, and $(\mathbf{s}^*, y^*)$ be the optimal allocation. The welfare loss is bounded by:
 $\sum_n c_n(s^a_n) + \alpha y^a \le \sum_n c_n(s^*_n) + \alpha y^* + \alpha\delta/N.$
\end{theorem}

Similarly to the price-taking case, the efficiency loss in the price-anticipating case decays to zero as $O(1/N)$, only with a larger constant. Also, as in the case of price-taking tenants, we again see that neither the tenants nor the operator suffers significant efficiency loss.



 \begin{theorem}
 \label{thm: diff-payment}
 Suppose that Assumption \ref{asn: cost_convexity}-\ref{asn: mc_lowerbound} hold. The cost of colo operator for price-anticipating tenants is smaller than the cost in the socially optimal case. Further, we have
  \begin{align*}
  \mathrm{cost}_o(p^*, y^*) -\frac{\alpha\delta}{N}\le\mathrm{cost}_o(p^a, y^a)\le \mathrm{cost}_o(p^*, y^*), \\
  \mathrm{cost}_o(p^a, y^a) -\frac{\alpha\delta}{N}\le\mathrm{cost}_o(p^t, y^t)\le \mathrm{cost}_o(p^a, y^a)
 \end{align*}
 \end{theorem}

Finally, let us end by considering the amount of on-site generation used in equilibrium.  
Here, in the worst-case, the equilibrium on-site generation for price-anticipating tenants can be arbitrarily worse than the socially optimal, i.e., the socially optimal can use no on-site generation while the equilibrium outcome uses only on-site generation.

 \begin{theorem}
 Suppose that Assumption \ref{asn: cost_convexity}-\ref{asn: mc_lowerbound} hold. For any $\eps >0$, $N \ge 1$, there exist cost functions $c_1, \ldots, c_N$, such that the on-site generation in the market equilibrium compared to the optimal is given by $y^a - y^* \ge \delta - \eps.$
 \label{thm: on-site_gap}
 \end{theorem}

This is a particularly disappointing result since a key goal of the mechanism is to obtain load shedding from the tenants.  However, the proof emphasizes that this is unlikely to occur in practice.  In particular, the worst-case scenario is that there exists a dominant (monopoly) tenant, which is unlikely in a multi-tenant
colo, that has a cost function asymptotically linear with unit cost roughly matching the on-site generation price $\alpha$. We confirm this in a case study in Section \ref{sec:simulation}.

\subsection{Discussion}
\label{sec: compare_models}

The main results for the price-taking and price-anticipating analyses are summarized in Table \ref{table: performance_guarantee}.  Note that simplified bounds are presented in the table, to ease interpretation, and the interested reader should refer to the theorems in \xref{sec: price-taking} and \xref{sec: price-anticipating} for the tightest bounds. Also, note that the benchmark for social cost we consider is an ideal, but not achievable, mechanism.

 \begin{table}[h]
 \centering
 \begin{tabular}{|c|c|c|c|}\hline
 Tenants & Price Ratio& Colo Saving  & Welfare Loss \\ \hline
 Price-taking & $[\frac{N-1}{N},~ 1 ]$ & $[0, ~ \alpha\delta / N ]$    & $[0, ~ \alpha\delta/ 2N]$  \\ \hline
 Price-anticipating & $[\frac{N-1}{N},~ 1 ]$ & $[0, ~ \alpha\delta / N]$ &   $[0, ~ \alpha\delta /N]$\\ \hline
 \end{tabular}
 \caption{Performance guarantee of \ouralg compared to the social optimal allocation.}
 \label{table: performance_guarantee}
 \end{table}



To summarize the results in Table \ref{table: performance_guarantee} briefly, note first that \ouralg always benefits the operator, since the price paid to tenants to reduce energy is always less than the socially optimal price, and the total cost incurred by operator for energy reduction is also less than that of the social optimal. Secondly, \ouralg also gives the tenants approximately the social optimal payment, since the operator's additional benefit is bounded above by $\alpha\delta / N$.  This naturally means that the loss in payment for tenants compared to the social optimal is also $\alpha\delta / N$, which approaches 0 as $N$ grows. Third, regardless of tenants being price-taking or price-anticipating, \ouralg is approximately socially cost-minimizing as the number of tenants grows.

However, while \ouralg is good in terms of operator, tenant, and social cost, it may not use the most environmentally friendly form of load reduction: in the worst case, the upper bound on the extra on-site generation that \ouralg uses is not decreasing with $N$. However, the analysis highlights that this worst-case occurs when there exists a dominant tenant with unit cost of energy reduction that is consistently just below the cost of diesel over a large range of energy reduction.  As our case study in \xref{sec:simulation} shows, this is unlikely to occur in practice.  So, \ouralg can be expected to use an environmentally friendly mix in most realistic situations.

\newcommand{\sumn}{\sum_{i=1}^n}
\newcommand{\gn}{\gamma_n}
\newcommand{\half}{\frac{1}{2}}
\section{Pricing Tenant Load Shedding in Voluntary \edr}\label{sec: voluntary_EDR}

We now turn from mandatory \edr to voluntary \edr and show how the analysis and design of \ouralg can be extended. Under voluntary \edr, a colo operator is offered a certain compensation rate for load reduction and can cut any amounts of energy \emph{at will} without any obligation. Voluntary \edr often supplements mandatory \edr, and both are widely adopted in practice \cite{pjm_emergency_demand_response_Performance,EDR_Market_Overview}.
Since the colo operator can freely decide on the amount of energy to cut based on the compensation rate
\cite{pjm_emergency_demand_response_Performance}, the amount of energy reduction responses from tenants is \emph{fully} elastic, differing
from mandatory \edr where the total energy reduction (including diesel generation if necessary)
needs to satisfy a constraint $\delta$.

In the following, we formulate the problem and generalize  \ouralg for the voluntary \edr setting. Furthermore, we illustrate that the efficiency
analysis, though more complicated, parallels that of mandatory \edr.



\subsection{Problem Formulation}

During a voluntary \edr event, the \lse offers a reward of $u$ for each unit of
energy reduction (or diesel generation if applicable). In our setting, the colo operator aims at maximizing its profit through extracting loads from tenants using parameterized supply function bidding, as considered for mandatory \edr.

A key difference with the case of mandatory \edr is that, since the reduction is voluntary, diesel generation need not be considered. In particular, if the reward offered the the \lse for reduction is larger than the cost of diesel, then the operator can contribute its whole diesel capacity and, if the reward is smaller than the cost of diesel, no diesel need be used.  Compared to the mandatory \edr setting, operator need to use more diesel generation when tenants' bids are high in order to meet the fixed reduction target $\delta$; in the voluntary \edr case, the operator can simply reduce the DR contribution by tenants when their bids are high. Thus, the optimization of diesel generation by the operator is separable from the optimization of tenant reduction.

This yields a situation where the net profit (from tenant reduction) received by the colo operator is:
\begin{equation}
	u\cdot d -  p\cdot d  \label{eqn: vdr-net-benefit}
\end{equation}
where $p$ is the unit price the colo operator pays to the tenants to solicit $d$ units of reduction in aggregate, which arises from $N$ tenants where tenant $i$ has reduction capacity $D_i$.

\paragraph{An overview of \ouralg}


It is straightforward to adapt \ouralg to this setting.  We outline its operation in four steps below, which parallel the steps in the case of mandatory \edr.

\begin{enumerate}
\item The colo operator receives the voluntary \edr reduction price $u$ and broadcasts the supply function $S(b_n, p)$ to tenants according to \begin{equation}
	S_i(b_i, p) = D_i - \frac{b_i}{p}, \label{eqn: vdr-supply-function}
\end{equation}
where $D_i$ is the capacity of tenant $i$ for reduction determined exogenously.
\item Participating tenants respond by placing their bids $b_n$ in order to maximize their own payoff;
\item The colo operator decides the total amount of reduction from tenants $d$ and market clearing price $p$ to maximize its utility. Given the bids $\mathbf{b} = (b_1, \ldots, b_n)$, if the operator decides to offer $d$ amount of energy reduction to the utility, then the market clearing price $p$ will be
\begin{equation}
	p = \frac{\sumn b_i}{\sumn D_i - d}. \label{eqn: vdr-price1}
\end{equation}
Hence to maximize the operator's profit, the operator will chooose $d$ such that
\begin{equation}
	d = \argmax_{0 \le d \le \sumn D_i } (u - p) d = \left(u - \frac{\sumn b_i}{\sumn D_i - d}\right) d.
	\label{eqn: profit-maximize-d}
\end{equation}
It follows from the first order optimality of \eqref{eqn: profit-maximize-d} that
\begin{equation}
	d = \sumn D_i - \sqrt{\frac{ (\sumn b_i) (\sumn D_i)}{u} }, \label{eqn: vdr-quantity}
\end{equation}
which gives that the price set by a  profit maximizing operator will be
\begin{equation}
	p = \sqrt{\frac{u\sumn b_i}{\sumn D_i }}. \label{eqn: vdr-price2}
\end{equation}

\item Voluntary \edr is exercised. $\forall n\in\mathcal{N}$, tenant $n$ sheds $S(b_n, p)$, and receives $pS(b_n, p)$ reward.
\end{enumerate}

\paragraph{Discussion}

The key difference in the operation of \ouralg for mandatory \edr and voluntary \edr is in the form of the supply function used.  In particular, we allow heterogeneity in the supply function for tenants in terms of their capacity $D_n$.  Recall, that in the case of mandatory \edr the desired reduction capacity $\delta$ was used.  This difference stems from the fact that the reduction target is flexible for voluntary demand response and creates significant challenges -- both in terms of efficiency, since it allows the chance of market power to emerge because of capacity differences, and for analysis, since it adds considerable complexity.

\subsection{Efficiency Analysis of \ouralg for Voluntary \edr}

Given the adaptation of \ouralg to the voluntary \edr setting, it is natural to ask how the efficiency of the mechanism changes when the operator has flexibility in the amount of response to provide to an \edr signal.  Intuitively, the increased flexibility leads to the possibility of more inefficiency, but how large is this effect?

We again quantify efficiency through a comparison with the (socially) optimal outcome. Assuming that each tenant has a cost $c_i(\cdot)$ associated with energy reduction that is convex, increasing, and $c_i(x) = 0, \forall x \le 0$ (Assumption \ref{asn: cost_convexity}). Then the allocation that maximizes social utility (the sum of operator's and tenants' utility) solves the following problem
\begin{subequations}
	\label{eqn: vdr-utility-maximization}
	\begin{align}
		\max_{d, \mathbf{s}} \quad&  ud - \sumn c_i(s_i) \\
		\text{subject to} \quad & \sumn s_i = d \\
		\quad & 0 \le s_i \le D_i .
	\end{align}
\end{subequations}

Finally, note that our analysis makes the following natural assumptions on the unit price $u$ and the marginal cost of each tenants.  Note that they are analogous to Assumption \ref{asn: cheap_on-site} and Assumption \ref{asn: mc_lowerbound}.

\begin{assumption}
	The market clearing price $p$ is lower than the price offered by the utility for any $d>0$, i.e., $u \ge \frac{\sumn b_i}{\sumn D_i}$.
		\label{asn: vdr-u-condition}
\end{assumption}

\begin{assumption}
	The marginal cost of each tenants satisfies $ \mz\Big|_{z=0} \ge \frac{\gn u}{2}, \forall n.$ \label{asn: vdr-mc-lbound}
\end{assumption}

Before moving to the main results, let us first define some notation. Let $\gn = \frac{D_n}{\sumn D_i}$, we have $\sum_n \gn = 1.$ Here $\gn$ behaves like ``market share'' of tenant $n$ in the voluntary DR market. In the EDR case, $\gn = 1/N $ for all $n$. Furthermore, define $\gamma = \max_n \gn$, as the ``dominant share'' in load reduction among the tenants, and $D = \max_n D_n$.

\subsection{Market Equilibrium Characterization}

As in the case of mandatory \edr, we consider both the cases price-taking and and price-anticipating tenants.  

\subsubsection{Price-taking Tenants} 

Given other tenants, each price-taking tenant $n$ optimizes the following cost over bidding value $b_n$,
\begin{align*}
	P_n(b_n, \mathbf{b}_{-n}) = pS_n(b_n, p) - c_n(S_n(b_n, p))
	= pD_n - b_n - c_n(D_n - \frac{b_n}{p})
\end{align*}
So, in a price-taking equilibrium $(\mathbf{b}, d, p)$, we must have $P_n(b_n; \mathbf{b}_{-n}) \ge P_n(\bar{b}_n; \mathbf{b}_{-n})$ hold for each tenant $n$ over all $\bar{b}_n \ge 0$. Also, the market clearing price must satisfy \eqref{eqn: vdr-price1} and the total reduction must satisfy \eqref{eqn: profit-maximize-d}. Using techniques similar to the proof of Theorem \ref{thm: price-taking-characterization}, we can completely characterize the the price-taking equilibrium of \ouralgvdr in voluntary \edr as follows:

\begin{theorem}
There exists a unique equilibrium of the game defined by $(P_1, \ldots, P_N)$ for \ouralgvdr. For such an equilibrium, the vector $\mathbf{s}^t$ defined by $s_n^t = S(p(\mathbf{b}^t), b_n^t)$ is the unique optimal solution to the following optimization:
 \begin{subequations}
 \label{eqn: mc0_taking}
 \begin{align}
 \label{eqn: mc1_taking}
\max \quad & ud - \frac{ud^2}{2\sum_n D_n} - \sum_n c_n(s_n) \\
 \label{eqn: mc2_taking}
 \text{s.t.}  \quad &  \sum_n s_n = d \\
 \label{eqn: mc3_taking}
 & d \ge 0,\  0 \le s_n \le D_n, \quad n = 1, \ldots, N,
 \end{align}
 \end{subequations}
 \label{thm: vdr_ne_taking}
\end{theorem}

\subsubsection{Price-anticipating Tenants} Given other tenants, each price-anticipating tenant $n$ optimizes the following cost over bidding value $b_n$,
\begin{align*} Q_n(b_n, \mathbf{b}_{-n}) = \pb S_n(b_n, p) - c_n(S_n(b_n, p))
 = \gn \sqrt{\Sigma_m b_m}\sqrt{u \sumn D_i} - b_n - c_n(D_n - \frac{b_n}{\Sigma_m b_m} \sqrt{\frac{\sumn D_i }{u}}),
\end{align*}
So, in a price-anticipating equilibrium $(\mathbf{b}, d, p)$, we must have $Q_n(b_n; \mathbf{b}_{-n}) \ge Q_n(\bar{b}_n; \mathbf{b}_{-n})$ for all $n$ over all $\bar{b}_{n}$. Also, the market clearing price must satisfy \eqref{eqn: vdr-price1} and the total reduction $d$ must satisfy \eqref{eqn: profit-maximize-d}. 

Using techniques similar to the proof of Theorem \ref{thm: modified_cost}, we can completely characterize the the price-anticipating equilibrium of \ouralgvdr in voluntary \edr as follows.

\begin{theorem}
There exists a unique equilibrium of the game defined by $(Q_1, \ldots, Q_N)$ for \ouralgvdr. For such an equilibrium, the vector $\mathbf{s}^a$ defined by $s_n^a = S(p(\mathbf{b}^a), b_n^a)$ is the unique optimal solution to the following optimization:
 \begin{subequations}
 \label{eqn: mc0_anticipating}
 \begin{align}
 \label{eqn: mc1_anticipating}
\max \quad & ud - \frac{ud^2}{2\sum_n D_n} - \sum_n \hat{c}_n(s_n) \\
 \label{eqn: mc2_anticipating}
 \text{s.t.}  \quad &  \sum_n s_n = d \\
 \label{eqn: mc3_anticipating}
 & d \ge 0,\  0 \le s_n \le D_n, \quad n = 1, \ldots, N,
 \end{align}
 \end{subequations}
 where, for $s_n \ge 0$,
 \begin{align}
 \hat{c}_n(s_n) = &\frac{1}{2}\left(s_n \frac{\gn u}{2} + c_n(s_n)\right)
  + \frac{1}{2}\int^{s_n}_0  \sqrt{ \left(\frac{\gn u}{2} - \mz \right)^2
  + 2 \mz \frac{ zu}{\Sigma_i D_i}} dz ,
 \label{eqn: vdr_modified_cost}
 \end{align}

 and for $s_n <0, \quad \hat{c}_n(s_n) = 0.$
 \label{thm: vdr_ne_anticipating}
\end{theorem}

Like in the case of mandatory \edr, the above characterization can be approximated using a modified cost function when $\gn$ is small, i.e., when there are a large number of firms and all firms have similar market shares.

\begin{lem}
	For $0 \le s_n \le D_n$, the modified cost in \eqref{eqn: vdr_modified_cost} can be upper and lower bounded by,
	\[
	c_n(s_n) \le \hat{c}_n (s_n)\le c_n(s_n) + s_n \frac{\gn u}{2},
	\]
	Furthermore, where the left or right derivatives are defined, we have
	\begin{subequations}
	\begin{align}
	&\frac{\partial^- c_n(s_n)}{\partial s_n} \le \frac{\partial^-\hat{c}_n(s_n)}{\partial s_n}
	\le \frac{\partial^+\hat{c}_n(s_n)}{\partial s_n} \le \frac{\partial^+ c_n(s_n)}{\partial s_n} + \frac{\gn u}{2}.
	\end{align}
	\end{subequations}
	\label{lemma: vdr_mc_bound}
\end{lem}

\subsection{Bounding Efficiency Loss}

We now use the characterization results of Theorem \ref{thm: vdr_ne_taking} and Theorem \ref{thm: vdr_ne_anticipating} to analyze the social efficiency of \ouralgvdr in the voluntary \edr setting for both price-taking and price-anticipating tenants.

\begin{theorem}
	For price taking tenants, the welfare loss of \ouralgvdr is bounded by
	$ud^t - \sum_n c_n(s_n^t) \ge ud^* - \sum_n c_n(s_n^*) - \frac{u {d^*}^2 }{2\sum_n D_n}.$
	Moreover, the bound is tight.
	\label{thm: vdr_loss_taking}
\end{theorem}

\begin{theorem}
	For price anticipating tenants, the welfare loss of \ouralgvdr is bounded by
	$ ud^a - \sum_n c_n(s_n^a) \ge ud^* - \sum_n c_n(s_n^*)  -  \frac{u}{2} \left(\Sigma_n D_n \gamma_n + \frac{{d^*}^2}{\Sigma_n D_n }\right).$
	\label{thm: vdr_loss_anticipating}
\end{theorem}

Theorem \ref{thm: vdr_loss_taking} highlights that the price-taking market equilibrium is efficient when the optimal energy reduction $d^*$ is small. This is due to the profit maximizing behavior of the operator: when the social optimal $d^*$ is large, the operator has greater opportunity to raise his profit by lowering the market price.

Comparing Theorem \ref{thm: vdr_loss_anticipating} with Theorem \ref{thm: vdr_loss_taking}, we can see that when tenants are price-anticipating, the additional welfare loss due to the price-anticipating behavior of tenants is a function of $\gn$, the market share of the tenants. It is easy to see the additional loss of social utility is minimized when $\gn = 1/N$ for all $n$, i.e., when the reduction capacity of each tenant is equal.

Additionally, we can obtain tight bounds on the market clearing price, energy reduction quantity, and operator's profit in a similar fashion as our analysis done for the mandatory \edr case using Theorem \ref{thm: vdr_ne_taking} and Theorem \ref{thm: vdr_ne_anticipating}. The results are summarized in Table \ref{table: vdr_comp_optimal} and Table \ref{table: vdr_comp_taking}. 

 \subsection{Market Clearing Price}
 \begin{prop}
 	 When tenants are price-taking, the operator running \ouralgvdr uses more on-site generation and pays a lower price for power reduction to its tenants than the social optimal. Specifically,
 	$d^t \le d^*$ and~  $(1 - \frac{d^*}{\sum_n D_n}) p^* \le p^t \le p^*.$
 	\label{prop: vdr_price_taking}
 \end{prop}

 By Lemma \ref{lemma: vdr_mc_bound}, we can characterize the the price markup under the supply function bidding mechanism:
  \begin{theorem}
 	  Let $(p^t, d^t)$ be the equilibrium price and total tenant energy reduction when tenants are price-taking, and $(p^a, d^a)$ be those when tenants are price-anticipating, then let $\gamma = \max_n \gn$, $D = \max_n D_n$, we have,
  $d^t \ge d^a \ge d^t - D/2$ and~ $p^t \le p^a \le \min( p^*, p^t + u\gamma/2).$
  \label{thm: vdr_price_diff}
  \end{theorem}

 \subsection{Operator's profit}
 Let $U_o(p, d) = (u - p) d$ be the operator's when the market clearing price is $p$ and the total demand response from tenants are $d$. From the price and vdr-quantity bounds provided in the previous sections, we can give bound on the utility of \ouralgvdr. 
  \begin{theorem}
  \label{thm: vdr_operator_profit}
  Suppose that Assumptions \ref{asn: cost_convexity}, \ref{asn: vdr-u-condition}, \ref{asn: vdr-mc-lbound} hold. The net utility for the colo operator of \ouralgvdr can be characterized by $0= U_o(p^*, d^*) \le U_o(p^a, d^a) \le U_o(p^t, d^t) \le \frac{u{d^*}^2}{\Sigma_n D_n}$, and furthermore, $U_o(p^t, d^t) \le U_o(p^a, d^a) + uD.$
  \end{theorem}

Table \ref{table: vdr_comp_optimal} shows that as the optimal reduction $d^*$ increases, there is more opportunity for the operator to profitably reduce market price and increase his own profit. Table \ref{table: vdr_comp_taking} shows further that, when tenants are price-anticipating, they will drive the market clearing price up, provide less energy reduction and reduce the operator's profit. However, all these additional losses can be bounded by linear functions of $\gamma$, the dominant share of the energy reduction capacity. Hence the loss due to price-anticipating behavior of tenants are minimized $D_1 = D_2 = \cdots = D_N$.



 \begin{table}
 \centering
 {\small
 \begin{tabular}{|c|c|c|c|}\hline
 Tenants & Price Ratio& Colo Extra Profit  & Welfare Loss \\ \hline
 Price-taking & $[1 - \frac{d^*}{\Sigma_n D_n},~ 1 ]$ & $[0, ~ u {d^*}^2/ \Sigma_n D_n ]$    & $[0, ~ u{d^*}^2 / 2 \Sigma_n D_n]$  \\ \hline
 Price-anticipating & $[1 - \frac{d^*}{\Sigma_n D_n},~ 1 ]$ & $[0, ~ u {d^*}^2/ \Sigma_n D_n ]$ &   $[0, ~ u (\Sigma_n D_n \gamma_n + {d^*}^2 /  \Sigma_n D_n ) /2 ]$\\ \hline
 \end{tabular}
 }
 \caption{Performance guarantee of \ouralgvdr compared to the social optimal allocation.}
 \label{table: vdr_comp_optimal}
 \end{table}

 \begin{table}
 \centering
 {\small
 \begin{tabular}{|c|c|c|c|}\hline
 Price Markup & Load Reduction   & Operator's cost \\ \hline
 $[0, ~ u\gamma/2]$ & $[-D / 2,~ 0 ]$ &  $[0, ~ uD]$  \\ \hline
 \end{tabular}
 }
 \caption{Performance guarantee of \ouralgvdr when tenants are price-anticipating compared to them being price-taking.}
 \label{table: vdr_comp_taking}
 \end{table}


\newcommand{\cent}{{\mathrm{c}\mkern-6.5mu{\mid}}}

\section{Case Study}\label{sec:simulation}

Our goal in this section is to investigate \ouralg in a realistic scenario.  Given the theoretical results in the prior sections, we know that \ouralg is efficient for both the operator and tenants when the number of tenants is large, but that it may use excessive on-site generation (in the worst case). Thus, two important issues to address in the case study are: \textit{How efficient is the pricing mechanism in small markets, i.e., when $N$ is small?  What is the impact of the pricing mechanism on on-site generation in realistic scenarios?} Additionally, the case study allows us to better understand when it is feasible to obtain load shedding from tenants, i.e., \emph{how flexible must tenants be in order to actively participate in a load shedding program?}

We discuss only on mandatory \edr in this section.  The results in the case of voluntary \edr are parallel.

\subsection{Simulation Settings}

We use trace-based simulations in our case study. Our simulator takes the tenants' workload trace and a trace of mandatory \edr signals from PJM as its inputs.  It then executes \ouralg (by emulating the bidding process and tenants' energy reduction for \edr),
and outputs the resulting equilibrium. The settings we use for modeling the colocation data center and the tenant costs follow.

\textbf{Colocation data center setup.}
We consider a colocation data center located in Ashburn, VA, which is a major data center market served by PJM Interconnection \cite{pjm}.
By default, there are three participating tenants interested in \edr, though we vary the number of participating tenants during the experiments.

Each participating tenant has 2,000 servers, and each server has an idle and peak power of $150$W and $250$W, respectively.
The default PUE of the colo is set to $1.5$ (typical for colo), and hence, whenever a tenant reduces 1kWh energy, the corresponding energy reduction at the colo level amounts to 1.5kWh. Thus, the maximum possible power reduction is 2.25MW (i.e., 1.5MW IT plus 0.75 non-IT).
We assume that the colo operator counts the extra energy reduction at the colo level as part of the tenants' contributions, and rewards the tenants accordingly.

The colo has an on-site diesel generator, which has cost $0.3$\$/kWh  estimated based on typical fuel efficiency \cite{wiki_diesel_cost}.

\begin{figure}[t]
        \centering
	  \subfigure[]{\label{fig:workload}\includegraphics[width=0.23\textwidth]{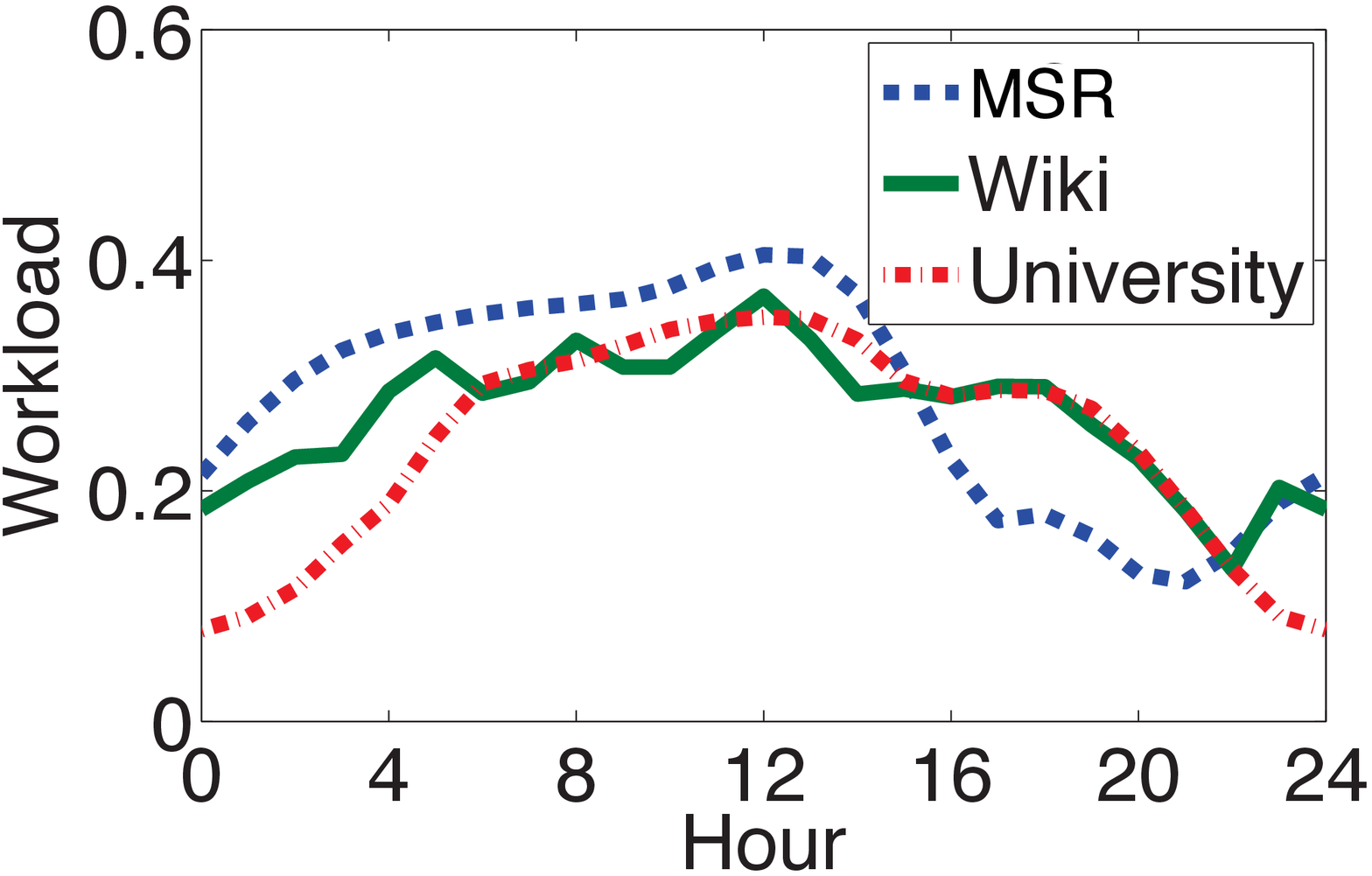}}
	 \subfigure[]{\label{fig:pjm_edr}\includegraphics[width=0.23\textwidth]{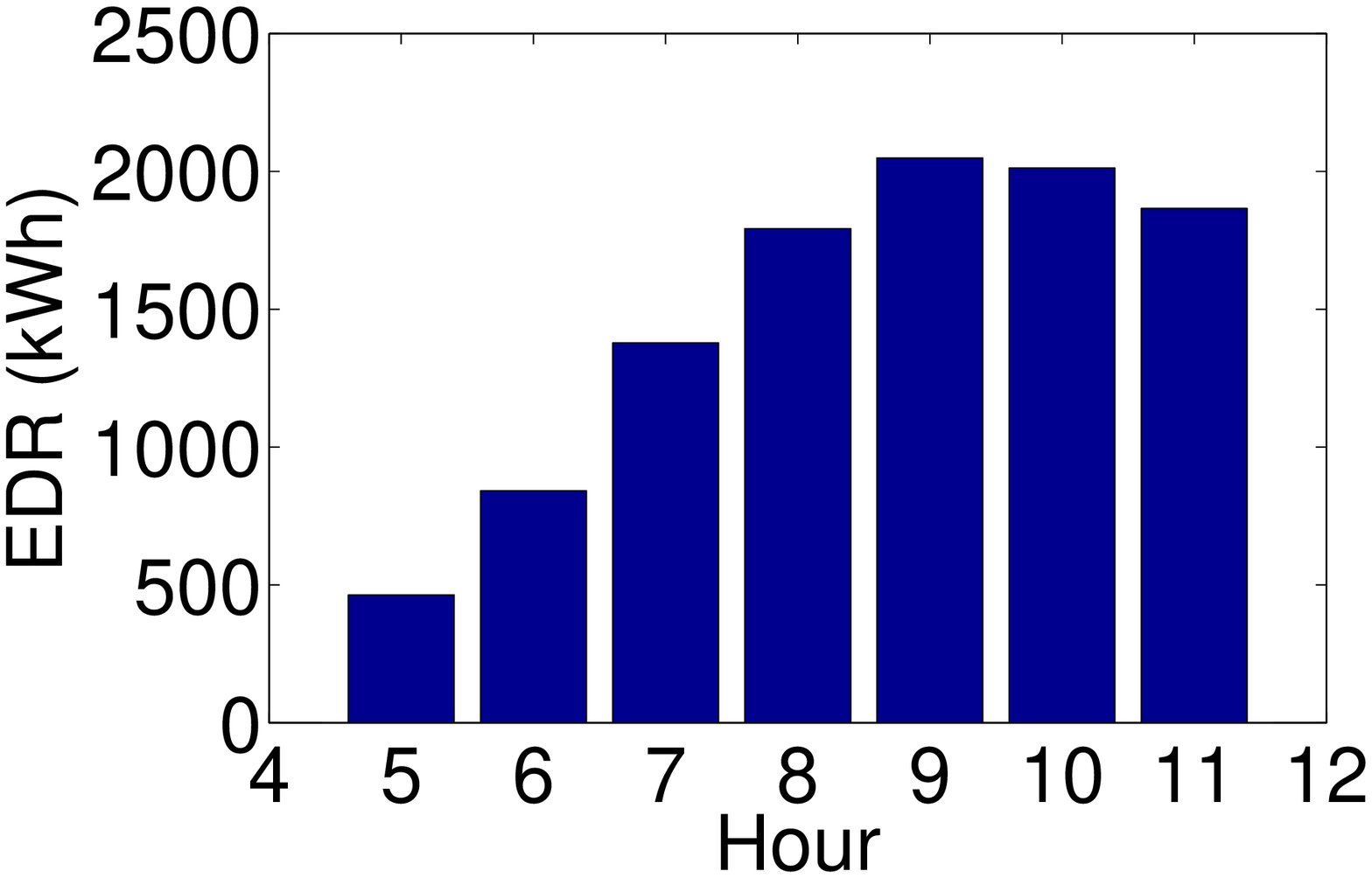} }
        \caption{\textbf{(a)} Workload traces. \textbf{(b)} Energy reduction
      for PJM's \edr on January 7, 2014 \cite{pjm}. }\label{fig:trace}
\end{figure}

For setting the energy reduction target received by the colo, we follow the \edrp signals issued by PJM Interconnection on January $7$, $2014$, when
many states in eastern U.S. experienced an extremely cold weather and faced electricity production shortage \cite{pjm}. Fig.~\ref{fig:pjm_edr} shows the total energy reduction requirement by PJM, which we further normalize and scale down such
that maximum energy reduction target for our considered colo is 900kWh.

\textbf{Tenant workloads characteristics.} We choose three representative types of workloads for participating tenants: tenant 1 is running delay-sensitive workloads (e.g., user-facing web service), tenant 2 is running delay-moderate workloads (e.g., enterprise's internal services), and tenant 3 is running delay-tolerant workload (e.g., back-end processing).

The workload traces for the three participating tenants were collected from logs of MSR \cite{thereska2009sierra}, Wiki \cite{urdaneta2009wikipedia}, and a public university (anonymous for review), respectively. Fig.~\ref{fig:workload} illustrates a snapshot of the traces, where the workloads are normalized with respect to each tenant's maximum service capacity.


The illustrated results us an average utilization for each tenant of 30\%, consistent with reported values from real systems \cite{Hoelzle_datacenter_book_2013}.  Our results are not particularly sensitive to this choice.

\begin{figure*}[t]
        \centering
	  \subfigure[Social cost]{\includegraphics[width=0.23\textwidth]{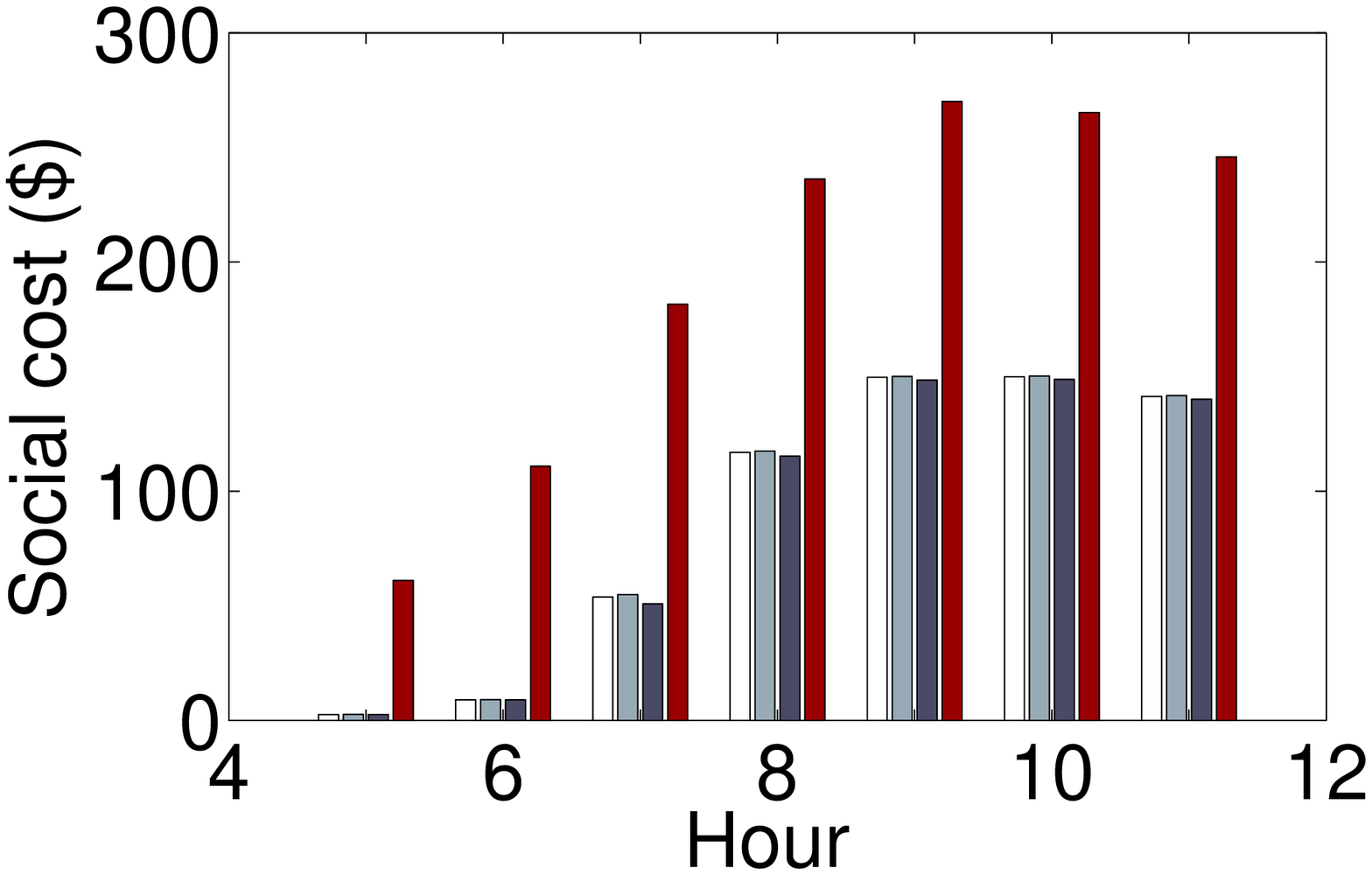}\label{fig:default:welfareLoss}}
	  \subfigure[Energy reduction]{\includegraphics[width=0.23\textwidth]{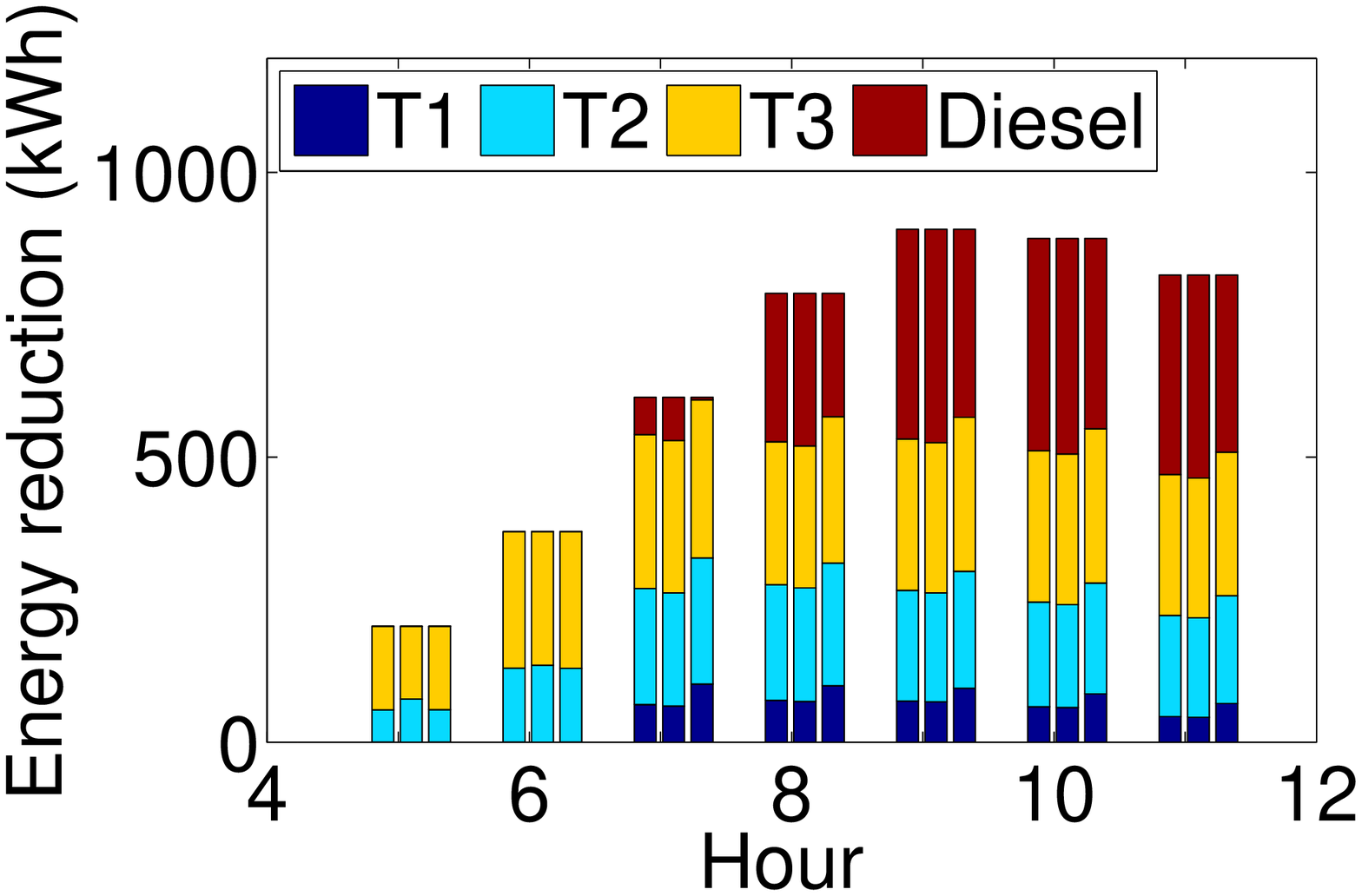} \label{fig:default:energy}}
	  \subfigure[Tenants' net profits]{\includegraphics[width=0.23\textwidth]{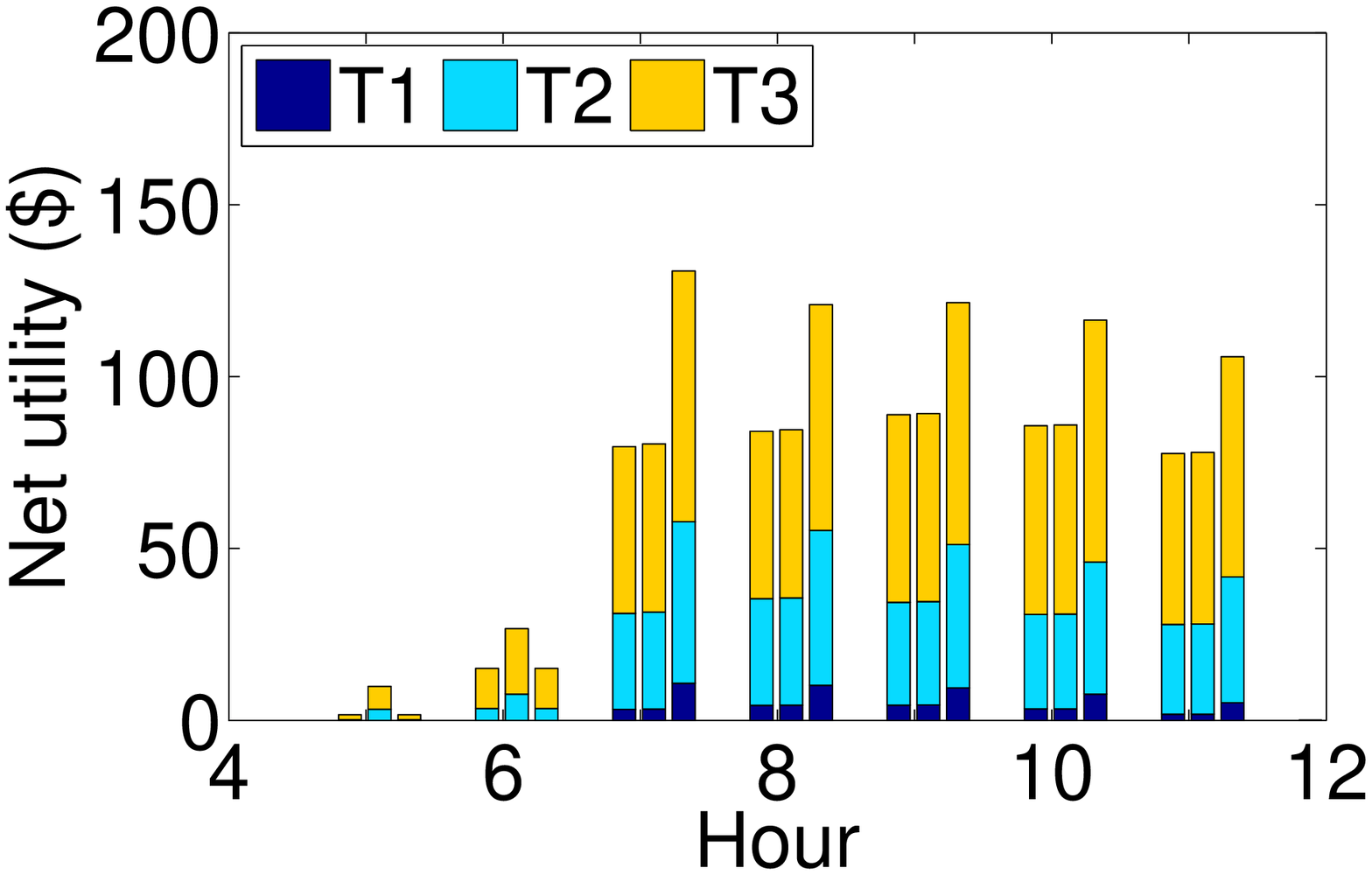}\label{fig:default:netUtilityTenant}}
	  \subfigure[Operator's total cost]{\includegraphics[width=0.23\textwidth]{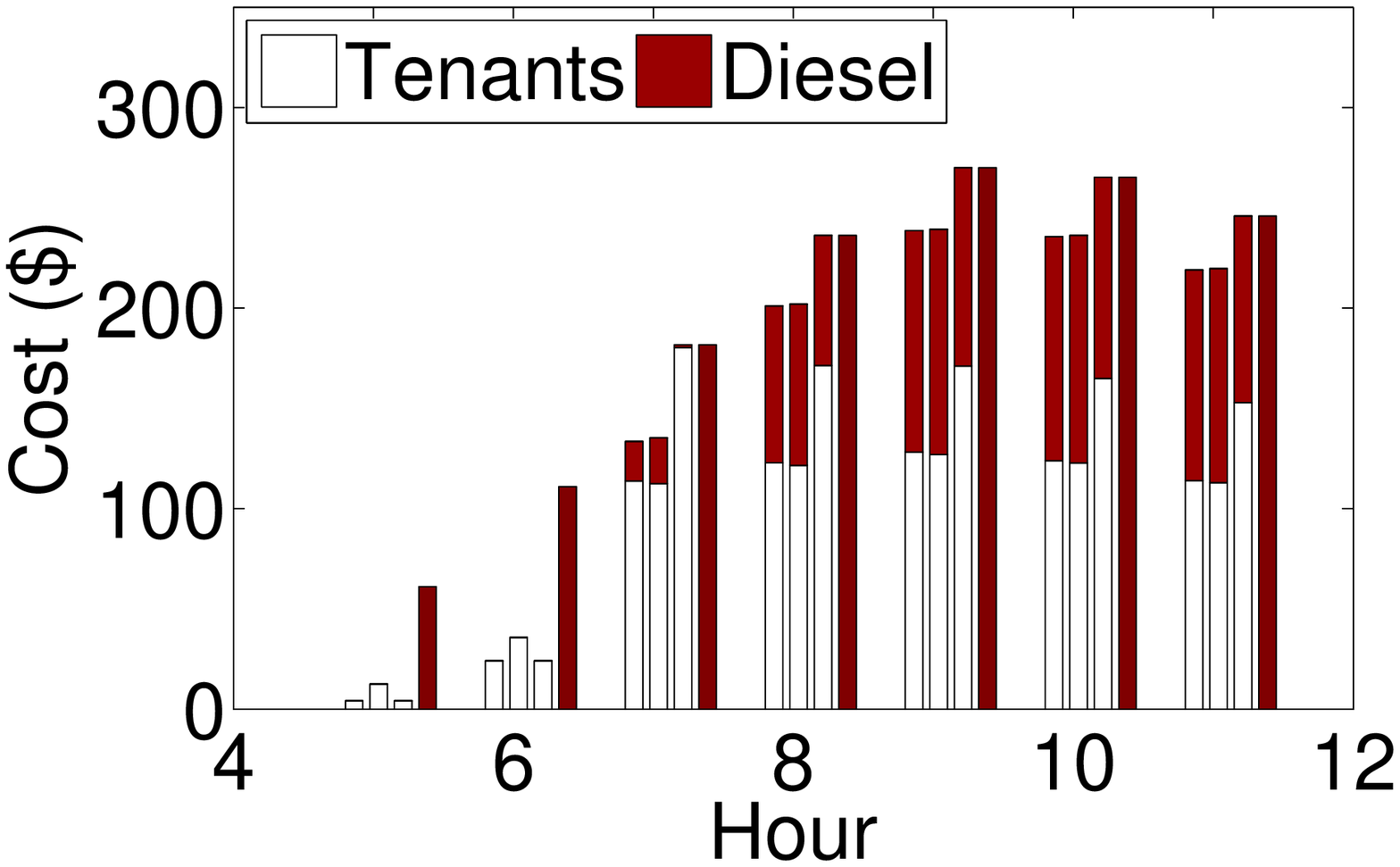}\label{fig:default:netUtilityDC}}\\
    \subfigure[Market clearing price]{\includegraphics[width=0.23\textwidth]{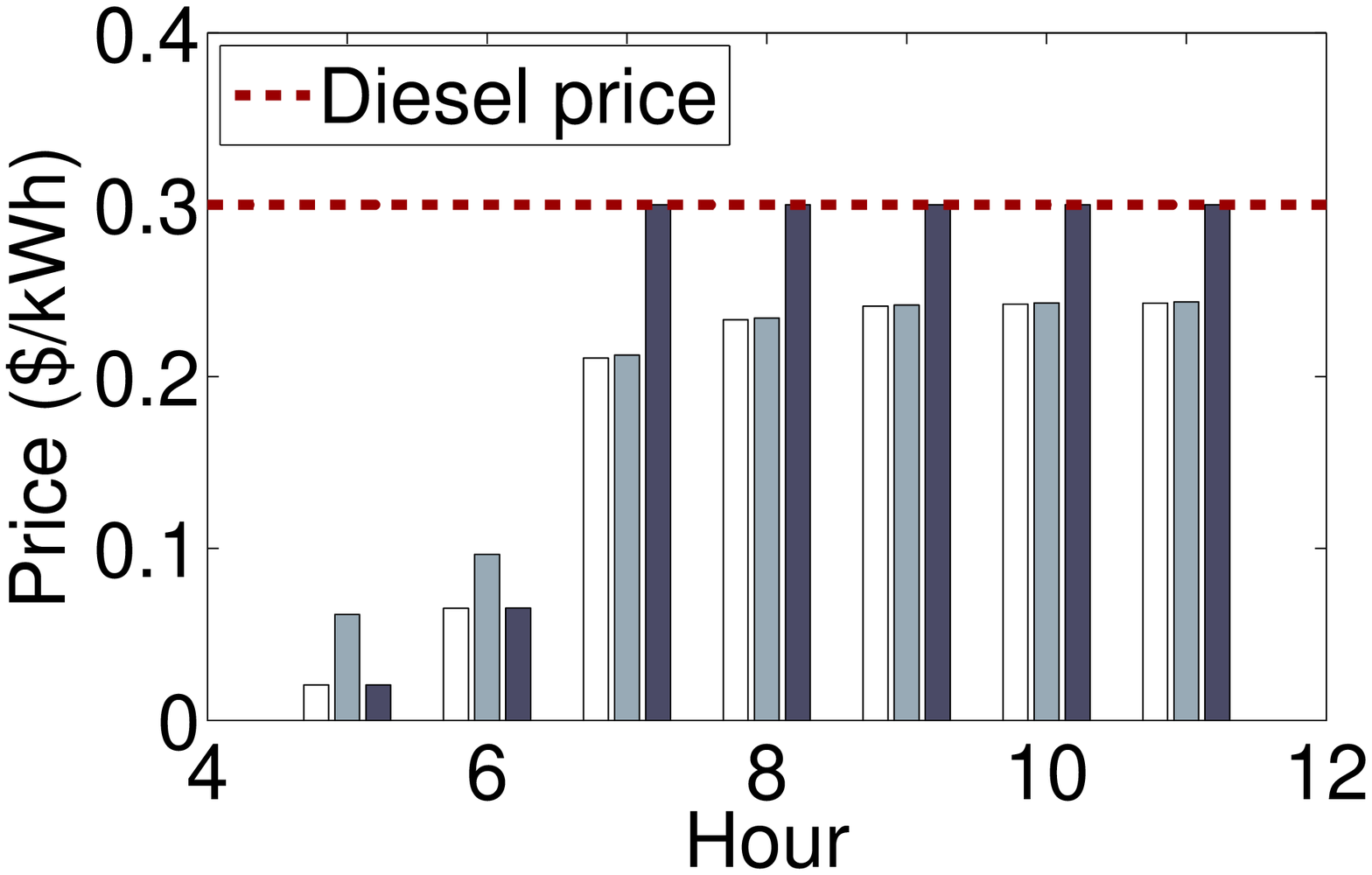} \label{fig:default:price}}
       \subfigure[Tenant 1's  utilization]{\includegraphics[width=0.23\textwidth]{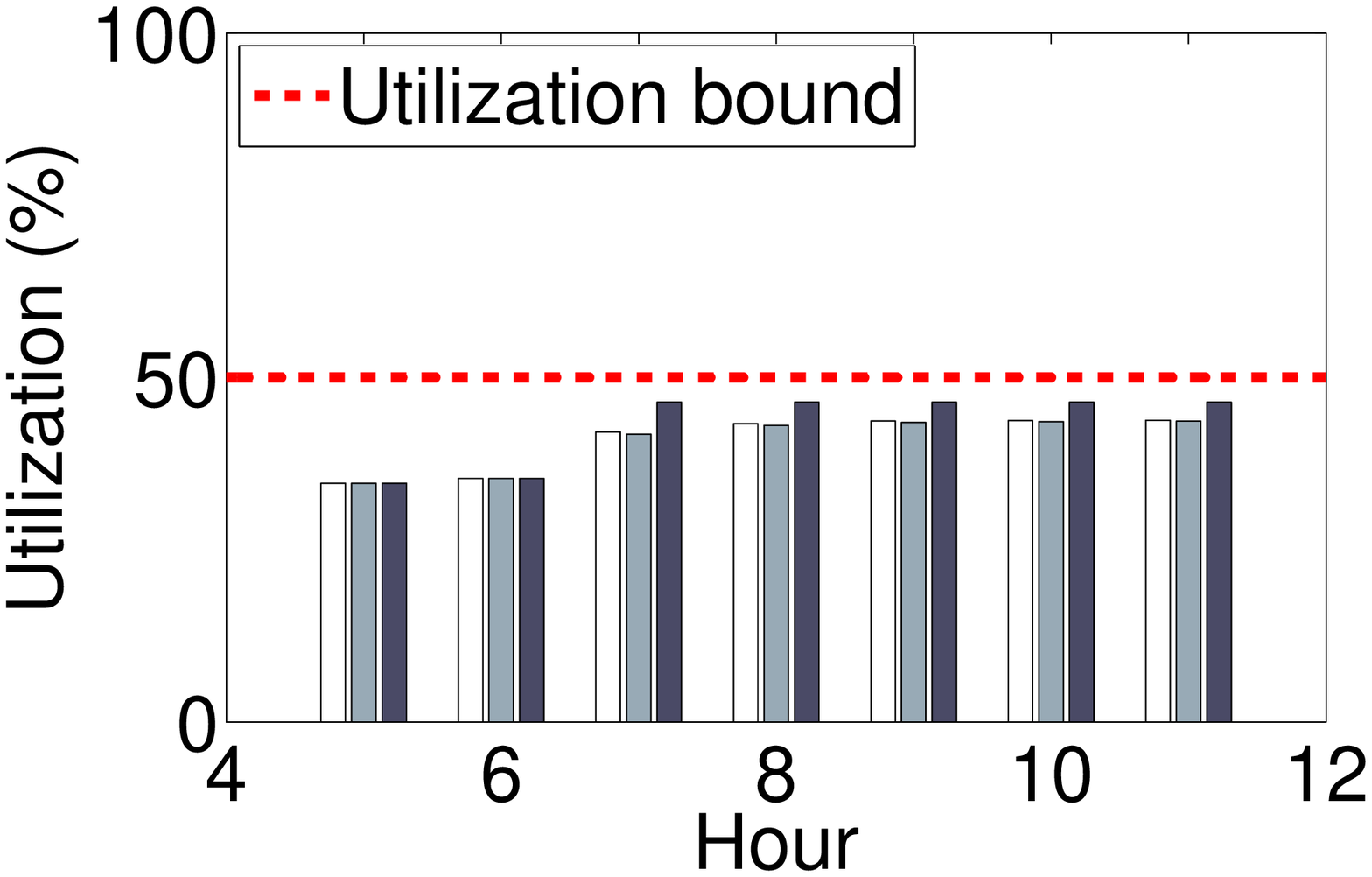} \label{fig:default:u1}}
       \subfigure[Tenant 2's utilization]{\includegraphics[width=0.23\textwidth]{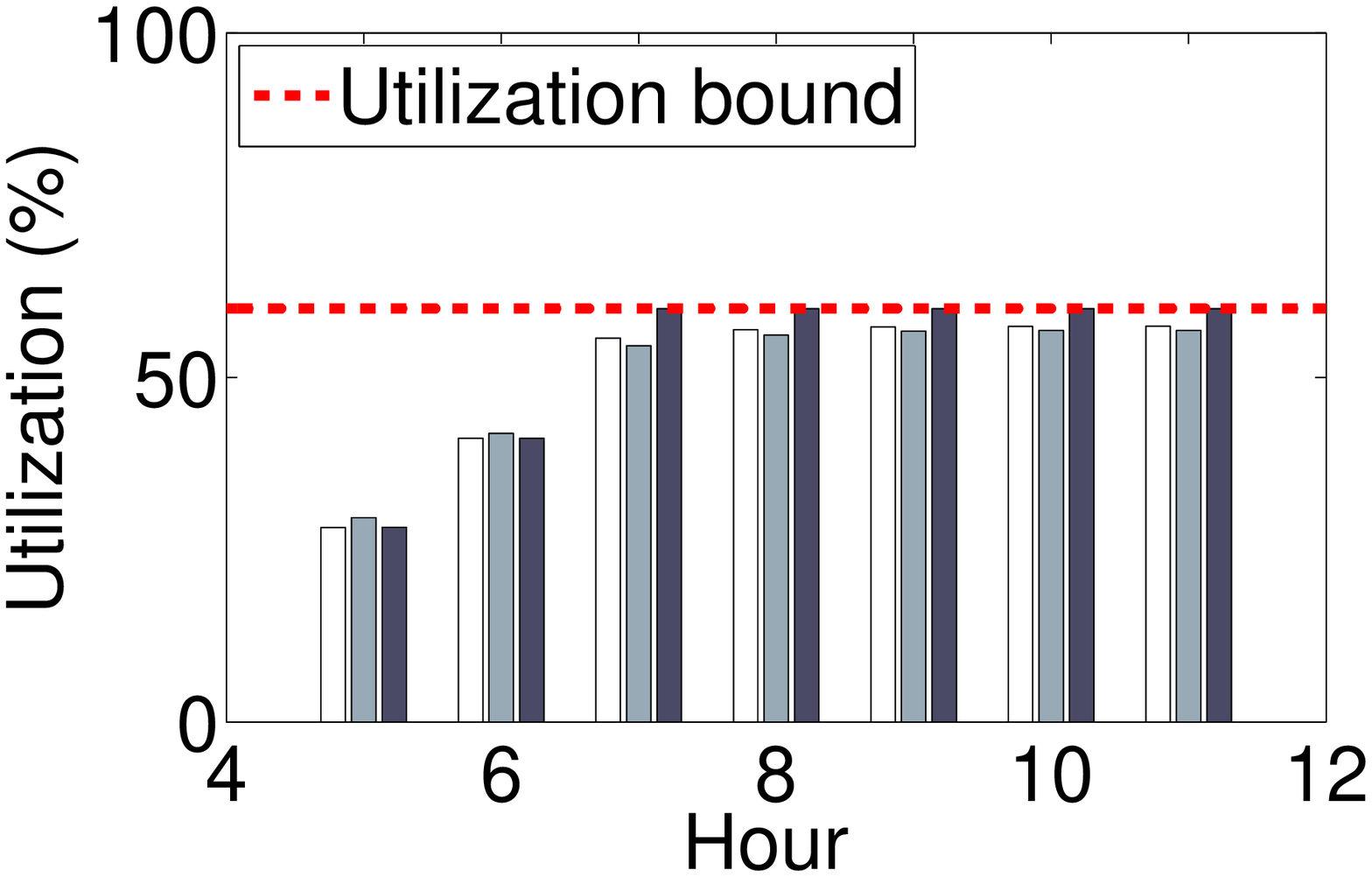} \label{fig:default:u2}}
	  \subfigure[Tenant 3's utilization]{\includegraphics[width=0.23\textwidth]{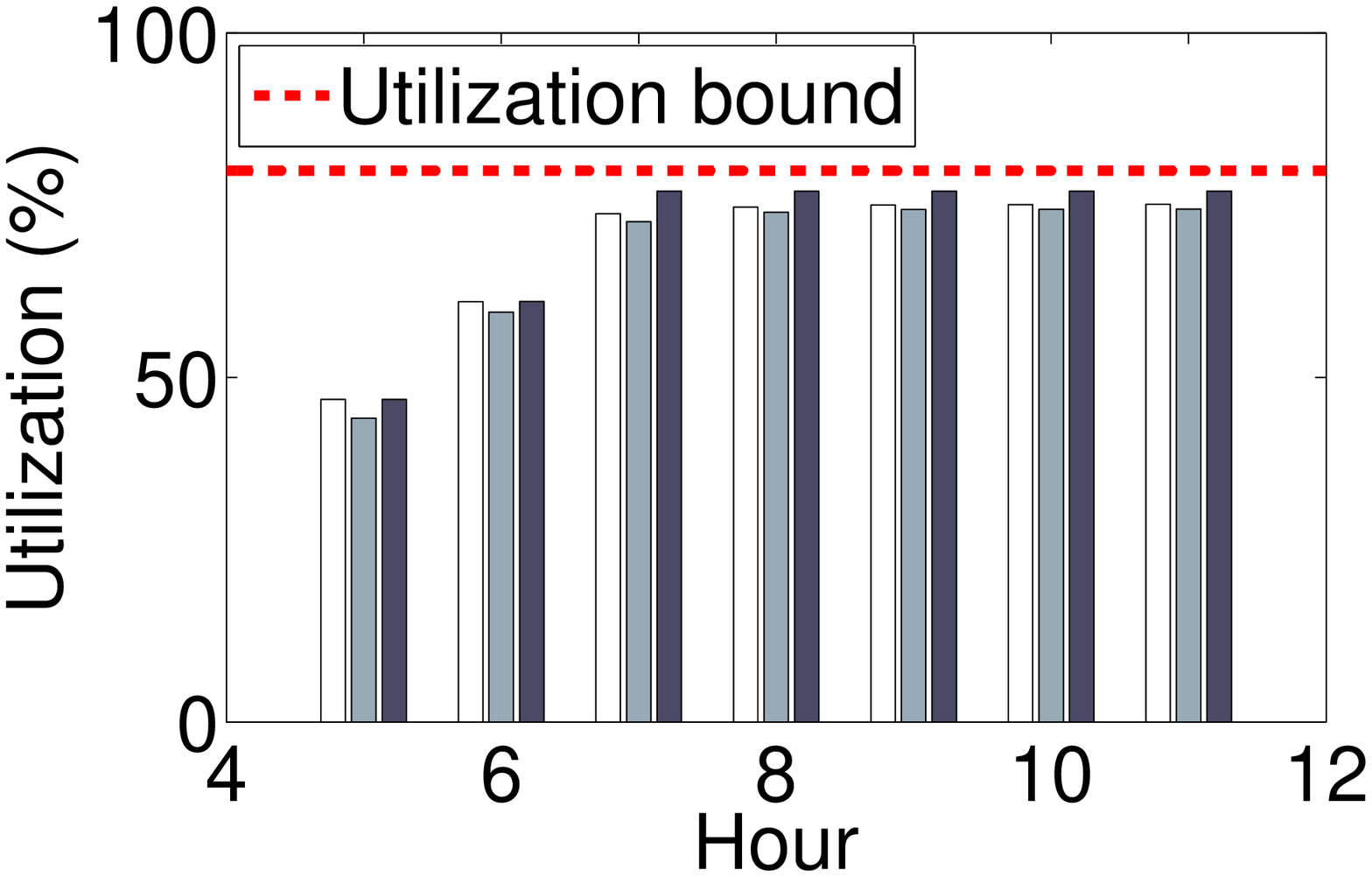}\label{fig:default:u3}}
        \caption{Performance comparison under default settings. Throughout this and later plots, the bars in each cluster are the price-taking, price-anticipating, socially optimal, and diesel only (if applicable) outcomes. }\label{fig:default}
\end{figure*}

There are various power management techniques, e.g., load migration/scheduling, that can be used for reducing tenants' server energy consumption. Here, as a concrete example, we consider that tenants dynamically turn on/off servers according to workloads for energy saving subject to SLA \cite{LinWiermanAndrewThereska}.  This power-saving technique has been widely studied \cite{LinWiermanAndrewThereska,Gandhi:2012:ADR:2382553.2382556} and also recently applied in real systems (e.g., Facebook's AutoScale \cite{Facebook_AutoScale_Energy_2014}).

When tenants save energy for \edr by turning off some unused servers, their application performance might be affected.  We adopt a simple model based on an M/G/1/Processor-Sharing queueing model, as follows.  For a tenant with $M$ servers each with a service rate of $\mu$, denote the workload arrival rate by $\lambda$. When $m$ servers are shut down, we model the 
total delay cost as
$\bar{c}(m) =\lambda \cdot \beta\cdot T \cdot \text{delay}(m)=\frac{\beta T}{\frac{1}{uM}-\frac{1}{M-m}}$,
where $u=\frac{\lambda}{\mu M}$ denotes the normalized workload arrival (i.e.,
 utilization without turning off servers),
$T$ is the duration of an \edr event, and $\beta$ is a cost parameter ({\$/time unit/job}). In our simulations, we set the cost parameter for tenant 1, tenant 2 and tenant 3 as 0.1, 0.03, 0.006, respectively, which
are already higher than those considered in the prior context of turning off servers for energy saving \cite{LinWiermanAndrewThereska}.  Note that we have experimented with a variety of other models as well and the results do not qualitatively change.

We use a standard model for energy usage \cite{Hoelzle_datacenter_book_2013} and take the energy reduction $s$ as linear in the number of servers shut down, i.e.,  $s = \theta\cdot m$, where $\theta$ is a constant decided by server's idle power  and $T$.
Then, it yields the following cost function for tenants's energy reduction $c(s) =\bar{c}(\frac{s}{\theta})-\bar{c}(0)$, where $\bar{c}(\,\cdot\,)$ is defined in the above paragraph.
Note that we have experimented with a variety of other forms, and our results are not sensitive to the details of this cost function.  

Finally, note that tenants typically have delay performance requirement which, based on the above queueing model, is  translated as an utilization upper bound. Such
translation is also common in real systems (e.g., default policy  for auto-scaling virtual machines \cite{Azure_AutoScaling_Rule}). In our simulation, we capture the performance constraint by setting utilization upper bounds for tenant 1, tenant 2, and tenant 3 as 0.5, 0.6, and 0.8, respectively.

\textbf{Efficiency benchmarks.} Throughout our experiments, we consider the price-taking, price-anticipating, and social optimal outcomes.  Additionally, we consider one other benchmark, \emph{diesel only}, which is meant to capture common practice today.  Under diesel only, the full \edr response is provided by the on-site diesel generator.  \textit{Throughout, our results are presented in grouped bar plots with the bars representing (from left to right) the price-taking, price-anticipating, social optimal, and diesel only (if applicable) outcomes. }

While other mechanisms (e.g., direct pricing \cite{Liu:2014:PDC:2591971.2592004},
auction \cite{Shaolei_Colo_TruthDR_Tech}) have been introduced in recent papers, we do not compare \ouralg with them here because \ouralg is already typically indistinguishable from the social optimal cost.

\subsection{Performance Evaluation}

We now discuss our main results, shown in Fig.~\ref{fig:default}.

\textbf{Social cost.} We first compare in
Fig.~\ref{fig:default:welfareLoss} the social costs incurred by different algorithms. Note that \ouralg is close to the social cost optimal under both price-taking
and price-anticipating cases even though there are only three participating
tenants.  Further, the resulting social costs in both the price-taking and price-anticipating scenarios are significantly lower than that of the diesel only outcome. This shows a great potential of tenants' IT power reduction
for \edr, which is consistent with the prior literature on owner-operated
data center demand response \cite{aikema2012data_ancillary_IGCC_2012,Liu:2014:PDC:2591971.2592004,AdamWierman_DataCenterDemandResponse_Survey_IGCC_2014}.

\textbf{Energy reduction contributions.}
Fig.~\ref{fig:default:energy} plots \edr energy reduction contributions from tenants and the diesel generator. As expected from analytic results, both price-taking and price-anticipating tenants tend to
contribute less to \edr (compared to the social optimal) because of their self-interested decisions.
In other words, given self-interested tenants, the colo operator needs more diesel generation than the social optimal. Nonetheless, the difference is fairly small, much smaller than predicted by the worst-case analytic results.  This highlights that worst-case results were too pessimistic in this case.  Of course, one must remember that all tenant reduction extracted is in-place of diesel generation, and so serves to make the demand response more environmentally friendly.

\textbf{Benefits for tenants and colocation operator.}
We show in Fig.~\ref{fig:default:netUtilityTenant} and
Fig.~\ref{fig:default:netUtilityDC} that both the tenants and the colo operator can benefit from \ouralg.
Specifically, Fig.~\ref{fig:default:netUtilityTenant} presents net profit (i.e., payment made by colo operator minus
performance cost) received by tenants, showing that all participating tenants receive positive net rewards.
While price-anticipating tenants can receive higher net rewards than when they are price-taking, the extra reward
gained is quite small. Similarly, Fig.~\ref{fig:default:netUtilityDC} shows cost saving for the colo operator, compared to the ``diesel only'' case .

\textbf{Market clearing price.}
Fig.~\ref{fig:default:price} shows the market clearing price. Naturally, when using \ouralg to incentivize tenants for \edr while
minimizing the total cost, the colo operator will not pay the tenants
at price higher than its diesel price (shown via the red horizontal line). We also note that the price under the price-anticipating case is higher than that under the price-taking case, because the price-anticipating tenants are more strategic. However, the price difference between price-anticipating and price-taking cases is quite small, which again confirms our analytic results.

\textbf{Tenant' server utilization.}
Tenants' server utilizations are shown in
Figs.~\ref{fig:default:u1}, \ref{fig:default:u2} and \ref{fig:default:u3}, respectively.  These illustrate that, while tenants reduce energy for \edr, their server utilizations still stay within their respective limits (shown via the red horizontal lines), satisfying performance constraints. This is because
tenants typically provision their servers based on the maximum possible workloads (plus a certain margin), while in practice their workloads are usually quite low, resulting
in a ``slackness'' that allows for saving energy while still meeting their performance requirements.

\subsection{Sensitivity Analysis}

To complete our case study, we investigate the sensitivity of the conclusions discussed above to the settings used. For each study, we only show results that are significantly different than those in Fig.~\ref{fig:default}.

\begin{figure}[t]
        \centering
	  \subfigure[Social cost]{\includegraphics[width=0.23\textwidth]{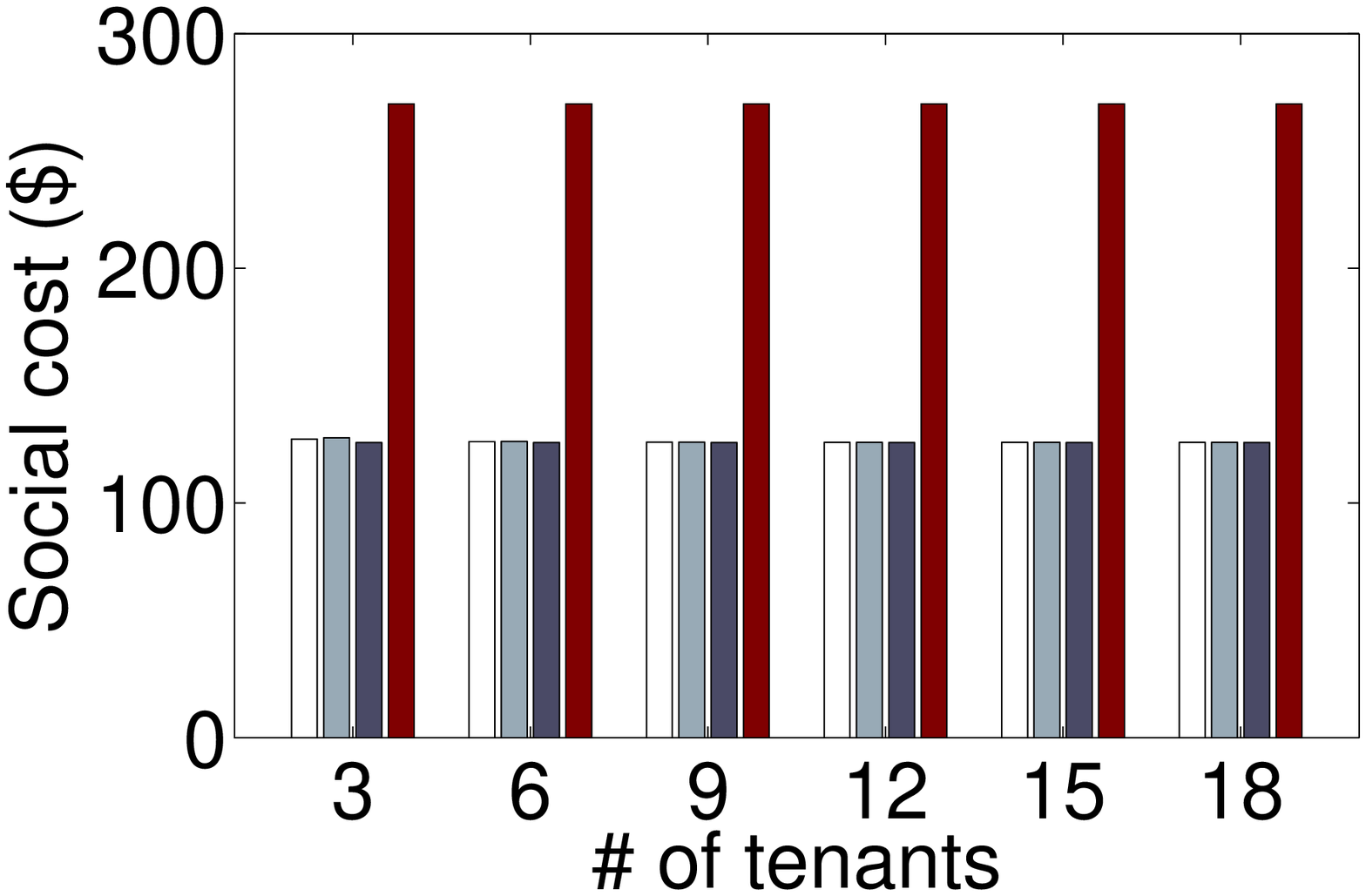}\label{fig:diffN:welfareLoss}}
       \subfigure[Market price]{\includegraphics[width=0.23\textwidth]{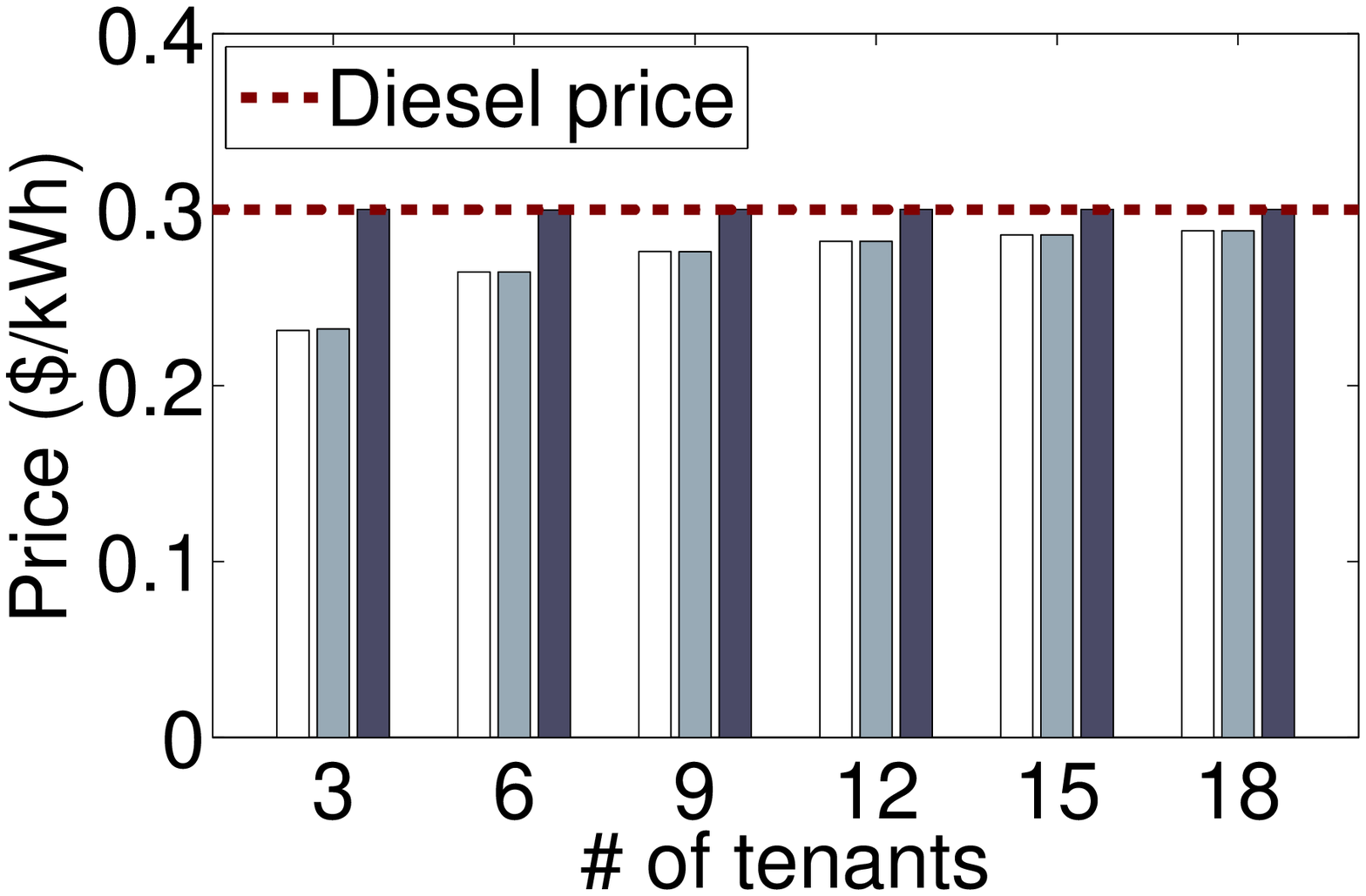} \label{fig:diffN:price}}
	  \subfigure[Tenants' net profits]{\includegraphics[width=0.23\textwidth]{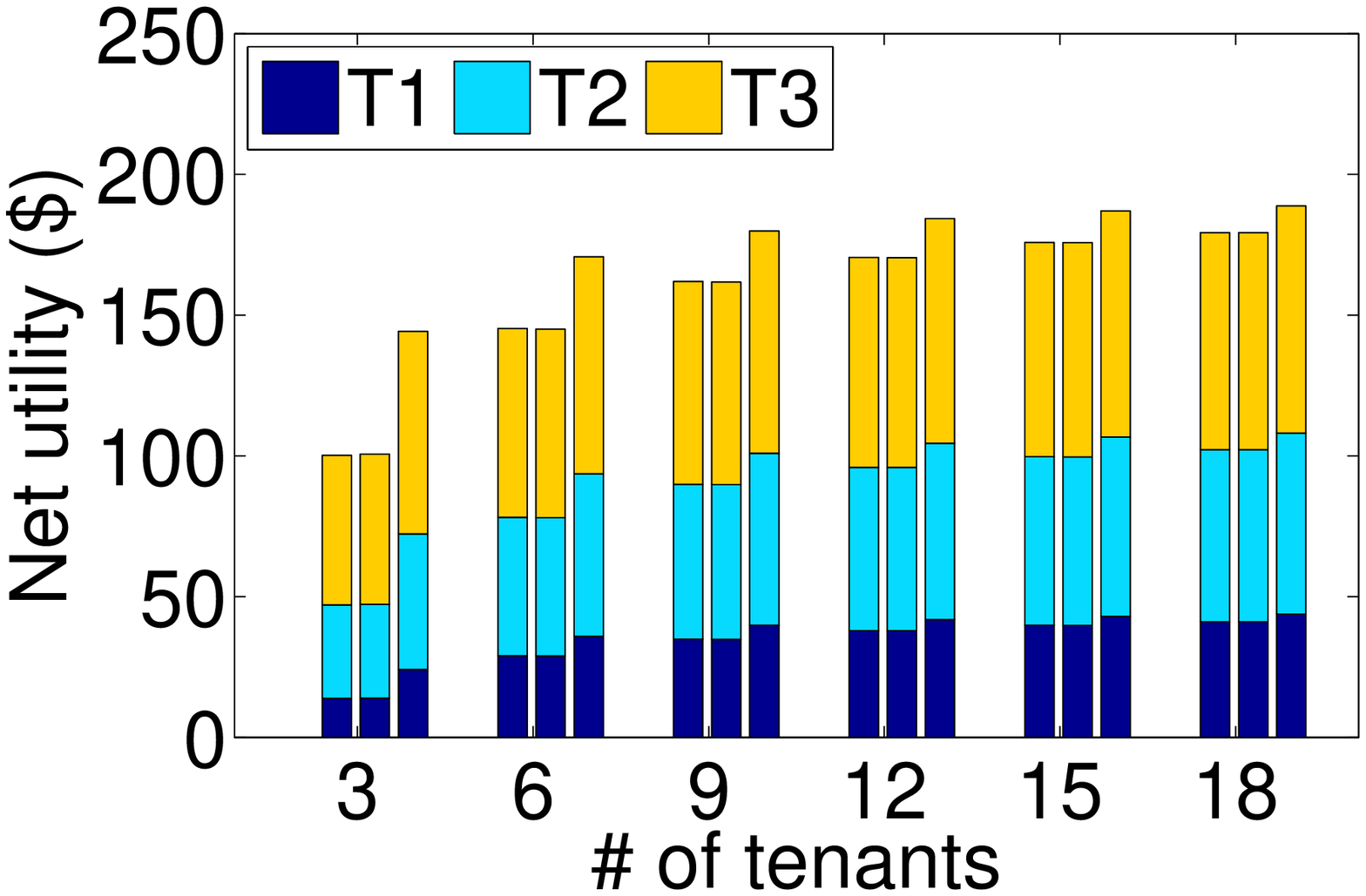}\label{fig:diffN:netUtilityTenant}}
	  \subfigure[Operator's total cost]{\includegraphics[width=0.23\textwidth]{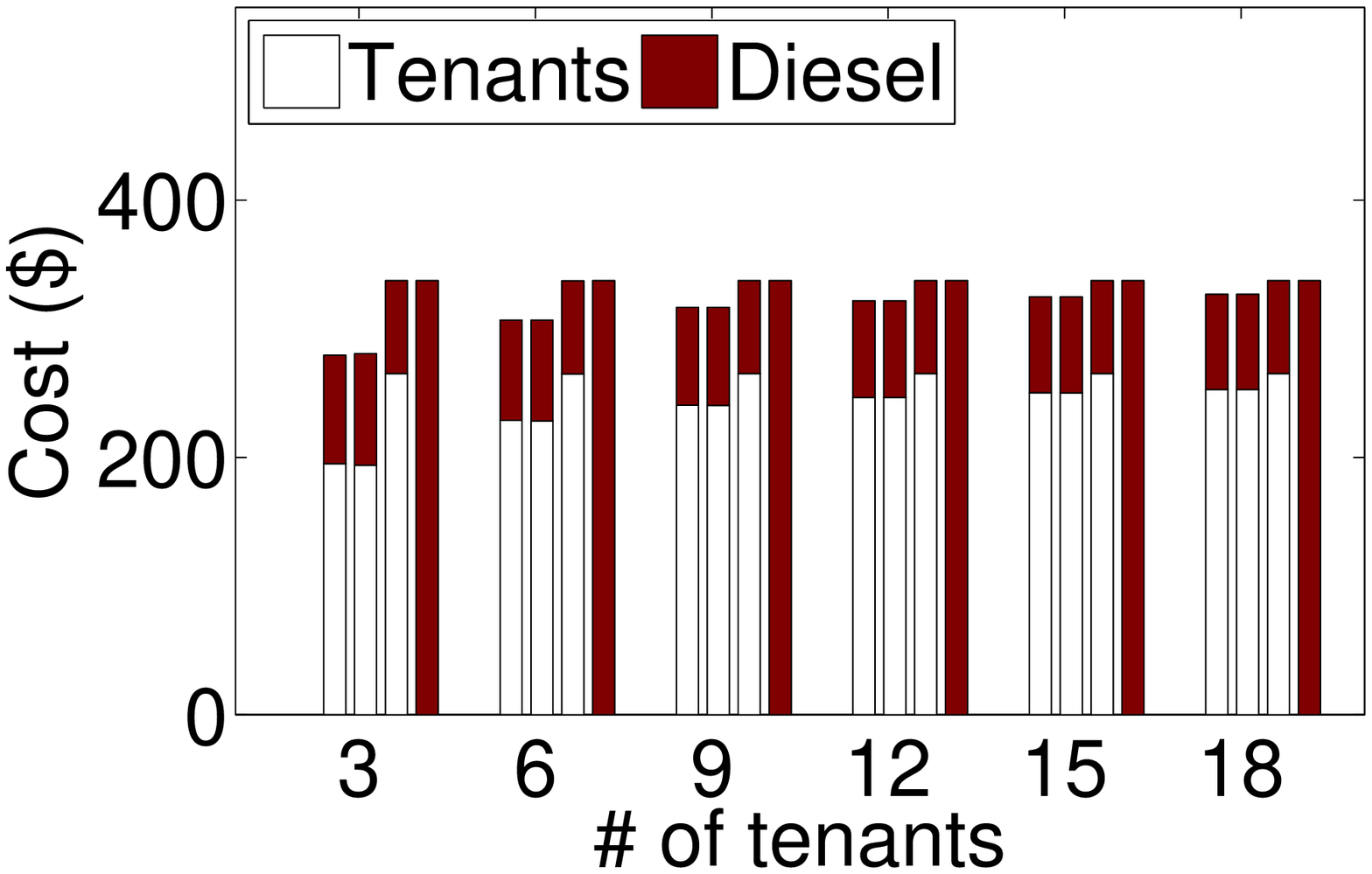}\label{fig:diffN:netUtilityDC}}
        \caption{Impact of number of tenants.}\label{fig:N}
\end{figure}

\textbf{Impact of the number of tenants.}
First, we vary the number of participating tenants and show the results in Fig.~\ref{fig:N}.
To make results comparable, we fix the \edr energy reduction requirement as well as total number of servers: tenant 1, tenant 2 and tenant 3 are each equally split into multiple smaller tenants, each having fewer servers.
We then aggregate replicas of the same tenant together for an easy viewing in the figures, e.g., ``tenant 1'' in the figures represent the whole group of tenants that are obtained by splitting the original tenant 1. 
One interesting observation is that as more tenants participate in \edr, the market becomes more ``competitive''.
Hence, each individual
tenant can only gain less net reward, but both the price and the aggregate net reward  become
higher (see Figs.~\ref{fig:diffN:price} and~\ref{fig:diffN:netUtilityTenant}).
Motivated by this, one might suggest a possible trick: a tenant may gain more utility by splitting its servers and pretending as multiple tenants. In practice, however, each tenant has only one account (for billing, etc.) which requires contracts and base fees, and thus pretending as multiple tenants is not possible in a colo.

\begin{figure}[t]
\centering
\subfigure[Social cost]{\includegraphics[width=0.23\textwidth]{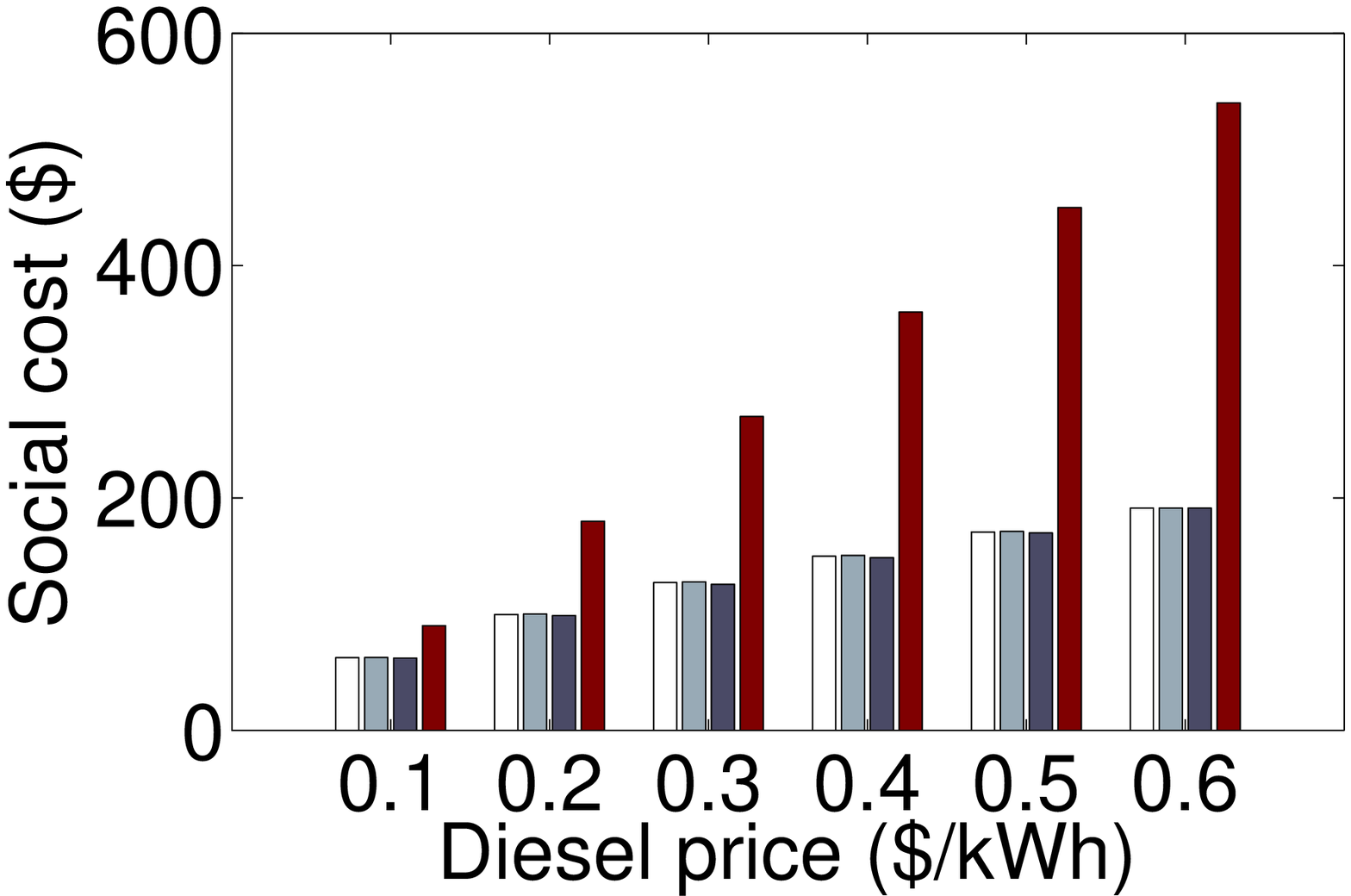}\label{fig:diffAlpha:welfareLoss}}
\subfigure[][Energy reduction]{\includegraphics[width=0.23\textwidth]{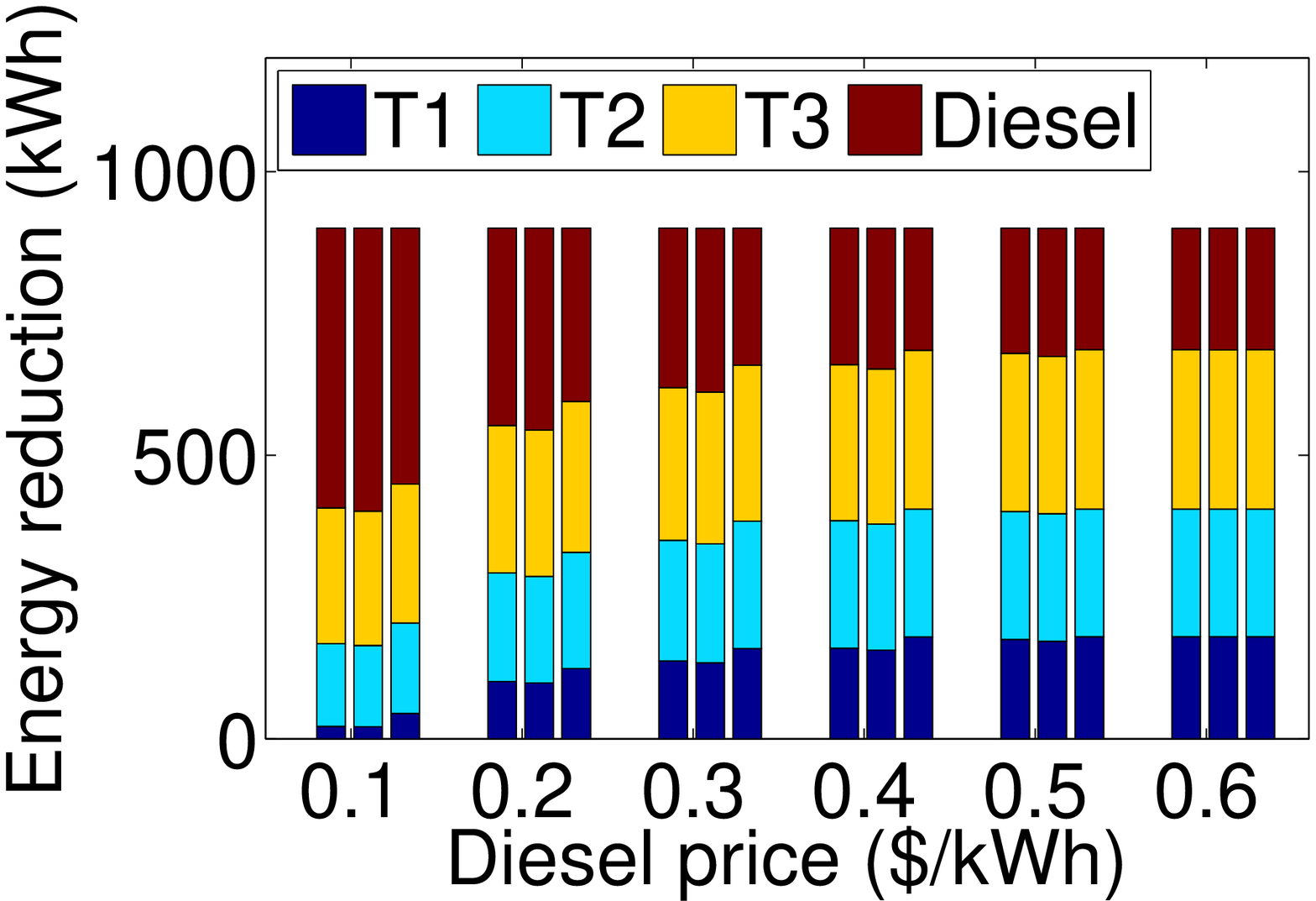} \label{fig:diffAlpha:energy}}
\subfigure[Market price]{\includegraphics[width=0.23\textwidth]{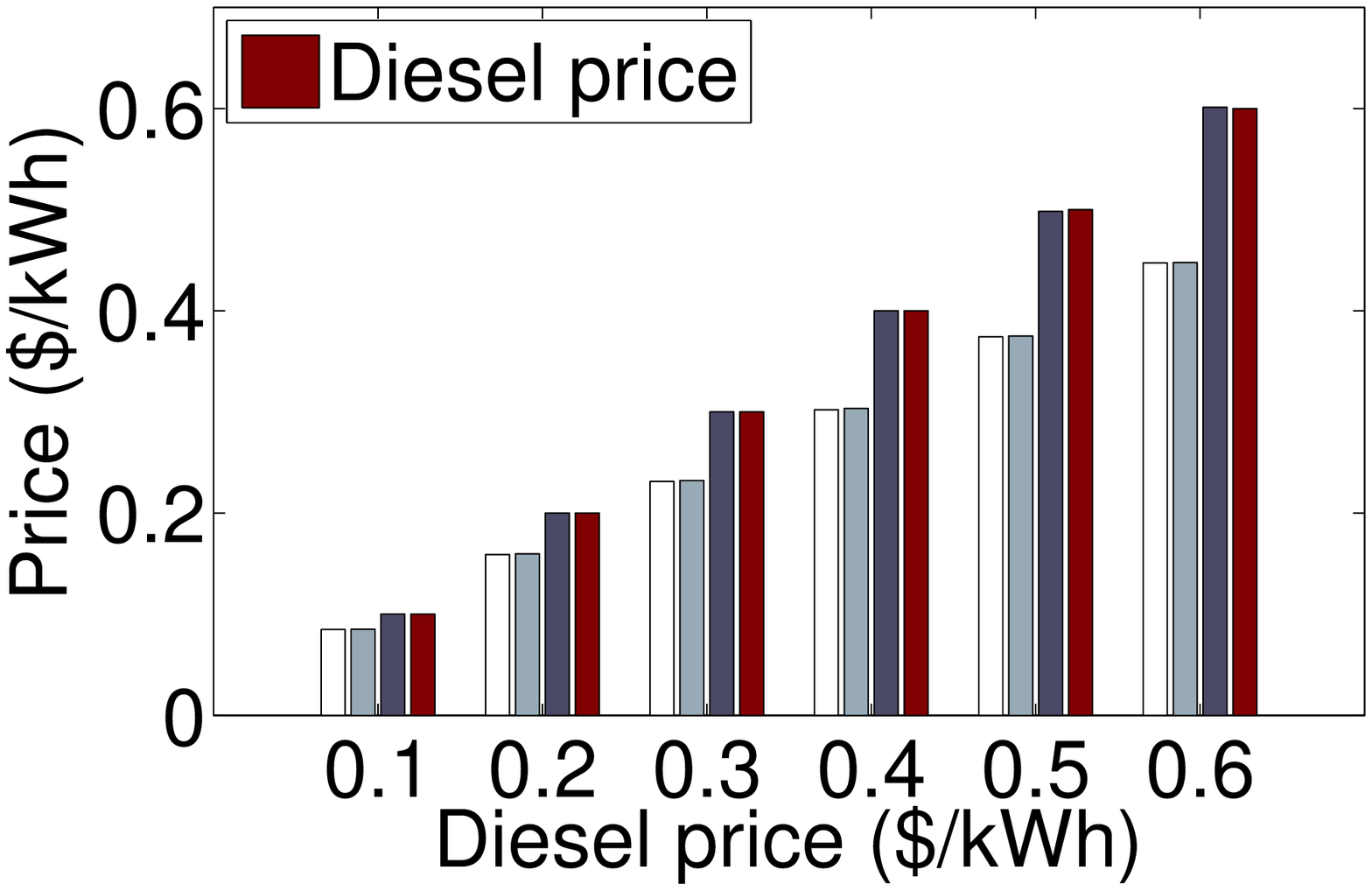} \label{fig:diffAlpha:price}}
\subfigure[Tenants' net profits]{\includegraphics[width=0.23\textwidth]{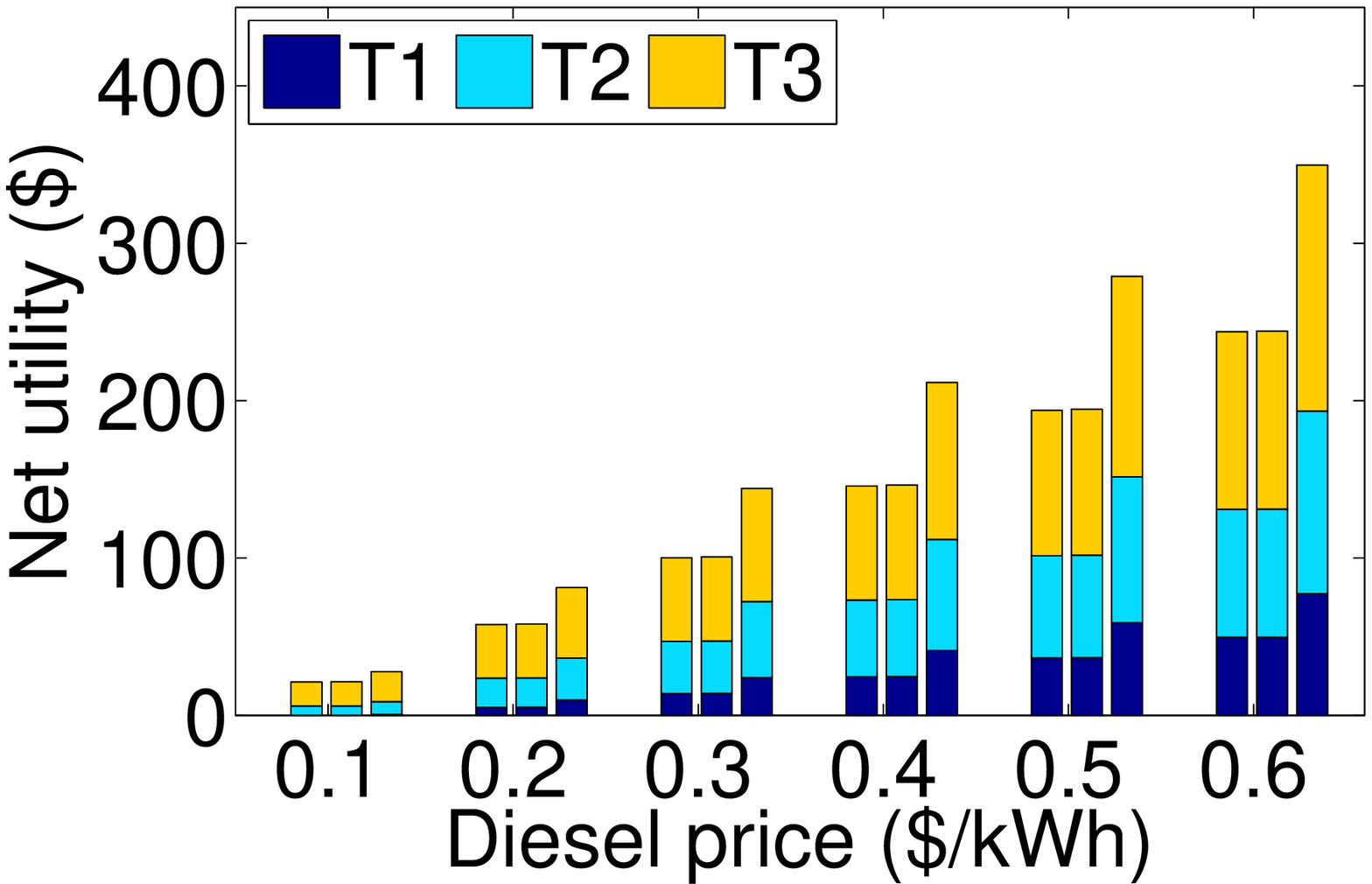}\label{fig:diffAlpha:netUtilityTenant}}
\caption{Impact of diesel price.}\label{fig:alpha}
\end{figure}

\textbf{Impact of the price of diesel.}
Fig.~\ref{fig:alpha} illustrates how our result changes as the diesel price varies. Intuitively, as
shown in Fig.~\ref{fig:diffAlpha:welfareLoss}, the social cost (which includes diesel cost as a key component) increases with the diesel price. We see from Figs.~\ref{fig:diffAlpha:energy} and \ref{fig:diffAlpha:price} that, when diesel price is very low (e.g.,  $0.1$\$/kWh), the colo operator is willing to use more diesel and offers a lower price to tenants.
As a result, tenants contribute less to \edr.
As the diesel price increases (e.g., from $0.2$\$/kWh to $0.3$\$/kWh), the colo operator increases the market price (but still below the diesel price) to encourage tenants to cut more energy for \edr. Nonetheless,
tenants' energy reduction contribution cannot increase arbitrarily due to their performance constraints.
Specifically, after the diesel price exceeds 0.4\$/kWh, tenants will not contribute more to \edr (i.e., almost all their IT energy reduction capabilities have been exploited), even
though the colo operator increases the reward. In this case, tenants simply receive higher net
rewards without further contributing to \edr, as shown in Fig.~\ref{fig:diffAlpha:netUtilityTenant}.

\begin{figure}[t]
\centering
\subfigure[Energy reduction]{\includegraphics[width=0.22\textwidth]{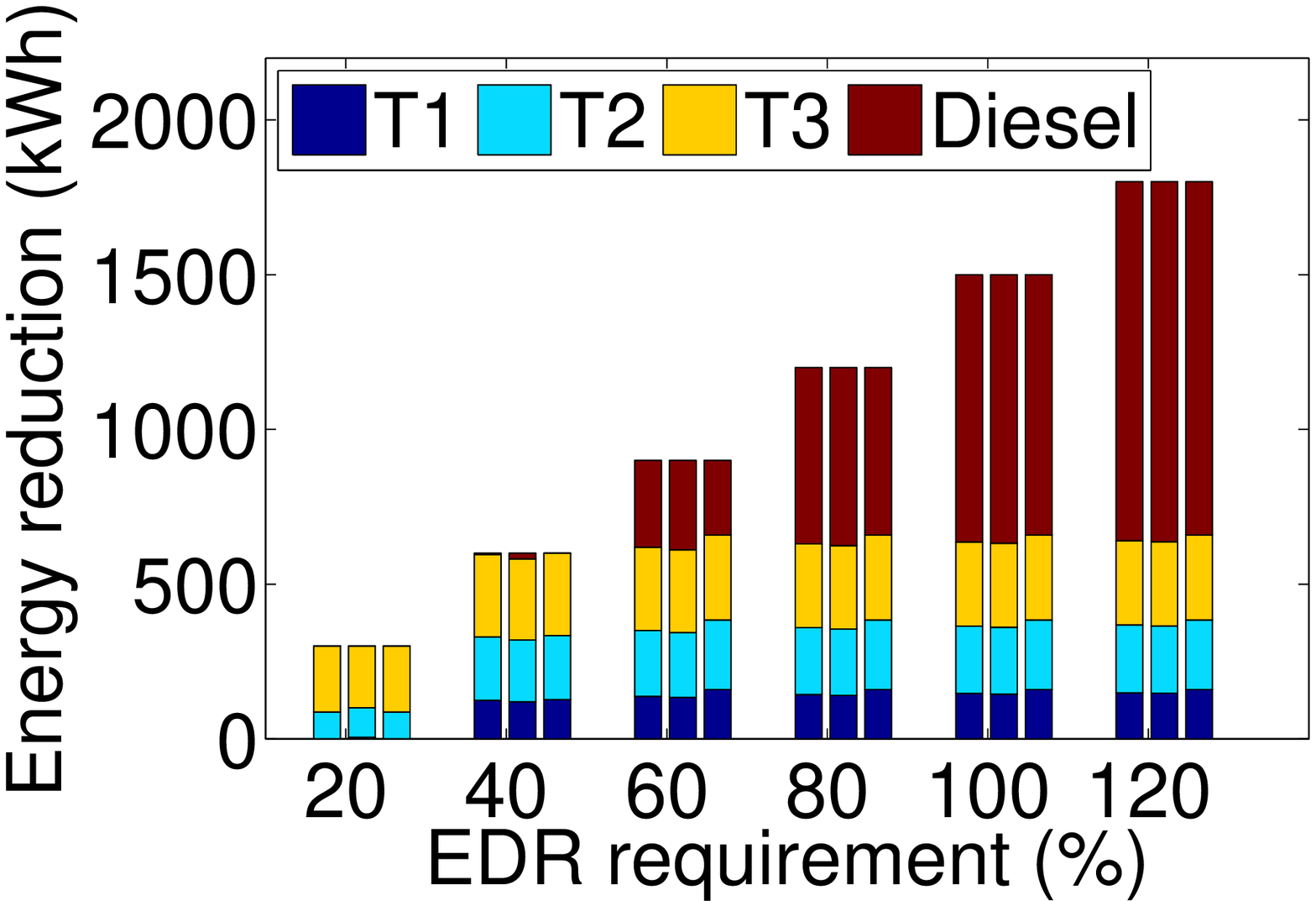} \label{fig:diffDelta:energy}}
\subfigure[Market price]{\includegraphics[width=0.22\textwidth]{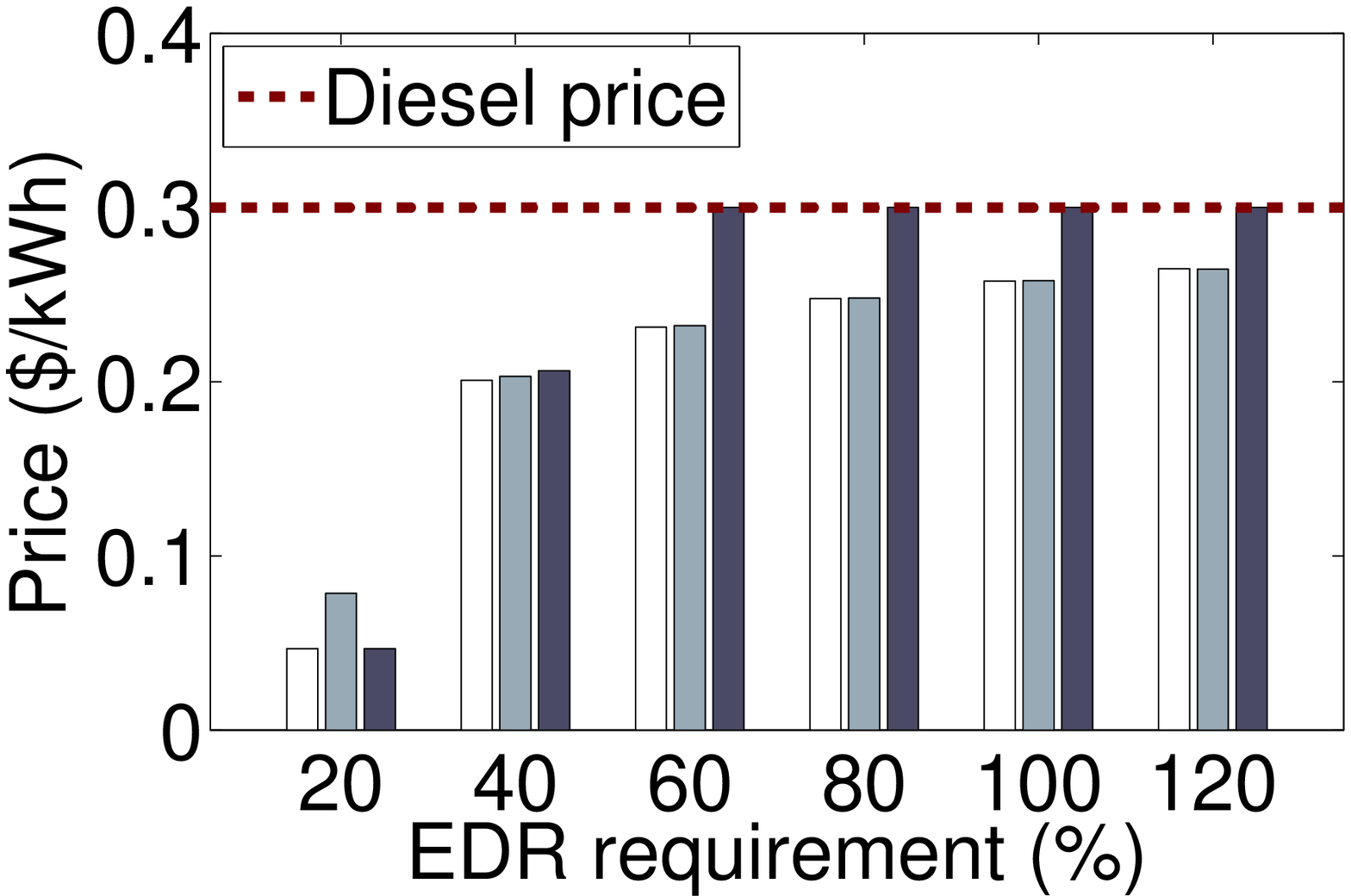} \label{fig:diffDelta:price}}
\caption{Impact of \edr energy reduction target.}\label{fig:delta}
\end{figure}

\textbf{Impact of \edr requirement.}
Fig.~\ref{fig:delta} varies the \edr energy reduction target, with the maximum reduction ranging from $20\%$ to $120\%$ of the colo's peak IT power consumption.
As the \edr energy reduction target increases, tenants' energy reduction for \edr also increases;
after a certain threshold, diesel generation becomes the main approach to \edr, while the increase in tenant's contribution is diminishing (even though
the colo operator increases the market price), because of tenants' performance requirements
that limit their energy reduction capabilities.

\begin{figure}[t]
\centering
\subfigure[Social cost]{\includegraphics[width=0.23\textwidth]{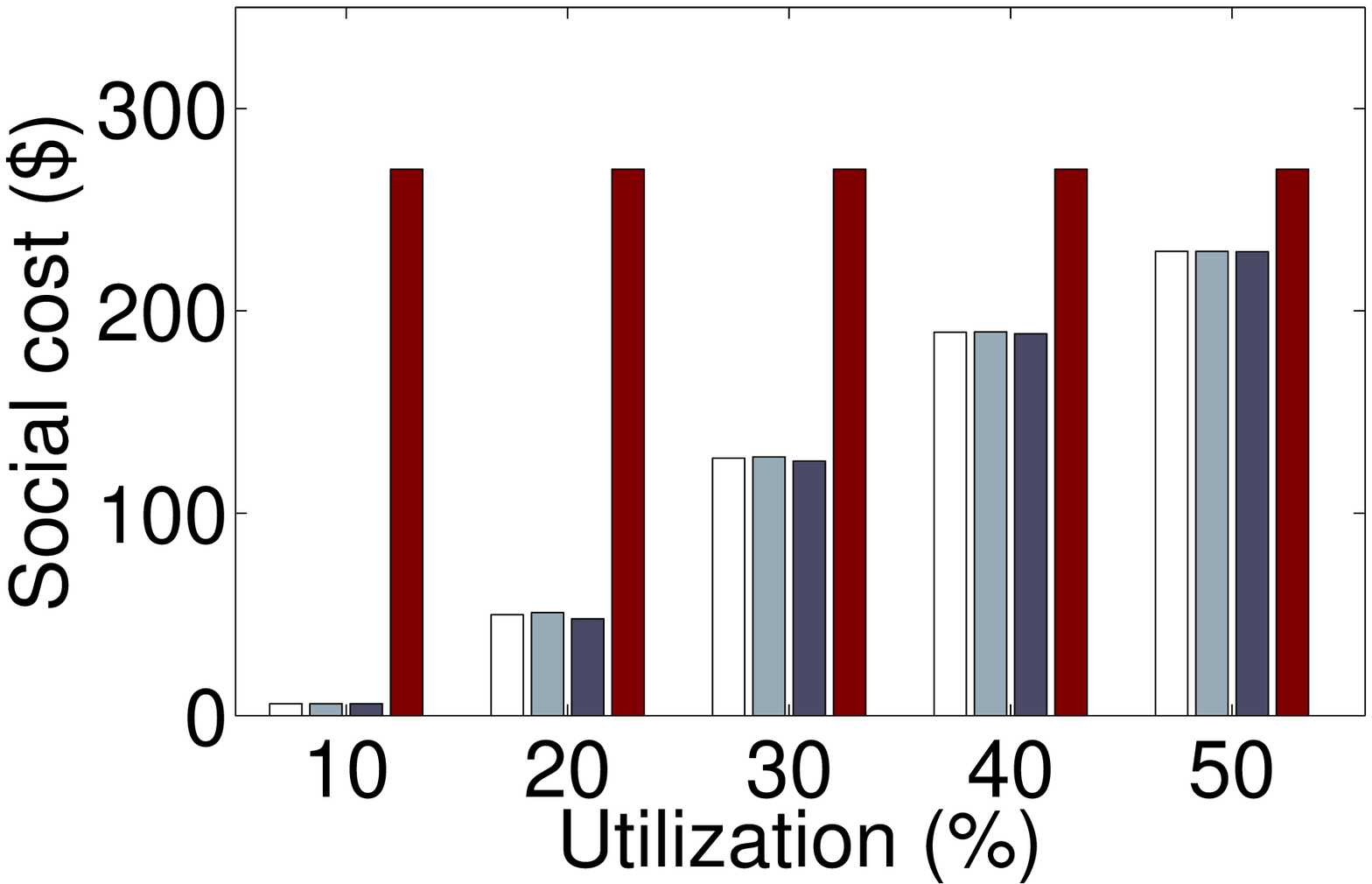}\label{fig:diffU:welfareLoss}}
\subfigure[Energy reduction]{\includegraphics[width=0.23\textwidth]{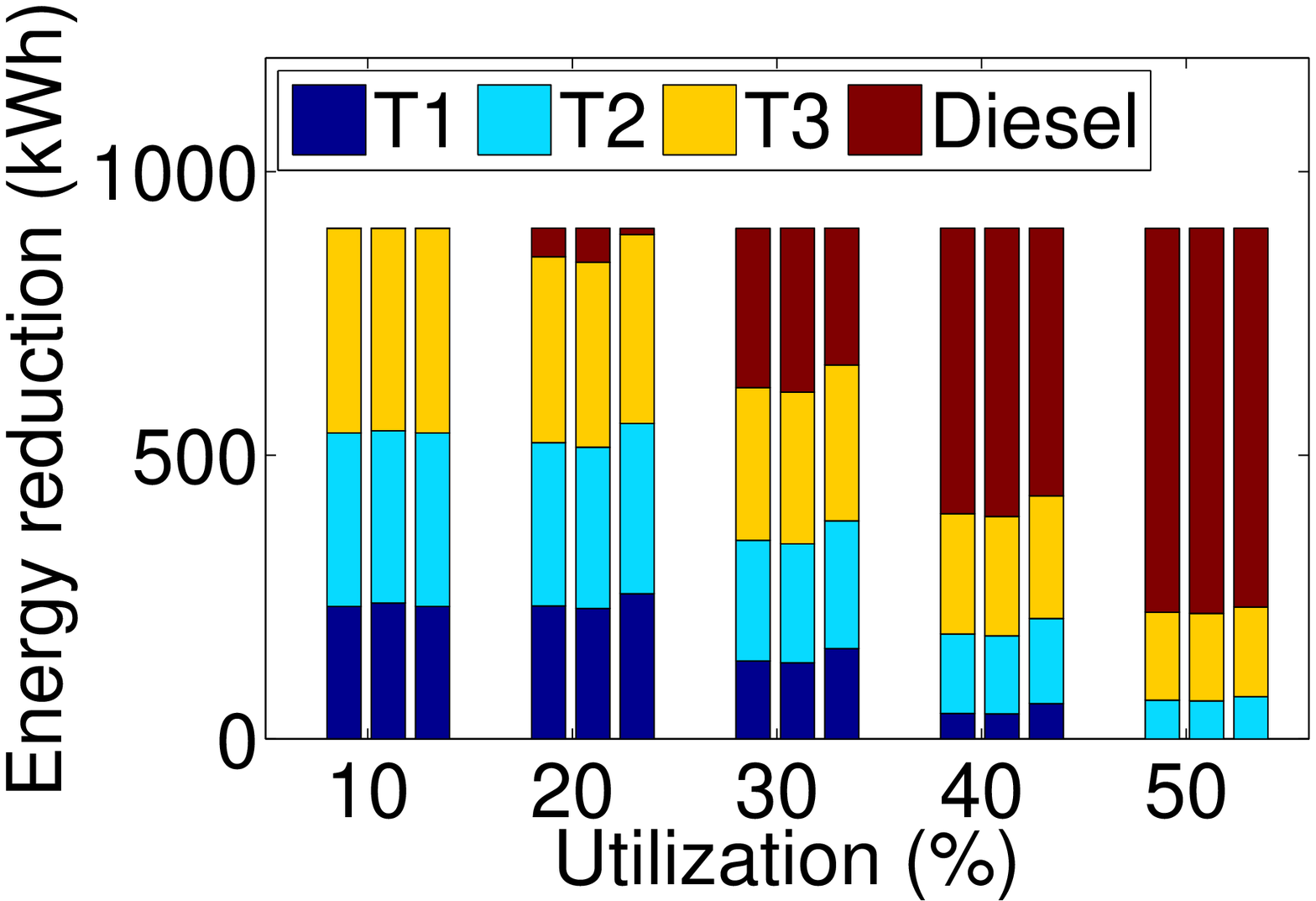} \label{fig:diffU:energy}}
\subfigure[Social cost]{\includegraphics[width=0.23\textwidth]{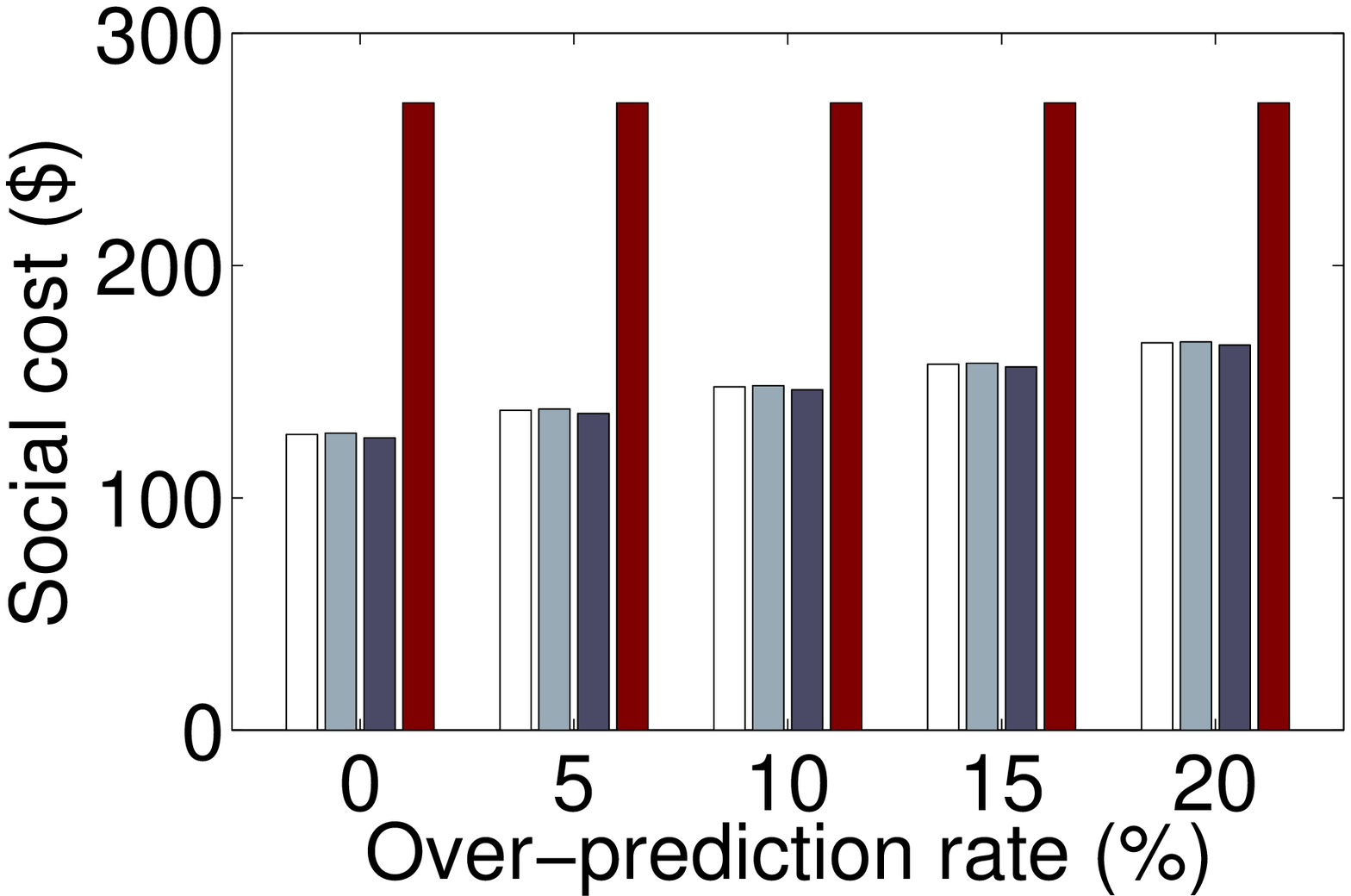}\label{fig:preError:welfareLoss}}
\subfigure[Market price]{\includegraphics[width=0.23\textwidth]{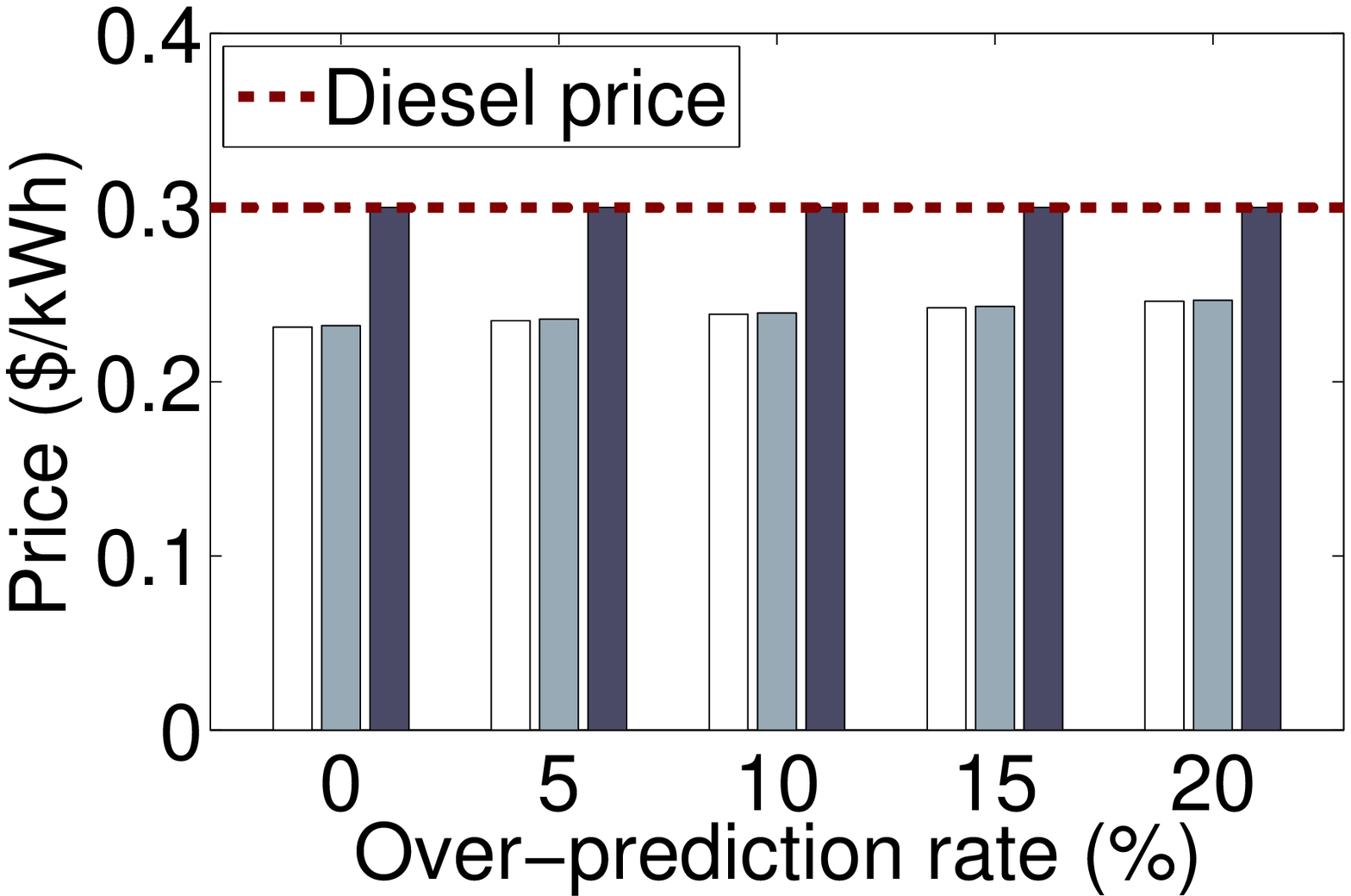} \label{fig:preError:price}}
\caption{Impact of tenants' workloads and the workload prediction errors.}\label{fig:u}
\end{figure}

\textbf{Impact of tenants' workloads.}
In Fig. \ref{fig:diffU:welfareLoss}-\ref{fig:diffU:energy}, we vary the tenants' workload intensity (measured in terms of the average server utilization
when all servers are active) from $10\%$ to $50\%$, while
still keeping the maximum utilization bounds to $50\%$,  $60\%$ and $80\%$ as
the performance requirements for the three tenants, respectively.
While it is straightforward that when tenants have more workloads, they tend
to contribute less to \edr, because they need to keep more servers active
to deliver a good performance. Nonetheless, even when their average
utilization without turning off servers is as high as 50\% (which is quite high in real systems,
considering that the average utilization is only around 10-30\% \cite{Hoelzle_datacenter_book_2013}),
tenants can still contribute more than 20\% of \edr energy reduction under \ouralg,
showing again the potential of IT power management for \edr.

\textbf{Impact of workload prediction error.}
In practice, tenants may not perfectly estimate their own workload arrival rates. To cope with possible traffic spikes, tenants can either keep more servers active as a backup or deliberately overestimate the workload arrival rate by a certain overestimation factor. We choose the later approach in our simulation.
Fig.~\ref{fig:preError:welfareLoss}-\ref{fig:preError:price} shows the result under workload prediction errors. We see that
both the social cost and market price are fairly robust against tenants' workload over-predictions.
For example, the social cost
increases by less than $10\%$, even when tenants overestimate their workloads
 by $20\%$ (which is already sufficiently high in practice, as shown in \cite{Gandhi:2012:ADR:2382553.2382556}).
Other results (e.g., tenants' net reward, colo operator's total cost)
are also only minimally affected,
thereby demonstrating the robustness of \ouralg against
tenants' workload over-predictions.



\section{Related Work}\label{sec:related_work}

Our work contributes both to the growing literature on data center demand response, and to the literature studying supply function equilibria. We discuss each in turn below.


Recently, data center demand response has received a growing amount of attention.  A variety of approaches have been considered, such as optimizing grid operator's pricing strategies for data centers \cite{Liu:2014:PDC:2591971.2592004}
and tuning computing (e.g., server control and scheduling) and/or non-computing knobs (e.g., cooling system) in data centers for various types of demand response programs
\cite{aikema2012data_ancillary_IGCC_2012,Wang:2014:ESG:2567529.2567556,DataCenterDemandResponse_CollaborativeOPtimization_10.1109/ICDEW.2013.6547462,AksanliRosing14_ProvidingRegulationServicesManagingDataCenterPeakPower,Chen_PowerControl_Regulation_CDC_2013}.
Field tests by LBNL also verify the practical feasibility of
data center demand response using a combination of existing power management techniques
(e.g., load migration) \cite{DataCenterDemandResponse_Report_Berkeley}.
These studies, however,  have all focused on large owner-operated data centers.

In contrast, to our best knowledge, colocation demand response has been investigated by only a few previous works.  The first is \cite{Shaolei_Colocation_ICAC_2014}, which proposes a simple mechanism, called iCODE, to
incentivize tenants' load reduction.  But, iCODE is purely based on ``best effort'' and does not include any energy reduction target (needed for \edr). More importantly, iCODE is designed without considering strategic behavior by tenants, and can be compromised by price-anticipating tenants  \cite{Shaolei_Colocation_ICAC_2014}. More relevant to the current work is \cite{Shaolei_Colo_TruthDR_Tech},
which proposes a VCG-type auction mechanism where colocation participation in \edr programs.  While the mechanism is approximately truthful, it
asks participating tenants to reveal their private cost information through complex bidding functions.  Further, the colocation operator may be forced to make arbitrarily high payments to tenants.  In contrast, our proposed solution provides a simple bidding space, protects tenants' private valuation,
and ensures that the colocation operator does not incur a higher cost for \edr than
the case tenant contributions. Thus, unlike \cite{Shaolei_Colo_TruthDR_Tech},
\ouralg benefits both colocation operator and tenants, giving both parties incentives to cooperate for \edr.

Finally, it is important to note that our approach builds on, and adds to, the supply function mechanism literature.  Supply function bidding (c.f. the seminal work by \cite{klemperer1989}) is frequently used in electricity markets due to its simple bidding language and the avoidance of the unbounded payments typical in VCG-like mechanisms.  Supply function bidding mechanisms have been extensively studied, e.g., \cite{day2002, baldick2004, green1992, green1996, anderson2008, vives2011}.  The literature primarily focuses on existence and computation of supply function equilibrium, sometimes additionally proving bounds on efficiency loss.  Our work is most related to \cite{johari2011}, which considers an inelastic demand $\delta$ that must be satisfied via extracting load shedding from consumers and proves efficient bounds on supply function equilibrium.  In contrast, our work assumes that the operator has an outside option (diesel) that can be used to satisfy the inelastic demand.  This leads to a multistage game between the tenants and the profit-maximizing operator, a dynamic which has not been studied previously in the supply function literature.

\section{Conclusion}

In this paper, we focused on ``greening'' colocation demand response by designing a pricing mechanism that can extract load reductions from tenants during \edr events.  Our mechanism, \ouralg, can be used in both mandatory and voluntary \edr programs and is easy put in place given systems available in colos today. The main technical contribution of the work is the analysis of the \ouralg mechanism, which is a supply function mechanism for an elastic setting, a setting for which efficiency results have not previously been attained in the supply function literature.  Our results highlight that \ouralg provides provably near-optimal efficiency  guarantees, both when tenants are price-taking and when they are price-anticipating. We also evaluate \ouralg using trace-based simulation studies and validate
that \ouralg  is both beneficial to the colo operator (by reducing costs), to the environment (by reducing diesel usage), and to the tenants (by providing payments for reductions).

{\scriptsize
\bibliographystyle{abbrv}
\bibliography{ref,ref_shaolei}
}

\newpage

\appendix


\newcommand{\lmmc}{\frac{
\partial^-\hat{c}_n(s_n)}{
\partial s_n}}
\newcommand{\rmmc}{\frac{
\partial^+\hat{c}_n(s_n)}{
\partial s_n}}
\newcommand{\cost}{\mathrm{cost}} \numberwithin{equation}{section}

\section{Price taking tenants}
\subsection{Proof of Proposition \ref{prop: ec1}} When tenants are price takers, they maximize the payout $P_n(b_n, p) = pS_n(b_n, p) - c_n(s_n)$ over the bid $b_n$. Note that $b_n \in [0, p\delta]$ as no tenant will bid beyond $p\delta$ otherwise the payout $P_n<0$. Hence $\mathrm{b} = (b_1, \ldots, b_n)$ is an equilibrium if and only if the following condition is satisfied
{\small \begin{subequations}
	\begin{align}
		\label{eqn: ne1} \frac{
		\partial^-c_n(s_n)}{
		\partial s_n} \le p, \quad & 0 \le b_n < p\delta, \\
		\label{eqn: ne2} \frac{
		\partial^+c_n(s_n)}{
		\partial s_n} \ge p, \quad & 0 < b_n \le p\delta.
	\end{align}
\end{subequations}
}
At least one feasible solution to \eqref{eqn: p1} exists because it is minimizing a continuous function over a compact set. Furthermore, \eqref{eqn: p1-2} - \eqref{eqn: p1-3} satisfy standard constraint qualification, hence for the Lagrangian
{\small\[L(\mathrm{s}, \mu) = \sum_{n} c_n(s_n) + \mu((\delta-y) - \sum_n s_n), \]}
there exists optimal primal dual pair $(\mathrm{s}, \mu)$, such that \eqref{eqn: p1-2} and \eqref{eqn: p1-3} are satisfied, and
{\small\begin{subequations}
	\begin{align}
		\label{eqn: opt1} \frac{
		\partial^-c_n(s_n)}{
		\partial s_n} \le {\mu}, \quad & s_n > 0, \\
		\label{eqn: opt2} \frac{
		\partial^+c_n(s_n)}{
		\partial s_n} \ge {\mu}, \quad & s_n \ge 0.
	\end{align}
\end{subequations}}
Given the optimal $(\mathrm{s}, \mu)$, let $p = {\mu}$, and $b_n = p(\delta - s_n)$, then \eqref{eqn: p1-2} implies $p$ satisfies \eqref{eqn: price}, and \eqref{eqn: opt1}-\eqref{eqn: opt2} implies \eqref{eqn: ne1} - \eqref{eqn: ne2}, hence an equilibrium exists.

Conversely, if $(\mathrm{b}, p)$ is an equilibrium and $p$ satisfies \eqref{eqn: price}, the resulting allocation $\mathrm{s}$ is optimal to \eqref{eqn: p1}. To see this, if $0 \le s_n < \delta - y$ for all $n$, \eqref{eqn: ne1}-\eqref{eqn: ne2} is equivalent to \eqref{eqn: opt1}-\eqref{eqn: opt2} if we set $\mu = p$, hence $(\mathrm{s}, \mu)$ is primal dual optimal pair for \eqref{eqn: p1}. If $s_n = (\delta - y)$, then $s_m = 0, \forall m \ne n$. In this case, we set $\bar{\mu} = \min\{ p,
\partial^+c_n(s_n)/
\partial s_n\}$, and we can check that $(\mathrm{s}, \bar{\mu})$ is the primal dual optimal solution for \eqref{eqn: p1}.

\subsection{Proof of Theorem \ref{thm: price-taking-characterization}} By Proposition \ref{prop: ec1}, when tenants are price-taking, for any $y$, the there is always an equilibrium, and the resulting $\mathbf{s}$ is always the optimal allocation to provide $(\delta - y)$ energy reduction.

Hence we only need to verify that the on-site generation level $y$ is the solution to \eqref{eqn: price-taking1}-\eqref{eqn: price-taking3}. Similar to the proof of Proposition \ref{prop: ec1}, by Assumption \ref{asn: cheap_on-site}, the first order optimality condition for the $y$ in \eqref{eqn: price-taking1}-\eqref{eqn: price-taking3} is $\frac{\alpha}{N\delta}(y+(N-1)\delta) = p.$
By Proposition \ref{prop: ec1}, $p$ satisfies the relation \eqref{eqn: price}, substitute the left-hand-side into \eqref{eqn: price} and solve for $y$, we have
$y = \sqrt{\frac{\Sigma_n b_n N\delta}{\alpha}} - (N-1)\delta.$
This is exactly the on-site generation $y$ that minimizes $\mathrm{cost}_o(\mathbf{b}, y)$ given in \eqref{eqn: on-site_gen}. Hence the datacenter will always pick $y$ that is optimal for \eqref{eqn: price-taking1}-\eqref{eqn: price-taking3}, together with Proposition \ref{prop: ec1}, an equilibrium exists, and the resulting allocation $(\mathbf{s}, y)$ is optimal for \eqref{eqn: price-taking1}-\eqref{eqn: price-taking3}.

\subsection{Proof of Proposition \ref{prop: ec2}} Since $y \ge 0$, it suffices to prove that whenever the optimal on-site generation is non-zero, $y^* >0$, $y^t \ge y^*$. From \eqref{eqn: edr1}, the Lagrangian of \pone is
\[ L(\mathrm{s}, y, \mu^*, \lambda^*) = \sum_n c_n(s_n) + \alpha y + \mu^*((\delta - y) - \sum_n s_n) - \lambda^* y. \]
By constraint qualification and the KKT conditions,
assuming $y^* >0$, then $\lambda = 0$, $\mu^* = \alpha$, hence the market clearing price in the optimal allocation should be $p^* = \alpha $.

Next, consider the market price for price taking tenants. From \eqref{eqn: price2},
\begin{equation}
	\label{eqn: price_t1} p^t = \frac{ \sum_{i\in\mathcal{N}}b^t_i}{(N-1)\delta+y^t} = \sqrt{\frac{(\Sigma_{i\in \mathcal{N}}b^t_i)\alpha}{N\delta}}.
\end{equation}
The second equality yields $ \sum_{i\in\mathcal{N}}b^t_i = \frac{((N-1)\delta+y^t)^2}{N\delta}\alpha$. Substitute this back to \eqref{eqn: price_t1},
\begin{align}
	\label{eqn: price_t2} p^t &= \frac{ \sum_{i\in\mathcal{N}}b^t_i}{(N-1)\delta+y^t}= \frac{(N-1)\delta+y^t}{N\delta}\alpha.
\end{align}
And note that $y_t \in [0, \delta]$ and $p^* = \alpha$, thus \eqref{eqn: price_t2} yields
$\frac{N-1}{N}p^* \le p^t \le p^*.$

Finally, from \eqref{eqn: price-taking0}, the Lagrangian of the price-taking characterization optimization is,
\begin{align*}
	L(\mathrm{s}, y, \mu^t, \lambda^t) = &\sum_n c_n(s_n) + \yterm
	+ \mu^t((\delta - y) - \sum_n s_n) - \lambda^t y.
\end{align*}
By examining the KKT condition and using a similar argument to the proof of Proposition \ref{prop: ec1}, we have $p^t = \mu^t$, also,
$\frac{\partial^{-}c_n(s^t_n)}{\partial s^t_n} \le p^t \le p^* \le \frac{\partial^{+}c_n(s^*_n)}{\partial s^*_n}.$
Thus, $\forall n, s^t_n \le s^*_n$. Since $y = \delta - \sum s_n$, $y^t \ge y^*$.

\subsection{Proof of Proposition \ref{prop: on-site_gen}} From the proof of Proposition \ref{prop: ec2}, we see that when $y^*>0$, $\lambda^* = 0$, and $\mu^* = \alpha$. Furthermore, we have $ \sum_n s_n < \delta$, but $s_n = \delta - \frac{b_n}{\mu^*}$. Hence
$ (N\delta - \frac{\Sigma_n b_n}{\alpha}) < \delta. $
Conversely, if \eqref{eqn: cheap_on-site} holds, then
$ \alpha (N-1)\delta < \sum_n b_n.$
But by Proposition \ref{prop: ec1} and \eqref{eqn: price}, we have
$ \sum_n b_n = (p^* (N-1)\delta + y). $
By combining the two equations above:
$ \alpha(N-1)\delta < p^*((N-1)\delta+y^*).$
However, from the proof in Proposition \ref{prop: ec1}, we have $p^* \le \alpha$, hence we must have $y^*>0$.

On the other hand, when the data center operator is profit maximizing, the cost to the operator
$\mathrm{cost}_o(\mathbf{b}, y) = \frac{(\Sigma_n b_n)(\delta-y)}{(N-1)\delta + y} + \alpha y$
is a convex function in $y$ over the domain $y \ge 0$. By first order condition, the cost is minimized when
\begin{align}
	y' = \sqrt{\frac{N\ \delta \Sigma_n b_n}{\alpha}} - (N-1)\delta, \label{eqn: yunc}
\end{align}
then $y = y'$ if and only if $y' \in [0, \delta]$. However,
$\Sigma_n b_n = \Sigma_n p(\delta - s_n)={p}((N-1)\delta + y) \le \alpha (N\delta), $
where the last inequality is because $y \le \delta$, and $p \le \alpha$, since operator always has the option to use on-site generation to get unit cost of energy reduction at $\alpha$. Hence we always have $y' \le \delta$. So, if $y >0$, by \eqref{eqn: yunc}, \eqref{eqn: cheap_on-site2} must hold, conversely, if \eqref{eqn: cheap_on-site2} holds, then by \eqref{eqn: yunc}, $y'>0$, so operator will use $y = y'$.

\subsection{Proof of Theorem \ref{thm: welfare_loss_taking}} Note that $(\mathbf{s}^*, y^*)$ is a feasible solution to \eqref{eqn: price-taking0}. By Theorem \ref{thm: price-taking-characterization}, we have
$\sum_n c_n(s^t_n) + \frac{\alpha}{2N\delta} (y^t + (N-1)\delta)^2	\le \sum_n c_n(s^*_n) + \frac{\alpha}{2N\delta} (y^* + (N-1)\delta)^2.$
Rearranging, we have
\begin{align*}
	&\sum_n c_n(s^t_n) +\alpha y^t - \left(\sum_n c_n(s^*) + \alpha y^*\right)
	\le  \frac{\alpha}{2N\delta}(y^t - y^*)\left(2\delta - (y^t + y^*)\right)\\
	=& \frac{\alpha}{2N\delta}[-(y^t-y^*)^2+2(\delta-y^*)(y^t-y^*)]
	\le \frac{\alpha}{2N\delta}[-(y^t-y^*-(\delta-y^*))^2 + (\delta-y^*)^2]\\
	= &\frac{\alpha}{2N\delta}(\delta-y^*)^2\le \frac{\alpha\delta}{2N}.
\end{align*}


\subsection{Proof of Theorem \ref{thm: colo_cost1_taking}} From Proposition \ref{prop: ec2}, we have $\frac{N-1}{N}\alpha \le p^t\le p^* = \alpha$, and $0 \le y^t \le \delta$, which yields
\begin{align*}
	&\mathrm{cost}_o^*(p^*, y^*)-\mathrm{cost}_o(p^t, y^t)
	= p^*(\delta-y^*)+\alpha y^* - \left(p^t(\delta-y^t)+\alpha y^t\right)=(\alpha-p^t)(\delta-y^t)
\end{align*}
Substituting the above bounds for $p^t$ and $y^t$ gives
$
	0\le \mathrm{cost}_o^*(p^*, y^*)-\mathrm{cost}_o(p^t, y^t) \le \frac{\alpha \delta }{N}.
$

\section{price-anticipating tenants}

\subsection{Proof of Theorem \ref{thm: modified_cost}} The proof proceeds in a number of steps. We first show that the payoff function $Q_n$ is a concave and continuous function for each firm $n$. We then establish necessary and sufficient conditions for $\mathbf{b}$ to be an equilibrium; these conditions look similar to the optimality conditions \eqref{eqn: ne1}-\eqref{eqn: ne2} in the proof of Proposition \ref{prop: ec1}, but for a ``modified'' cost function defined according to \eqref{eqn: modified_cost}. We then show the correspondence between these conditions and the optimality conditions for the problem \eqref{eqn: mc1}-\eqref{eqn: mc3}. This correspondence establishes existence of an equilibrium, and uniqueness of the resulting allocation.
\begin{enumerate}
	[label= Step \arabic*:, leftmargin=0cm,itemindent=.5cm,labelwidth=
	\itemindent,labelsep=0cm,align=left]
	\item \emph{ If $\mathbf{b}$ is an equilibrium, and Assumption \ref{asn: cheap_on-site} is satisfied, at least one coordinate of $\mathbf{b}$ is positive.}
	
	By Assumption \ref{asn: cheap_on-site}, $0 < \alpha < \frac{\Sigma_n b_n}{(N-1)\delta}$, hence at least one coordinate of $\mathbf{b}$ must be positive.
	
	\item \emph{ The function $Q_n(\bar{b}_n; \mathbf{b}_{-n})$ is concave and continuous in $\bar{b}_n$, for $\bar{b}_n \ge 0$. } From \eqref{eqn: cost_anticipating} and by plugging $\pb$ into $s_n$ in \eqref{eqn: supply_function}, we have \label{step: concave}
	\begin{align*}
		Q_n(\bar{b}_n; \mathbf{b}_{-n}) &= \sqrt{\frac{(\Sigma_{m \ne n} b_m + \bar{b}_n)\alpha\delta}{N}} - \bar{b}_n - c_n\left(\delta - \frac{\bar{b}_n}{\sqrt{\Sigma_{m \ne n} b_m + \bar{b}_n}} \sqrt{\frac{N\delta}{\alpha}}\right).
	\end{align*}
	
	When $\Sigma_{m\ne n} b_m+ \bar{b}_n >0$, the function $\bar{b}_n / \sqrt{\Sigma_{m \ne n} b_m + \bar{b}_n}$ is a strictly concave function of $\bar{b}_n$ (for $\bar{b}_n \ge 0$). Since $c_n$ is assumed to be convex and nondecreasing (and hence continuous), it follows that $Q_n(\bar{b}_n, \mathbf{b}_{-n})$ is concave and continuous in $\bar{b}_n$, for $\bar{b}_n \ge 0$.
	
	It is easy to show that for $s_n$ to be positive, we need $b_n \le \overline{b_n}$ where
	$\overline{b_n} = \brange.$
	\item \emph{In an equilibrium, $0 \le b_n \le \overline{b_n}, \forall n$.}
	
	Tenant $n$ would never bid more than $\bar{b}_n$ given $\mathbf{b}_{-n}$. If $b_n > \overline{b_n}$, then
	$S(\pb, b_n) = \delta - \frac{b_n}{\sqrt{b_n + \Sigma_{m \ne n} b_m}}\frac{N\delta}{\alpha} <0.$
	so the payoff $Q_n(b_n; \mathbf{b}_{-n})$ becomes negative; on the other hand, $Q_n(\overline{b_n}; \mathbf{b}_{-n}) = 0$.
	
	We specify the following condition when marginal cost of production is not less than the price:
	\begin{equation}
		\forall n, \quad \lmc \le \pb, \quad s_n >0. \label{eqn: p>mc}
	\end{equation}
	This condition is satisfied when tenants are price-taking, in the next step, we show that \eqref{eqn: p>mc} also holds in an equilibrium outcome when tenants are price-anticipating.
	
	\item \emph{ The vector $b$ is an equilibrium if and only if \eqref{eqn: p>mc} is satisfied, at least one component of $\mathbf{b}$ is positive, and for each $n$, $b_n \in [0, \overline{b_n}],$ and the following conditions hold:
	{ \begin{subequations}
		\begin{align}
			\label{eqn: ne3}
			&\text{if }\ 0<b_n \le \overline{b_n}, \quad \frac{1}{2} \left( \rmc + \frac{\alpha}{2N} \right)
			 + \frac{1}{2}\sqrt{ \left(\rmc - \frac{ \alpha}{2N} \right)^2 + \rmc \frac{2 s_n\alpha}{N\delta}} \ge p(\mathbf{b}), \\
			\label{eqn: ne4}
			&\text{if }\ 0 \le b_n < \overline{b_n}, \quad \frac{1}{2} \left( \lmc + \frac{\alpha}{2N} \right)
			+ \frac{1}{2}\sqrt{ \left(\lmc - \frac{ \alpha}{2N} \right)^2 + \lmc \frac{2 s_n\alpha}{N\delta}} \le p(\mathbf{b}).
		\end{align}
	\end{subequations}}
	} By Step 2, $Q_n(b_n; \mathbf{b}_{-n})$ is concave and continuous for $b_n \ge 0$. By Step 3, $b_n \in [0, \overline{b_n}]$. $b_n$ must maximize $Q_n(b_n; \mathbf{b}_{-n})$ over $0 \le b_n \le \overline{b_n}$, and satisfy the following first order optimality conditions:
	{\begin{align*}
		\frac{
		\partial^+Q_n(b_n; \mathbf{b}_{-n})}{
		\partial b_n} &\le 0 , \quad \text{if } \ 0 < b_n \le \overline{b_n}; \\
		\frac{
		\partial^-Q_n(b_n; \mathbf{b}_{-n})}{
		\partial b_n} &\ge 0 , \quad \text{if } \ 0 \le b_n < \overline{b_n};
	\end{align*}}
	Recalling the expression for $\pb$ given in \eqref{eqn: price2}, we have
	{ \begin{align*}
		\frac{1}{2\sqrt{\Sigma_m b_m}} \sqrt{\frac{\alpha\delta}{N}} - 1 + \lmc \frac{1}{p(\mathbf{b})} (1 - \frac{b_n}{2 \Sigma_m b_m}) \le 0,
		\quad \text{if }\ 0 \le b_n < \overline{b_n}; \\
		\frac{1}{2\sqrt{\Sigma_m b_m}} \sqrt{\frac{\alpha\delta}{N}} - 1 + \rmc \frac{1}{p(\mathbf{b})} (1 - \frac{b_n}{2 \Sigma_m b_m}) \ge 0,
		\quad \text{if }\ 0 < b_n \le \overline{b_n}.
	\end{align*}}
	We now note that by \eqref{eqn: price2} and \eqref{eqn: supply_function}, we have :
	$\frac{1}{\sqrt{\Sigma_m b_m}} = \frac{1}{\pb} \sqrt{\frac{\alpha}{N\delta}},$ and $\frac{b_n}{\sqrt{\Sigma_m b_m}} = (\delta - s_n) \sqrt{\frac{\alpha }{N\delta}}. $
	
	Substituting these two equations into the above, we have
	\begin{subequations}
		\begin{align}
			\label{eqn: p>mc1} \frac{1}{2\pb} \frac{\alpha}{N} -1 + \lmc \frac{1}{\pb}\left(1 - \frac{1}{2\pb} \frac{\alpha}{N} \frac{\delta - s_n}{\delta} \right) \le 0. \\
			\label{eqn: p>mc2} \frac{1}{2\pb} \frac{\alpha}{N} -1 + \rmc \frac{1}{\pb}\left(1 - \frac{1}{2\pb} \frac{\alpha}{N} \frac{\delta - s_n}{\delta} \right) \ge 0.
		\end{align}
	\end{subequations}
	To show \eqref{eqn: p>mc} holds, we divide into two cases, when $N \ge 2$, by rearranging \eqref{eqn: p>mc1}, we have
	\begin{align*}
		\lmc \frac{1}{\pb} &\le \frac{2N\pb - \alpha }{ 2N\pb - \alpha \frac{\delta - s_n}{\delta} } \le 1. \\
	\end{align*}
	This is because by Assumption \ref{asn: cheap_on-site}, $2N\pb - \alpha >0$ when $N\ge 2$. Also, we have
	$2N\pb - \alpha\frac{\delta-s_n}{\delta} \ge 2N\pb - \alpha .$
	Hence \eqref{eqn: p>mc} holds for $N \ge 2$.
	
	When $N = 1$, we can simplify \eqref{eqn: p>mc1} further to
	\begin{align*}
		\frac{1}{2\pb} {\alpha} -1 + \lmc \frac{1}{2\pb} \le 0, \
		\Rightarrow \pb \ge \frac{1}{2}\left(\alpha + \lmc\right) \ge \lmc.
	\end{align*}
	The last inequality is because $\alpha \ge \lmc$, otherwise $\pb > \alpha$, but profit maximizing operator will not pay for price more than $\alpha$, contradiction. Hence \eqref{eqn: p>mc} must hold for all $N$. After multiplying through \eqref{eqn: p>mc1}-\eqref{eqn: p>mc2} by $\pb$ and rearranging, we have two quadratic inequalities in terms of $\pb$.	
	Solving the inequalities lead to two sets of conditions of $\pb$ that satisfy the first order optimality conditions, they are:
	\begin{subequations}
		\begin{align}
			\label{eqn: p=mc2} & \text{if }\ 0 \le b_n < \overline{b_n}, \quad \frac{1}{2} \left( \lmc + \frac{\alpha}{2N} \right) \pm \half \sqrt{ \left(\lmc - \frac{ \alpha}{2N} \right)^2 + 4 \lmc \frac{ s_n\alpha}{2N\delta}} &\le \pb\\
			\label{eqn: p=mc1} & \text{if }\ 0 < b_n \le \overline{b_n}, \quad \frac{1}{2} \left( \rmc + \frac{\alpha}{2N} \right) \pm \half \sqrt{ \left(\rmc - \frac{ \alpha}{2N} \right)^2 + 4 \rmc \frac{ s_n\alpha}{2N\delta}} &\ge \pb
		\end{align}
	\end{subequations}
	However, only the conditions with plus sign satisfies \eqref{eqn: p>mc}, the conditions with minus sign violates \eqref{eqn: p>mc} because since
	\[\forall s_n >0, \quad \pb \le \frac{\alpha}{2N} \le \frac{
	\partial^+c_n(0)}{
	\partial s_n} < \frac{
	\partial^-c_n(s_n)}{
	\partial s_n}.\]
	Hence we discard the conditions with minus sign and note that \eqref{eqn: p=mc1}-\eqref{eqn: p=mc2} corresponds to \eqref{eqn: ne3}-\eqref{eqn: ne4}.
	
	Conversely, suppose that $\mathbf{b}$ has at least one strictly positive component, that $0 \le b_n \le \overline{b_n}$, and that $\mathbf{b}$ satisfies \eqref{eqn: p>mc} and \eqref{eqn: ne3}-\eqref{eqn: ne4}. Then we may simply reverse the argument: by Step 2, $Q_n(b_n; \mathbf{b}_{-n})$ is concave and continuous in $b_n \ge 0$, and in this case the conditions \eqref{eqn: ne3}-\eqref{eqn: ne4} imply that $b_n$ maximizes $Q_n(b_n; \mathbf{b}_{-n})$ over $0 \le b_n \le \overline{b_n}$. Since we have already shown that choosing $b_n > \overline{b_n}$ is never optimal for firm $n$, we conclude that $\mathbf{b}$ is an equilibrium, and it is easy to check that in this case condition \eqref{eqn: p>mc} is satisfied.
	
	\item \emph{ If Assumption \ref{asn: cheap_on-site} holds, then the function $\hat{c}_n(s_n)$ defined in \eqref{eqn: modified_cost} is continuous, and strictly convex and strictly increasing over $s_n \ge 0$, with $\hat{c}(s_n) = 0$ \ for $s_n \le 0$.}
	
	$\hat{c}_n(s_n)$ is continuous on $s_n >0$ by continuity of $c_n$ and on $s_n <0$ by definition. We only need to show that $\hat{c}_n(0) = 0$, this is because when $s_n=0$, $c_n(s_n) = 0, s_n\frac{\alpha}{2N} = 0$, and integrating from 0 to $s_n$ is 0. Hence $\hat{c}_n(s_n) = 0$ for $s_n \le 0$.
	
	For $s_n \ge 0$, we simply compute the directional derivatives of $\hat{c}_n$:
	\begin{align*}
		\frac{
		\partial^+\hat{c}_n(s_n)}{
		\partial s_n} &= \frac{1}{2} \left(\frac{\alpha}{2N} + \rmc \right)
		+ \frac{1}{2}\sqrt{\left(\frac{\alpha}{2N} - \rmc\right)^2 + 2\rmc\frac{s_n\alpha}{N\delta}}, \\
		\frac{
		\partial^-\hat{c}_n(s_n)}{
		\partial s_n} &= \frac{1}{2} \left(\frac{\alpha}{2N} + \lmc \right)
		+ \frac{1}{2}\sqrt{\left(\frac{\alpha}{2N} - \rmc\right)^2 + 2\rmc\frac{s_n\alpha}{N\delta}}.
	\end{align*}
	Since $c_n$ is strictly increasing and convex, for $0 \le s_n < \bar{s}_n$, we will have
	\[ 0 \le \frac{
	\partial^+\hat{c}(s_n)}{
	\partial s_n} < \frac{
	\partial^-\hat{c}(\bar{s}_n)}{
	\partial s_n} \le \frac{
	\partial^+\hat{c}(\bar{s}_n)}{
	\partial s_n}.\]
	This guarantees that $\hat{c}_n$ is strictly increasing and strictly convex over $s_n \ge 0$.
	
	\item \emph{ There exists a unique vector $\mathbf{s} \ge 0, y\ge 0$ and at least one scalar $\rho > 0$ such that:
	\begin{subequations}
		\begin{align}
			\label{eqn: mopt1} & \frac{1}{2} \left( \rmc + \frac{\alpha}{2N} \right)
			 + \frac{1}{2}\sqrt{ \left(\rmc - \frac{ \alpha}{2N} \right)^2 + \rmc \frac{2 s_n\alpha}{N\delta}} \ge \rho, \quad \text{if }\ s_n \ge 0; \\
			\label{eqn: mopt2} & \frac{1}{2} \left( \lmc + \frac{\alpha}{2N} \right)
			 + \frac{1}{2}\sqrt{ \left(\rmc - \frac{ \alpha}{2N} \right)^2 + \rmc \frac{2 s_n\alpha}{N\delta}} \le \rho, \quad \text{if }\ s_n >0; \\
			\label{eqn: mopt3} & \frac{\alpha}{N\delta}\left(y + (N - 1)\delta\right) = \rho; \\
			\label{eqn: mopt4} &\sum_n s_n = (\delta - y).
		\end{align}
	\end{subequations}
	The vector $\mathbf{s}$ and $y$ is then the unique optimal solution to \eqref{eqn: mc1}-\eqref{eqn: mc3}. }
	
	By Step 5, since $\hat{c}_n$ is continuous and strictly over the convex, compact feasible region for each $n$, we know that \eqref{eqn: mc1}-\eqref{eqn: mc3} have a unique optimal solution $\mathbf{s}, y$. As in the proof of Proposition \ref{prop: ec1}, form the Lagrangian
	\begin{align*}
		L(\mathbf{s}, y; \rho) &= \sum_n \hat{c}_n(s_n) + \yterm
		+ \rho((\delta-y) -\sum_n s_n).
	\end{align*}
	By assumption \ref{asn: cheap_on-site}, $y>0$, and by the fact that $\hat{c}_n(s_n) = 0$ for $s_n \le 0$, $s_n \ge 0$. there exists a Lagrange multiplier $\rho$ such that $(\mathbf{s}, y, \rho)$ satisfy the stationarity conditions which corresponds to 	%
	\eqref{eqn: mopt1}-\eqref{eqn: mopt3} when we expand the definition of $\hat{c}_n(s_n)$, together with the constraint \eqref{eqn: mopt4}. The fact that $\rho>0$ follows by \eqref{eqn: mopt3} as $y >0$.
	
	\item \emph{ If $\mathbf{s} \ge 0, y \ge 0$ and $\rho > 0$ satisfy \eqref{eqn: mopt1}-\eqref{eqn: mopt4}, then the triple $(\mathbf{b}, \rho, y)$ defined by $b_n = (\delta - s_n)\rho$ is an equilibrium as defined in \eqref{def: ne1} and \eqref{def: ne2}.}
	
	First observe that with this definition, together with \eqref{eqn: mopt4} and the fact that $s_n \ge 0$, we have $b_n \ge 0$ for all $n$. Furthermore, we can show $b_n \le \overline{b_n}$, since $s_n \ge 0$, $b_n \le \rho \delta$, but by \eqref{eqn: mopt3}-\eqref{eqn: mopt4}, we have
	\begin{align}
		\label{eqn: rho} \rho = \frac{\alpha}{N\delta}(y+(N-1)\delta)
		 =\frac{\alpha}{N\delta} (N\delta - \sum_n s_n)
	\end{align}
	Substitute the definition $s_n = \delta - \frac{b_n}{\rho}$ into \eqref{eqn: rho}, we have
	\begin{equation}
		\rho = \frac{\alpha}{N\delta} \frac{\Sigma_n b_n}{\rho} \Rightarrow \rho = \sqrt{\frac{\Sigma_n b_n \alpha}{N\delta}}. \label{eqn: rho2}
	\end{equation}
	Substituting \eqref{eqn: rho2} into $b_n \le \rho\delta$, we have
	$b_n \le \sqrt{\frac{(\Sigma_{m\ne n} b_m + b_n) \alpha\delta}{N}},$
	Solving this inequality we have $b_n \le \overline{b_n}$.
	
	Finally, at least one component of $\mathbf{b}$ is strictly positive, since otherwise we have $s_{n1} = s_{n2}=\delta$ for some $n1\ne n2$, in which case $\Sigma_n s_n >\delta,$ which contradicts \eqref{eqn: mopt4}. (or $s_n=\delta$, $y=0$, contradicting our assumption that $y>0$.)
	
	By Step 4, to check that $\mathbf{b}$ is an equilibrium, we must only check the stationarity conditions \eqref{eqn: ne3}-\eqref{eqn: ne4}. We simply note that under the identification $b_n = \rho(\delta - s_n)$, using \eqref{eqn: rho2} and \eqref{eqn: mopt3}, we have
	\begin{align*}
		y = \sqrt{\frac{\Sigma_n b_n N\delta}{\alpha}} - (N-1)\delta ; \quad
		\rho =\frac{ \Sigma_n b_n}{(N-1)\delta + y} = \pb.
	\end{align*}
	Substitute $\pb$ into \eqref{eqn: mopt1} will correspond to \eqref{eqn: ne3}, and \eqref{eqn: mopt2} implies \eqref{eqn: ne4} and \eqref{eqn: p>mc} because $\lmc \le \rmc$. Thus $(\mathbf{b}, \rho, y)$ is an equilibrium.
	
	\item \emph{ If ($\mathbf{b}, \pb, y)$ is an equilibrium, then there exists a scalar $\rho \ge 0$ such that the vector $\mathbf{b}$ defined by $s_n = S(\pb, b_n)$ satisfies \eqref{eqn: mopt1}-\eqref{eqn: mopt4}.}
	
	We simply reverse the argument of Step 7. Since $\mathbf{b}$ is an equilibrium bids, by \eqref{def: ne2} and $s_n = S(\pb, b_n)$, we have $\sum_n s_n = (\delta - y)$, i.e., \eqref{eqn: mopt4} is satisfied. By Step 4, $\mathbf{b}$ satisfies \eqref{eqn: ne3}-\eqref{eqn: ne4}. Since $y>0$ by Assumption \ref{asn: cheap_on-site}, $0\le s_n < \delta$ for all $n$, let
	\begin{align*}
		\rho = \max \Big\{ \pb, &\half\left(\lmc + \frac{\alpha}{2N}\right) + \half\sqrt{(\rmc-\frac{\alpha}{2N})^2 + \rmc\frac{2s_n\alpha}{N\delta} } \Big\}.
	\end{align*}
	In this case $\rho >0$ and $0 \le b_n \le \overline{b_n}$ for all $n$, so \eqref{eqn: ne4} implies \eqref{eqn: mopt2} by definition of $\rho$, and \eqref{eqn: mopt1} holds by \eqref{eqn: ne3} and the fact that $
	\partial^-c_n(s_n) \le
	\partial^+ c_n(s_n)$ (by convexity).
	
	\item \emph{ There exists an equilibrium $\mathbf{b}$, and for any equilibrium that price is greater than marginal cost, the vector $\mathbf{s}$ defined by $s_n = S(\pb, b_n)$ is the unique optimal solution of \eqref{eqn: mopt1}-\eqref{eqn: mopt4}. }
	
	The conclusion is now straightforward. Existence follows from Steps 6 and 7. Uniqueness of the resulting production vector $\mathbf{s}$, and the fact that $\mathbf{s}$ is an optimal solution to \eqref{eqn: mc1}-\eqref{eqn: mc3}, follows by Steps 6 and 8.
\end{enumerate}

\subsection{Proof of Lemma \ref{lem: mc_bound}} We exploit the structure of the modified cost $\hat{c}_n$ to prove the result. Note that, for all $n$, $ s_n \ge 0$, if we define $G_n(s_n) = \int^{s_n}_0 \sqrt{(\frac{
\partial^+c_n(z)}{
\partial z} - \frac{\alpha}{2N})^2 + \frac{
\partial^+c_n(z)}{
\partial z} \frac{2z\alpha}{N\delta}} dz$, then
\begin{align}
	G_n(s_n) 
	&\ge \int^{s_n}_0 \sqrt{\left(\frac{
	\partial^+c_n(z)}{
	\partial z} - \frac{\alpha}{2N}\right)^2} dz =c_n(s_n) - s_n\frac{\alpha}{2N} \notag.
\end{align}
First inequality is because $z\ge 0$, last equality is because by convexity and Assumption \ref{asn: mc_lowerbound}, we have
 $\mz \ge \frac{
\partial^+c_n(0) }{
\partial s_n} \ge \frac{\alpha}{2N}.$

Hence we have
	$\hat{c}_n(s_n) = \frac{1}{2}\left(c_n(s_n) + s_n\frac{\alpha}{2N}\right) + \frac{1}{2}G_n(s_n) \ge c_n(s_n)$.

On the other hand, notice that $s_n \le \delta$, we have:
\begin{align}
	G_n(s_n) 
	&\le \int^{s_n}_0 \sqrt{\left(\frac{
	\partial^+c_n(z)}{
	\partial z} - \frac{\alpha}{2N}\right)^2 + \frac{
	\partial^+c_n(z)}{
	\partial z} \frac{2\delta\alpha}{N\delta}} dz \notag \\
	&= \int^{s_n}_0 \sqrt{\left(\frac{
	\partial^+c_n(z)}{
	\partial z} + \frac{\alpha}{2N}\right)^2 } dz = c_n(s_n) + s_n \frac{\alpha}{2N} \notag.
\end{align}
Hence we have
	$\hat{c}_n(s_n) = \frac{1}{2}\left(c_n(s_n) + s_n\frac{\alpha}{2N}\right) + \frac{1}{2}G_n(s_n) \le c_n(s_n) + s_n\frac{\alpha}{2N}.$ The bounds for the left and right derivatives can be obtained from taking the left (or right) derivatives at the bounds of $G_n(s_n)$.

\subsection{Proof of Theorem \ref{thm: diff-price-anticipate}} Firstly we will prove one side of the inequality $p^t \le p^a, y^t \le y^a.$ Recall that by the examinging the Lagrangians of the optimizations in  Proposition \ref{prop: ec2} in and Theorem \ref{thm: modified_cost}, we have $p^t \ge \partial^-c_n(s_n^t)/\partial s_n$, $p^t \le \partial^+c_n(s_n^t)/\partial s_n$, $p^a \ge \partial^-\hat{c}_n(s_n^a)/\partial s_n$, $p^a \le \partial^+\hat{c}_n(s_n^a)/\partial s_n$, at the domain where the left or right derivative is defined, and $ p^t = \frac{\alpha}{N\delta}(y^t + (N-1)\delta), p^a = \frac{\alpha}{N\delta}(y^a + (N-1)\delta).$ 
If $y^t > y^a$, then 
$p^t > p^a$. Also, because the total energy reduction $\delta$ is constant, we have $\sum_n s_n^t < \sum_n s_n^a $.

Hence there exist $s_r > 0$ such that $s_r^a > s_r^t$ for some $r \in \{1, \ldots, N\}$. Therefore, by strict convexity of $c_n$ (Assumption \ref{asn: cost_convexity}):
{ \begin{equation}
	p^t \le \frac{
	\partial^+ c_r(s_r^t)}{\partial s_r} < \frac{\partial^- c_r(s_r^a)}{\partial s_r}. \label{eqn: pt_ub}
\end{equation}}
However, by Lemma \ref{lem: mc_bound} we have
$	\frac{
	\partial^- \hat{c}_r(s_r)}{
	\partial s_r} 
	\ge  \frac{
	\partial^-c_r(s_r)}{
	\partial s_r}.
$ Hence, we have
{ \begin{align}
	p^a \ge \frac{\partial^- \hat{c}_r(s^a_r)}{\partial s_r} \ge \frac{\partial^- c_r(s_r^a)}{\partial s_r}. \label{eqn: pa_lb}
\end{align}}
Combining \eqref{eqn: pt_ub} and \eqref{eqn: pa_lb}, we have $p^t < p^a$, contradiction. Hence we have $y^t \le y^a$, and $p^t \le p^a$.

Next we show the other side of the inequality $p^a \le p^t + \frac{\alpha}{2N}, y^a \le y^t + \frac{\delta}{2},$ by the previous part, we have $\sum_n s_n^a \le \sum_n s^t_n$.

Let $n = \argmax_m (s^t_m - s_m^a)$, clearly $s^t_n \ge s^a_n$, otherwise $\sum_n s^t_n < \sum_n s^a_n$, contradiction.

If $s^t_n = s^a_n,$ then $\forall m, s^t_m = s^a_m$, and $y^t = y^a$, then $p^t = p^a$.

If $s^t_n > s_n^a$, then by strict convexity of $c_n$ (assumption \ref{asn: cost_convexity}), and the fact that $s_n^a \ge 0, s^t_n > 0$, we have
\begin{equation}
	\frac{\partial^+ \hat{c}_n(s^a_n)}{s_n} < \frac{\partial^- c_n(s^t_n)}{s_n} \le p^a. \label{eqn: pt_lb}
\end{equation}
Also, by Lemma \ref{lem: mc_bound}, we have
$
	\frac{\partial^+ \hat{c}_n(s_n)}{\partial s_n} 
	\le  \frac{\partial^+c_n(s_n)}{
	\partial s_n} + \frac{\alpha}{2N},
$
this gives us
\begin{equation}
	p^a \le \frac{\partial^+ \hat{c}_n(s^a_n)}{\partial s_n} \le \frac{\partial^+ c_n(s^a_n)}{\partial s_n} + \frac{\alpha}{2N}. \label{eqn: pa_ub}
\end{equation}
Combining \eqref{eqn: pt_lb} and \eqref{eqn: pa_ub}, we have
$p^a < p^t + \frac{\alpha}{2N}.$
Hence we have
\begin{align*}
	\frac{\alpha}{N\delta}(y^a + (N -1)\delta) < \frac{\alpha}{N\delta}(y^t + (N-1)\delta) + \frac{\alpha}{2N}, \Rightarrow
	y^a  < y^t + \frac{\delta}{2}.
\end{align*}

\subsection{Proof of Theorem \ref{thm: on-site_gap}} Given any $\eps >0$, let $\eps' = \frac{1}{2}\eps$. Consider the following set of cost function:
{\[ c_1(s_1) =
\begin{cases}
	\frac{\alpha}{2N}s_1 , & \text{ if } s_1 < \eps' ; \\
	\alpha(1- \frac{3\eps'}{2N\delta})s_1+C_1 , & \eps' \le s_1 \le \delta - \eps'; \\
	2\alpha s_1 + C_2, & s_1 > \delta - \eps'
\end{cases}
\]}
where $C_1, C_2$ are constants that make $c_1$ continuous\footnote{$C_1 = -\alpha\eps'(\frac{ (2N-1)\delta - 3\eps'}{2N\delta} )$, and $C_2 = -\frac{\alpha}{N\delta}(N\delta^2 +\delta\eps' - 3\eps')$}, then $c_1$ is piece-wise linear and convex. Also, $\forall m \ne 1, c_m(s_m) = 2\alpha s_m$. It is easy to see that $s_1^* = \delta - \eps'$ and $y^* = \eps'$ is the optimal allocation.

Let $s^a_1 = \eps', y^a = \delta - \eps',$ and $\forall m\ne 1, s^a_m = 0$, we claim that $(\mathbf{s}^a, y^a)$ is the unique optimal solution to \eqref{eqn: mc1}-\eqref{eqn: mc3}. To see this, let $\rho = \alpha(1 - \eps / (N\delta))$, then,
\begin{subequations}
	\begin{align}
		\frac{\alpha}{N\delta}(y^a + (N-1)\delta) &= \rho; \quad  \sum_n s^a_n = \delta - y^a; \\
		\frac{\partial^- \hat{c}_1(s^a_1)}{\partial s_1} &\le \rho ; \quad
		\frac{\partial^+ \hat{c}_1(s^a_1)}{\partial s_1} \ge \rho ; \quad
		\frac{\partial^+\hat{c}_m(0)}{\partial s_m}  \ge \rho , \quad \forall m \ne 1.		 
	\end{align}
\end{subequations}
where the second inequality is because if we let $H_n$ be the term under square root for $\rmmc$, then
\begin{align*}
	H_n &= \sqrt{\left(\rmc - (\frac{\alpha}{2N} - \frac{\alpha}{N}\frac{s_n}{\delta})\right)^2 + (\frac{\alpha^2}{N^2}\frac{(\delta+s_n)(\delta-s_n)}{\delta^2})} \\
	&\ge \rmc - (\frac{\alpha}{2N} - \frac{\alpha}{N}\frac{s_n}{\delta}).
\end{align*}
Note that $\mrmc = \frac{1}{2} (\rmc + \frac{\alpha}{2N}) + \frac{1}{2}H_n$. Hence we have
$\frac{\partial^+ \hat{c}_1(s^a_1)}{\partial s_1} \ge \frac{\partial^+c_1(s_1^a)}{\partial s_1} + \frac{\alpha s_1}{2N\delta} = \rho.$
These conditions correspond to \eqref{eqn: mopt1}-\eqref{eqn: mopt4}, so we conclude that $(\mathbf{s^a}, y^a)$ is the unique optimal solution to \eqref{eqn: mc1}-\eqref{eqn: mc3}. Hence $y^a - y^* = \delta - 2\eps' = \delta - \eps.$

\subsection{Proof of Theorem \ref{thm: welfare_loss2}} As $(\mathbf{s}^*, y^*)$ is a feasible solution to \eqref{eqn: mc0}, by Theorem \ref{thm: modified_cost}, we have
\begin{align}
	\label{eqn: cost_compare1} \sum_n \hat{c}_n(s_n^a) + \frac{\alpha}{2N\delta}(y^a + (N -1)\delta)^2
	 \le \sum_n \hat{c}_n(s^*_n) + \frac{\alpha}{2N\delta}(y^* + (N -1)\delta)^2 .
\end{align}
Rearranging, we have
$\sum_n \hat{c}_n(s^a_n) + \alpha y^a - \left(\sum_n \hat{c}_n(s_n^*) + \alpha y^*\right)
	\le  \frac{\alpha}{N}\left( (y^a - y^*) (1 - \frac{y^a+y^*}{2\delta})\right). $
By Corollary \ref{cor: diff-price-anticipate2} and the fact that $y^* \le \delta, y^a \le \delta,$ both terms in the brackets are positive, hence right-hand-side expression is maximized when $y^* \rightarrow 0^+$ and $y^a = \delta$, hence
\begin{align}
	\label{eqn: cost_compare2} \left( \sum_n \hat{c}_n(s^a_n)+ \alpha y^a\right) - \left(\sum_n \hat{c}_n(s_n^*) + \alpha y^*\right) \le \frac{\alpha\delta}{2N}.
\end{align}
However, by Lemma \ref{lem: mc_bound}, we have
$\sum_n \hat{c}_n(s^*_n) \le \sum_n c_n(s^*_n) + \frac{\alpha}{2N}(\sum_n s_n) \le \sum_n c_n(s^*_n) + \frac{\alpha\delta}{2N}$; and
$\sum_n \hat{c}_n(s^a_n) \ge \sum_n c_n(s^a_n). $
Substituting the above relations into \eqref{eqn: cost_compare2} and rearranging, we have the desired result.

\subsection{Proof of Theorem \ref{thm: diff-payment}} First, we compare the cost by operator between the price-taking and price anticipating cases, by definition \eqref{eqn: operator_cost} and rearranging, we have 
$  \mathrm{cost}_o(p^a, y^a) - \mathrm{cost}_o(p^t, y^t) =  (p^a - p^t) \left({\delta - y^t} \right) + \left(\alpha - {p^a}\right) (y^a - y^t).$
By the fact that $p^a=\frac{\alpha}{N\delta}(y^a + (N-1)\delta)$ (shown in Theorem \ref{thm: diff-price-anticipate}) and the fact that $0 \le y^a \le \delta$, we have
\begin{equation}
	\alpha\left(\frac{N-1}{N}\right) \le p^a \le \alpha . \label{eqn: pa_range}
\end{equation}
By the upper bound of $p^a$ in \eqref{eqn: pa_range} and the upper bounds of $p^t, y^t$ in Theorem \ref{thm: diff-price-anticipate}, we have
\begin{equation}
	\mathrm{cost}_o(p^a, y^a) - \mathrm{cost}_o(p^t, y^t) \ge 0. \label{eqn: anticipating_v_taking}
\end{equation}
Similarly, using the lower bound of $p^a$ in \eqref{eqn: pa_range} and the upper bounds of $p^a, y^a$ in Theorem \ref{thm: diff-price-anticipate}, we have
\begin{align*}
	 \mathrm{cost}_o(p^a, y^a) - \mathrm{cost}_o(p^t, y^t)
	\le  \left(\frac{\alpha}{2N} \right)\cdot \left({\delta}\right) + \left(\alpha \cdot \frac{1}{N}\right)\left(\frac{\delta}{2} \right)= \frac{\alpha\delta}{N}.
\end{align*}

Second, we compare the cost by the operator to the social optimal. Since the energy reduction goal $\delta$ is the same, by Proposition \ref{prop: ec2} and Corollary \ref{cor: diff-price-anticipate2}, we have $p^t \le p^*$ and $p^a \le p^*$. Hence we have $\mathrm{cost}_o(p^t, y^t) \le \mathrm{cost}_o(p^a, y^a) \le \mathrm{cost}_o(p^*, y^*).$
Furthermore,
\begin{align}
	&\cost_o(p^*, y^*) - \cost_o(p^t, y^t) = \alpha\delta - (p^t(\delta - y^t) + \alpha y^t)\notag \\
	= &(\alpha - p^t)(\delta - y^t) = \alpha \left( \frac{\delta - y^t}{N\delta} \right)(\delta - y^t) \le \frac{\alpha\delta}{N}. \label{eqn: taking_v_optimal}
\end{align}
Lastly by \eqref{eqn: anticipating_v_taking} and \eqref{eqn: taking_v_optimal}, we have
$\cost(p^*, y^*) -\cost(p^a, y^a) \le \cost(p^*, y^*) - \cost(p^t , y^t) \le \frac{\alpha\delta}{N}.$

\section{Results for \ouralgvdr}
\subsection{Proof Sketch of Theorem \ref{thm: vdr_ne_taking}}
Theo proof is similar to that of Theorem \ref{thm: price-taking-characterization}, which uses Proposition \ref{prop: ec1}, note that in the VDR case, we can change $N\delta$ in the proof of Theorem \ref{thm: price-taking-characterization} to $\Sigma_n D_n$, and interpret the variable $y$ as $\Sigma_n D_n - d $, $\alpha$ as $u$ and $\gn$ as $1/N$ in the proof of Theorem \ref{thm: price-taking-characterization}. 

\subsection{Proof Sketch of Theorem \ref{thm: vdr_ne_anticipating}}
Theo proof is similar to that of Theorem \ref{thm: modified_cost}. note that in the VDR case, we can change $N\delta$ in the proof of Theorem \ref{thm: price-taking-characterization} to $\Sigma_n D_n$, and interpret the variable $y$ as $\Sigma_n D_n - d $, $\alpha$ as $u$ and $\gn$ as $1/N$ in the proof of Theorem \ref{thm: modified_cost}. 

\subsection{Proof of Lemma \ref{lemma: vdr_mc_bound}}
	For the bound on the magnitude of the modified cost, we exploit the structure of the modified cost $\hat{c}_n$ to prove the result. Note that, for all $n$, $ s_n \ge 0$, if we define $G_n(s_n) = \int_{0}^{s_n}\sqrt{\left( \frac{\partial^+c_n(z)}{\partial z} - \frac{\gn u}{2} \right)^2 + 2\frac{\partial^+c_n(z)}{\partial z} \frac{zu}{\Sigma_i D_i}}$, then
\begin{align}
	G_n(s_n) 
	&\ge \int^{s_n}_0 \sqrt{\left( \frac{\partial^+c_n(z)}{\partial z} - \frac{\gn u}{2} \right)^2} dz =c_n(s_n) - s_n\frac{u\gn}{2} \notag.
\end{align}
First inequality is because $z\ge 0$, last equality is because by convexity and Assumption \ref{asn: vdr-mc-lbound}, we have
 $\mz \ge \frac{
\partial^+c_n(0) }{
\partial s_n} \ge \frac{u\gn}{2}.$

Hence we have
	$\hat{c}_n(s_n) = \frac{1}{2}\left(c_n(s_n) + s_n\frac{u\gamma}{2}\right) + \frac{1}{2}G_n(s_n) \ge c_n(s_n)$.

On the other hand, notice that $s_n \le D_n$, we have:
\begin{align}
	G_n(s_n) 
	&\le \int^{s_n}_0 \sqrt{\left(\frac{\partial^+c_r(s_n)}{\partial s_n} - \frac{u \gn}{2}\right)^2 + 2\frac{\partial^+c_n(s_n)}{\partial s_n}\gn u} dz = c_n(s_n) + s_n \frac{\alpha}{2N} \notag.
\end{align}
Hence we have
	$\hat{c}_n(s_n) = \frac{1}{2}\left(c_n(s_n) + s_n\frac{\alpha}{2N}\right) + \frac{1}{2}G_n(s_n) \le c_n(s_n) + s_n\frac{u\gn}{2}.$ The bounds for the left and right derivatives can be obtained from taking the left (or right) derivatives at the bounds of $G_n(s_n)$.

\subsection{Proof of Theorem \ref{thm: vdr_loss_taking}}

	We can combine \eqref{eqn: vdr-quantity} with \eqref{eqn: vdr-price2} to eliminate the $\sqrt{\sumn b_i}$ term to get a relation between market price and the vdr-quantity decided by the profit maximizing operator:
	\begin{equation}
		p = \frac{u}{\sumn D_i} (\sumn D_i - d)
		\label{eqn: price_d}
	\end{equation}

	By the characterization theorem, we have 
	$ ud^* - \frac{u{d^*}^2}{2\sum_n D_n} - \sum_n c_n(s_n^*) \le ud^t - \frac{u{d^t}^2}{2\sum_n D_n} - \sum_n c_n(s_n^t). $
	Rearranging, we have 
	\begin{align*} 
		ud^* - \sum_n c_n(s_n^*) &\le ud^t - \sum_n c_n(s_n^t) + \frac{u ({d^*}^2 - {d^t}^2)}{2\sum_n D_n} \\
		&\le ud^t - \sum_n c_n(s_n^t) + \frac{u\sum_n {d^*}^2}{2\sum_n D_n} 
	\end{align*}
where the last inequality is due to the fact that $d^t \ge 0.$
%
%

\subsection{Proof of Theorem \ref{thm: vdr_loss_anticipating}}
	By Theorem \ref{thm: vdr_ne_anticipating}, we have 
		$ud^a - \frac{u{d^a}^2}{2} - \sum_n \hat{c}_n(s_n^a) \ge ud^a - \frac{u{d^*}^2}{2} - \sum_n \hat{c}_n(s_n^*)$
	Using Lemma \ref{lemma: vdr_mc_bound}, and rearranging, we have 
	\begin{align*}
		& ud^a - \sum_n c_n(s_n^a) \\
	\ge & ud^* - \sum_n c_n(s_n^*) - \frac{u({d^*}^2 - {d^a}^2)}{2\sum_n D_n} - \sum_n s_n^*\frac{\gn u}{2} \\ 
	\ge & ud^* - \sum_n c_n(s_n^*) - \frac{u \sum_n D_n }{2} - \sum_n D_n \frac{\gn u}{2}\\
	=	& ud^* - \sum_n c_n(s_n^*) - \frac{u}{2}\sum_n D_n(1 + \gn).
	\end{align*}
	where the first inequality is because $c_n(s_n^a) \le \hat{c}_n(s_n^a)$, and $\hat{c}_n(s_n^*) \le c_n(s_n^*) + s_n^* \frac{\gn u}{2}$, and the second inequality is becuase $s_n^* \le D_n$.

\subsection{Proof of Proposition \ref{prop: vdr_price_taking}}
	The Lagrangian of the welfare maximization problem \eqref{eqn: vdr-utility-maximization} is 
	\[ L(\mathbf{s}, d; \mu, \bar{\lambda}, \underline{\lambda}) = ud - \sumn c_i(s_i) + \mu(\sumn s_i - d) + \sumn \underline{\lambda}_is_i + \sumn \bar{\lambda}_i(D_i - s_i). \]
	By constraint qualification, the optimal primal dual solutions $(\mathbf{s}, y; \mu)$ satisfies the KKT conditions 
	\begin{align*}
		& \mu^* = u, \\
		& \lmc \le \mu^*, \text{ if } 0 < s_n \le D_n \\
		& \rmc \ge \mu^*, \text{ if } 0 \le s_n < D_n. 
	\end{align*}
	Hence the market clearing price in the optimal allocation should be $p^* = u$.
	Now consider the market clearing price for price taking tenants, from \eqref{eqn: price_d}, we know that 
	$p^t = u - \frac{u d^t}{\sum_n D_n} \le u = p^*$. Similarly, by Theorem \ref{thm: vdr_ne_taking} and looking at the Lagrangian of \eqref{eqn: mc0_taking}, we have $ \frac{\partial^- c_n(s_n^t)}{\partial s_n} \le p^t$ for all $0< s_n^t \le D_n$, hence for all $n$, such that $s_n^t >0$ and $s_n^* < D_n$, we have 
	\[ \frac{\partial^- c_n(s_n^t)}{\partial s_n} \le p^t \le p^* \le \frac{\partial^+ c_n(s_n^*)}{\partial s_n}, \]
	hence $s_n^t \le s_n^*$ for all such $n$, on the other hand, if $s_n^t = 0$ or $s_n^* = D_n$, we also have $s_n^t \le s_n^*$, hence $d^t = \sum_n s_n^t \le \sum_n s_n^* = d^*.$  Finally, by the fact that $d^t \le d^*$ and \eqref{eqn: price_d}, we have 
	\[ p^t = u - \frac{u d^t}{\sum_n D_n} \ge u - \frac{u d^*}{\sum_n D_n} = \left(1 - \frac{d^*}{\sum_n D_n}\right)p^*. \]

\subsection{Proof of Theorem \ref{thm: vdr_price_diff}}

	 Firstly we will prove one side of the inequality $p^t \le p^a, d^t \ge d^a.$ 
	 We can prove this by contradiction. Suppose $d^t < d^a$, then by \eqref{eqn: price_d}, $p^t > p^a$. Also, $\sum_n s_n^t < \sum_n s_n^a $.

	 Hence there exist $s_r^a > 0$ such that $s_r^a > s_r^t$ for some $r \in \{1, \ldots, N\}$. Therefore, by the stationarity of the Lagrangian of \eqref{eqn: mc0_taking} and strict convexity of $c_n$ (Assumption \ref{asn: cost_convexity}): 
	 \begin{equation}
	 	p^t \le \frac{
	 	\partial^+ c_r(s_r^t)}{\partial s_r} < \frac{\partial^- c_r(s_r^a)}{\partial s_r} \label{eqn: vdr_pt_ub} 
	 \end{equation}
	 However, by the stationarity of the Lagrangian of \eqref{eqn: mc0_anticipating} and Lemma \ref{lemma: vdr_mc_bound}, we have
	 \begin{align}
	 	p^a \ge \frac{\partial^- \hat{c}_r(s_r^a)}{\partial s_r} \ge \frac{\partial^- c_r(s_r^a)}{\partial s_r}. 
		\label{eqn: vdr_pa_lb} 
	 \end{align}
	 Combining \eqref{eqn: pt_ub} and \eqref{eqn: pa_lb}, we have $p^t < p^a$, contradiction. Hence we have $y^t \le y^a$, and $p^t \le p^a$.

	 Next we show the other side of the inequality $p^a \le p^t + \frac{u \gamma }{2}, d^a \le d^t - \frac{D}{2},$ by the previous part, we have $\sum_n s_n^a \le \sum_n s^t_n$.

	 Let $n = \argmax_m (s^t_m - s^a_m)$, clearly $s^t_n \ge s^a_n$, otherwise $\sum_n s^t_n < \sum_n s^a_n$, contradiction.

	 If $s^t_n = s^a_n,$ then $\forall m, s^t_m = s^a_m$, and $d^t = d^a$. By \eqref{eqn: price_d}, $p^t = p^a$.

	 If $s^t_n > s^a_n$, then by stationary condition of the Lagrangian of \eqref{eqn: mc0_taking} and strict convexity of $c_n$ (assumption \ref{asn: cost_convexity}), and the fact that $s^a_n \ge 0, s^t_n > 0$, we have 
	 \begin{equation}
	 	\frac{\partial^+ c_n(s_n^a)}{\partial s_n} < \frac{\partial^- c_n(s^t_n)}{s_n} \le p^t. \label{eqn: pt_lb} 
	 \end{equation}
	 Also, by Lemma \ref{lemma: vdr_mc_bound} and stationary condition of Lagrangian of \eqref{eqn: mc0_anticipating}, we have  definition of 	  	       \begin{equation}
	 	p^a \le \frac{\partial^+\hat{c}_n(s_n^a)}{\partial s_n} \le \frac{\partial^+\hat{c}_n(s_n^a)}{\partial s_n} + \frac{\gn u}{2}. \label{eqn: pa_ub} 
	 \end{equation}
	 Combining \eqref{eqn: pt_lb} and \eqref{eqn: pa_ub}, we have
	 $p^a < p^t + \frac{\gn u}{2} \le p^t + \frac{u\gamma}{2}.$ 
	 Substitute the above relation into \eqref{eqn: price_d}, we have 
	 \begin{align*}
	 	u - \frac{ud^a}{\sum_n D_n} &< u - \frac{ud^t}{\sum_n D_n} + \frac{u\gamma}{2}, \\
	 	d^a & > d^t - \frac{D}{2}, 
	 \end{align*}
	 the last inequality is because $D = \max_n D_n = (\sum_n D_n) \gamma.$
\subsection{Proof of Theorem \ref{thm: vdr_operator_profit}}

	Firstly, by theorem \ref{thm: vdr_price_diff}, $d^a \le d^t$, and $p^a \ge p^t$, hence $U_o(p^a, d^a) \le U_o(p^t, d^t)$.  Furthermore, 
	\begin{align}
		&U_o(p^t, d^t) - U_o(p^a, d^a) = (u-p^t)d^t - (u-p^a)d^a \notag \\
		\label{eqn: cost_o1} = & (u-p^t)(d^t - d^a) + d^a(p^a - p^t). 
	\end{align}
	By theorem \ref{thm: vdr_price_diff}, we have $d^t \le d^a + D/2$, $p^a \le p^t + u\gamma/2$, and by the fact that $d^a \le \sum_n D_n$, we have 
	\begin{equation}
		 U_o(p^t, d^t) - U_o(p^a, d^a) \le u\cdot\frac{D}{2} + (\sum_n D_n)\frac{u\gamma}{2} = uD.
	\end{equation}
	%
	%

\end{document}